\documentclass[aps,rmp, reprint]{revtex4-2}

\newcommand{\tw}{\ensuremath{t_\mathrm{w}}\xspace}
\newcommand{\twone}{\ensuremath{t_\mathrm{w1}}\xspace}
\newcommand{\twtwo}{\ensuremath{t_\mathrm{w2}}\xspace}

\newcommand{\Tc}{\ensuremath{T_\mathrm{g}}\xspace}
\newcommand{\Tg}{\ensuremath{T_\mathrm{g}}\xspace}
\newcommand{\Tm}{\ensuremath{T_\mathrm{m}}\xspace}
\newcommand{\teff}{\ensuremath{t^\mathrm{eff}_\mathrm{w}}\xspace}

\newcommand{\xiZ}{\ensuremath{\xi_\mathrm{Zeeman}}\xspace}
\newcommand{\Ns}{\ensuremath{N_\text{s}}\xspace}
\newcommand{\Ds}{\ensuremath{D_\text{s}}\xspace}
\newcommand{\ximN}{\ensuremath{\xi_\mathrm{micro}^\mathrm{native}}\xspace}
\newcommand{\xim}{\ensuremath{\xi_\mathrm{micro}}\xspace}
\newcommand{\ximZFC}{\ensuremath{\xi_\mathrm{micro}^\mathrm{ZFC}}\xspace}
\newcommand{\ximTRM}{\ensuremath{\xi_\mathrm{micro}^\mathrm{TRM}}\xspace}
\newcommand{\ximT}{\ensuremath{\xi_\mathrm{micro}^\perp}\xspace}
\newcommand{\ximP}{\ensuremath{\xi_\mathrm{micro}^\parallel}\xspace}
\newcommand{\qEA}{\ensuremath{q_\text{\tiny EA}}\xspace}
\newcommand{\zetaJ}{\ensuremath{\zeta^\mathrm{jump}}\xspace}

\newcommand{\Tmeas}{\ensuremath{T_\mathrm{m}}\xspace}

\newcommand{\disorderav}[1]{\ensuremath{\overline{#1}}}

\newcommand{\dd}{\text{d}}

\newcommand{\changes}[1]{{#1}}
\newcommand{\changesbis}[1]{{#1}}

\usepackage{bm}
\usepackage{array}
\usepackage{xcolor}
\usepackage{float}
\usepackage{graphicx}
\usepackage{amsmath}
\usepackage{amssymb}
\usepackage{amsfonts}
\usepackage{color}
\usepackage{xspace}
\usepackage{multirow}
\usepackage{dcolumn}
\usepackage{hyperref}
\hypersetup{
    colorlinks=true,
    linkcolor=blue,
    filecolor=magenta,      
    urlcolor=blue,
    citecolor=blue,
    pdfpagemode=FullScreen,
}

\begin{document}

\title{Spin-glass dynamics: experiment, theory and simulation}

\author{E.~D.~Dahlberg}\affiliation{School of Physics and Astronomy, The University of Minnesota, Minneapolis, Minnesota 55455, USA}

\author{I.~González-Adalid Pemartín}\affiliation{Istituto per le Applicazioni del Calcolo, CNR, Rome, Italy}

\author{E.~Marinari}\affiliation{Dipartimento di Fisica, Sapienza Università di Roma, and CNR-Nanotec, Rome Unit, and INFN, Sezione di Roma, 00185 Rome, Italy}

\author{V.~Martin-Mayor}\affiliation{Departamento de Física Teórica, Universidad Complutense, 28040 Madrid, Spain}
 
\author{J.~Moreno-Gordo}\affiliation{Departamento de Didáctica de las Ciencias Experimentales y de las Matemáticas, Universidad de Extremadura, 10004 Cáceres, Spain}

\author{R.L.~Orbach}
\affiliation{Texas Materials Institute, The University of Texas at Austin, Austin, Texas  78712, USA}\email{Contact author: orbach@utexas.edu}

\author{I.~Paga}\affiliation{Department of Computing Sciences,Bocconi University, 20136 Milano, Italy}

\author{G.~Parisi}\affiliation{Dipartimento di Fisica, Sapienza Università di Roma, and CNR-Nanotec, Rome Unit, and INFN, Sezione di Roma, 00185 Rome, Italy}

\author{F.~Ricci-Tersenghi}\affiliation{Dipartimento di Fisica, Sapienza Università di Roma, and CNR-Nanotec, Rome Unit, and INFN, Sezione di Roma, 00185 Rome, Italy}

\author{J.J.~Ruiz-Lorenzo}\affiliation{Departamento de Física, Universidad de Extremadura, 06006 Badajoz, Spain}\affiliation{Instituto de Computación Científica Avanzada (ICCAEx), Universidad de Extremadura, 06006 Badajoz, Spain}

\author{D.~Yllanes}
\affiliation{Fundación ARAID, Diputación General de Aragón, 50018 Zaragoza, Spain}
\affiliation{Instituto de Biocomputación y Física de Sistemas Complejos (BIFI) and Zaragoza Scientific Center for Advanced Modeling (ZCAM), 50018 Zaragoza, Spain}
\email{Contact author: david.yllanes@bifi.es}

\date{\today{}}

\begin{abstract}
The study of spin-glass dynamics, long considered the paradigmatic complex system, has reached important milestones. The availability of high-quality single crystals has allowed the experimental measurement of spin-glass coherence lengths of almost macroscopic dimensions, while the advent of special-purpose massive computers ---the Janus Collaboration--- enables dynamical simulations that approach experimental time and length scales.
This review provides an account of the quantitative convergence of these two avenues of research, with precise experimental measurements
of the expected scaling laws and numerical reproduction
of classic experimental results, such as memory and rejuvenation.
The article opens with a brief review of the defining spin-glass properties, randomness and frustration, and their experimental consequences. These apparently simple characteristics are shown to generate rich and complex physics. Models are introduced that enable quantitative dynamical descriptions, either analytically or through simulations.  The many theoretical pictures of the low-temperature phase are reviewed.  After a summary of the main numerical results
in equilibrium, paying particular attention to the concept of temperature chaos, this review examines off-equilibrium dynamics in the absence of a magnetic field and shows how it can be related to the structure of the equilibrium spin-glass phase through the fluctuation-dissipation relations. The nonlinear response at a given temperature is then developed, including experiments and scaling in the vicinity of the spin-glass transition temperature $\Tg$.  The consequences of temperature change ---including temperature chaos, rejuvenation, and memory--- are reviewed. The interpretation of these phenomena requires identifying several length scales relevant to dynamics, which, in turn, generate new insights. Finally, issues for future investigations are introduced, including what is to be ``nailed down'' theoretically, why the Ising Edwards-Anderson model is so successful at modeling spin-glass dynamics, and experiments yet to be undertaken.
\end{abstract}


\maketitle
\makeatletter
\def\l@subsubsection#1#2{}
\makeatother
\tableofcontents{}

\section{Introduction (and a little history)}
\label{sect:intro}
\changes{The study and broad applications of spin-glass dynamics have advanced remarkably since the last major review~\cite{binder:86}, nearly forty years ago.  Subsequent advances have been summarized  by~\citet{fischer:93,young:98,castellani:05,kawamura:15,mydosh:15,charbonneau:23,altieri:24,vincent:24}.  This work aims to bring the experiments and theoretical accomplishments of the last decade up to date  and to point to future opportunities.}  
The give and take between experiment, theory, and simulations in the field constitutes a perfect example of the “three legs of scientific research.”  The accomplishments of the Uppsala and Saclay groups were responsible for the surge of interest in the field in the late 20th and early 21st century.  The following advent of powerful computational facilities custom-designed for Ising spin glasses produced simulations, both in and out of equilibrium, that have enabled direct comparisons with experimental results. These in turn have inspired new experiments that have, in combination with simulations, led to a deep quantitative understanding of spin-glass dynamics.

\changes{An important theme in our discussion will be the distinction between equilibrium and nonequilibrium results. This distinction looks somewhat artificial 
from an experimental point of view, because a temperature exists ---the so-called glass temperature $\Tg$--- such that thermal equilibrium is essentially unreachable for any $T<\Tg$. As we shall discuss at length, and at variance with the open debate for other glass formers~\cite[see, \emph{e.g.},][]{cavagna:09,biroli:21}, in the case of spin glasses the physical origin of the slow dynamics below $\Tg$ is known to be a thermodynamic phase transition. It is, therefore, possible to formulate a theory for a proper spin-glass phase for $T<\Tg$~\cite{parisi:23}. This equilibrium phase is  reachable in equilibrium simulations only for samples of fairly small linear size $L$ and not at all in experiments to date. Nevertheless, the knowledge of equilibrium spin-glass structures will be important for our endeavor, because the off-equilibrium dynamics can be viewed as an endless journey toward them.}

\changes{We shall compare in some detail the results 
obtained experimentally with the outcome of simulations 
for the extremely simplified Edwards-Anderson model. The 
conceptual framework justifying this leap will be the 
renormalization group and the resulting property of
universality~\cite{wilson:75,zinn-justin:05,parisi:88}. 
Given a small number of constraints (essentially the space 
dimension and some crucial symmetries), and provided that
the correct scaling variables are employed, quantitative 
agreement can be reached between experimental results 
and the outcome of toy models representative of the appropriate \emph{universality class}. 
In this context, the 
glassy coherence length, $\xi$, will be crucial to define 
the natural units for the parameters.}

\subsection{What is a spin glass?}\label{subsect:intro-1}
Over the years since the original experiments of \textcite{cannella:72}, the combination of randomness mixed with frustration \cite{toulouse:77} has been identified as the essential property of a spin glass \cite{anderson:70}. Simply defined, frustration means that not all exchange-coupled spin orientations can be energetically satisfied for any configuration.  While each property {\it by itself} results in a different class (\emph{e.g.}, an antiferromagnet in a triangular lattice is fully frustrated), the combination of the two is defining for spin glasses so far as we know.  Spin-glass behavior is found in both metals and insulators with competing ferro- and anti-ferromagnetic couplings.  The RKKY oscillating coupling \cite{ruderman:54, kasuya:56, yosida:57} in dilute magnetic alloys \cite[\emph{e.g.}, CuMn,][]{goodenough:55,kanamori:57} and competing exchange couplings in insulators \cite[\emph{e.g.}, CdCr$_{1.7}$In$_{0.3}$S${_4}$,][]{vincent:24} are stereotypical examples.

What is so wonderful about this ``simple'' set of requirements is that \cite{anderson:89}: ``the key result here is the beautiful revelation of the structure of the randomly `rugged landscape' that underlies many complex optimization problems [\dots] Physical spin glasses and the SK model \cite{sherrington:75} are only a
jumping-off point for an amazing cornucopia of wide-ranging
applications of the same kind of thinking.''

The purpose of this review is to explore the dynamics of the spin-glass state.  The richness of the ensuing consequences will prove remarkable.  What will be even more impressive is the applicability of these results to a great variety of physical and even sociological systems. As noted by \citet{parisi:23}, there is an interplay between disorder and fluctuations in physical systems from atomic to planetary scales.  Theories based on multiple equilibria are present in many disciplines: the ``punctuated equilibria'' of evolutionary theory \cite{EldredgeGould1972}; in ecosystems, glaciations and geological eras; memory through associative neural networks \cite{Hopfield1982}; the physics of structural glasses \cite{Goldstein1969}; the multiple equilibria of economics~\cite{Dybvig2022}, etc.

The utility of solutions of the spin-glass problem is precisely this.  Over the past five decades, from their experimental properties, together with somewhat simplified theoretical models (though with highly complex mathematics!), the field has amassed Anderson's ``cornucopia'' of concepts, solutions, and insights. The authors of this review hope that, by exhibiting these results, the reader will be able to understand and attack other complex systems of interest over a wide variety of applications.

\changes{It is useful to quote some of the fields where these techniques are already of paramount importance. The interested reader can easily check that the list was already long 40 years ago \cite{mezard:87}, and today it is rather impressive.  We will not quote all individual contributions, which would lead us to repeat here the index of \citet{charbonneau:23}, without great gain. Optimization theory probably needs to be mentioned first. The importance of the spin-glass approach in this field has become paramount  \cite[for a very remarkable analysis see, for example,][]{gamarnik:25}. The theories of communication and information are very close by. Glasses (even quantum) and the theory of jamming are also important in this list. The study of polymers has seen spin-glass theory playing a crucial role. Random lasers are a relatively new subject that was very prominent in the motivation of the Nobel prize to Giorgio Parisi. Crucial applications are also found in the biological world, on many spatial and time scales: neural networks, molecules, active matter and ecological systems are among these. Lastly we quote economics and finance, where in many contexts spin-glass analogies are crucial.}

\subsection{Experimental milestones}\label{subsect:intro-2}
\begin{figure}[tb]
    \centering
    \includegraphics[width=\linewidth]{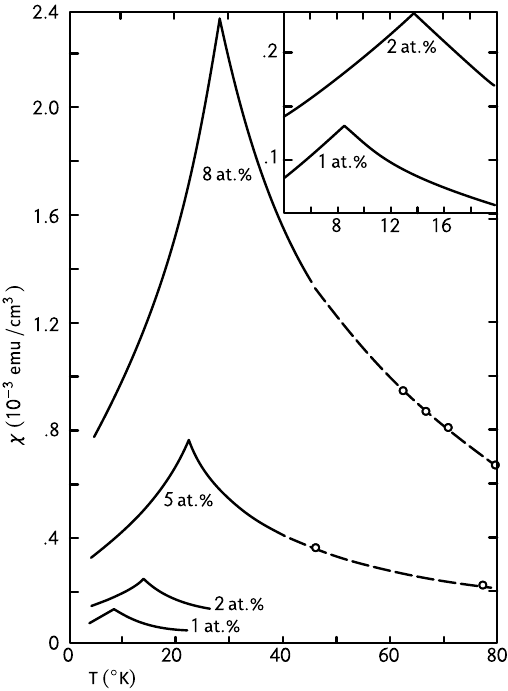}
    \caption{ Susceptibility for AuFe $1\leq {\text {c}}\leq 8$~at.\% in the region of low magnetic fields.  The data were taken every $0.25$~K around the peak and every $0.5$ or 1~K elsewhere. The scatter of the points is of the order of the thickness of the lines.  The open circles indicate isolated points taken at higher temperatures.  Reproduced from Fig. 9 of \citet{cannella:72}.}
    \label{fig:1}
\end{figure}

Interest in the magnetic alloys now known as spin glasses began with the work of Cannella and Mydosh, which showed a cusp in the ac susceptibility of dilute Au:Fe alloys~\cite[][see Fig.~\ref{fig:1}]{cannella:72,cannella:73}.  \changes{Others followed with similar findings \cite{mulder:81, mulder:82,huser:83,nagata:79}}. 
The sharpness of the cusp indicated what might be a phase transition, while earlier dc magnetization measurements on similar alloys only showed a rounded transition  \cite[see, for example,][]{lutes:64}.

\begin{figure}[tb]
    \centering
    \includegraphics[width=\linewidth]{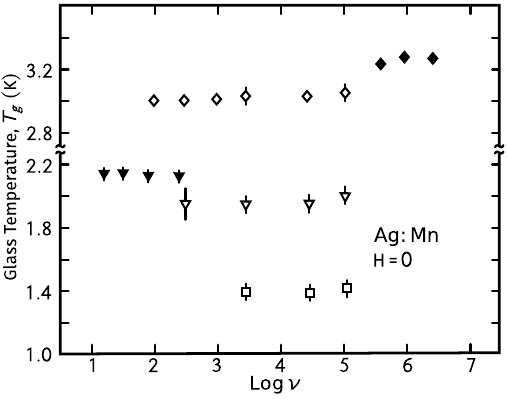}
    \caption{The mean ac spin-glass temperature, $\Tg$, vs the logarithm of the measuring frequency for different concentrations in AgMn samples.  $\blacklozenge$, 5,000 ppm, HF (0.4 to 2.8 MHz); $\Diamond$, 5,000 ppm MF (93 to 109 kHz); $\triangledown$, 3,000 ppm MF (93 to 109 kHz); $\blacktriangledown$, 3,000 ppm LF (16 to 160 Hz); and $\square$, 2,000 ppm MF (93 Hz to 109 kHz).  The vertical error bars are equal to two standard deviations of the \Tg measurements at a particular frequency. Reproduced from Fig. 2 of \citet{dahlberg:79}.}
    \label{fig:2}
\end{figure}

\begin{figure}[tb]
    \centering
    \includegraphics[width=\linewidth]{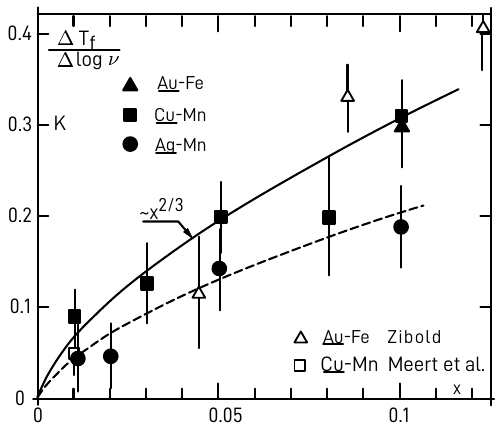}
    \caption{The absolute variation of $T_\text{f}$ (equivalent to what we call \Tg in this review) per decade of time, reported as a function of \changesbis{$x$, the percentage fraction of Mn}, for CuMn (\changesbis{$1\% <x < 10\%$}).  It can be represented by an $x^{2/3}$ law. Reproduced from Fig.~2 of \citet{tholence:81}.}
    \label{fig:tholence-81}
\end{figure}

A simple picture described the system as merely a distribution of independent magnetic domains, so that the cusp was an artifact, caused by the freezing of a portion of the sample at a given temperature and frequency~\cite{tholence:74,Tholence-1977,Ford-1982}. In such a case, a variation in the measurement frequency would significantly change the temperature of the cusp. This question was addressed by the UCLA group~\cite{dahlberg-1978, dahlberg:79}. They measured the frequency dependence of the cusp temperature and found the maximum shift to be of the order of 50~mK from 16 Hz to 2.8 MHz (Fig.~\ref{fig:2}).  This was strong evidence that the cusp was the signature of some type of phase transition.  \changes{In their work, a small frequency dependence, of the order of 50~mK over three decades in frequency, was observed in the most extensively studied 3000~ppm AgMn sample.  At the time it was believed to be associated with critical slowing down~\cite{souletie:85} as was observed in many second-order magnetic transitions.  Further work by \citet{tholence:81} found in more concentrated samples a more significant frequency dependence (see  Fig.~\ref{fig:tholence-81}).  This was also the first use of $\Delta T_\text{g}/\Delta \ln\, \nu$ as a measure of the shift of $T_\text{g}$ with frequency.  This frequency dependence is yet another defining characteristic of a spin-glass transition (see below the first of the Mydosh criteria for distinguishing a spin glass).}

\begin{figure}[tb]
    \centering
    \includegraphics[width=\linewidth]{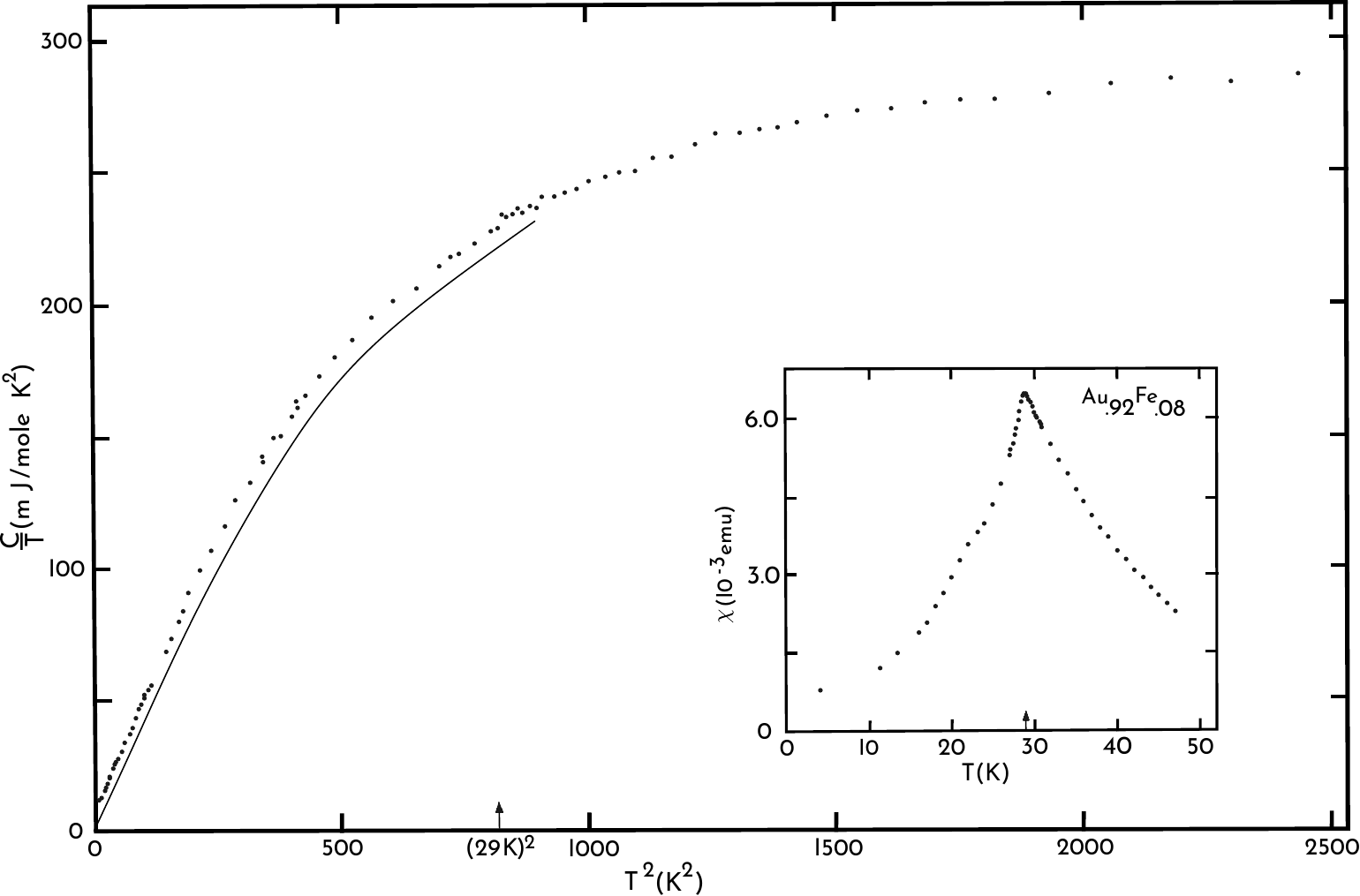}
    \caption{Specific heat of Au$_{0.92}$Fe$_{0.08}$ in the temperature region 3-50 K is shown as a plot of $C/T$ vs $T^2$.  The solid curve is the calculated nonmagnetic contribution to the specific heat of the alloy between 0 and 30 K.  The inset shows the susceptibility results of the same sample, which were provided by S. A. Werner.  Reproduced from Fig.  2 of \citet{wenger:75} and \citet{wenger:76}.}
    \label{fig:3}
\end{figure}

In the early years, prior to this study in an extended frequency range, the dc magnetic measurements had not revealed critical behavior.  The search was on for another signature.  It seemed obvious that heat capacity should reveal some sort of thermodynamic signature. However, heat-capacity measurements by \citet{wenger:75,wenger:76} showed no evidence (see Fig. \ref{fig:3}). 

 Other measurements were made in hopes of observing something indicating critical behavior.  They included neutron scattering, resistivity, Hall, thermopower, ultrasound, muon spin resonance, ultrasound, and Mossbauer measurements, but none provided the same strong signature as the ac susceptibility \cite[see][for an in depth discussion of each of these probes]{Ford-1982}. In 1991, it became clear that a thermodynamic phase transition had taken place at \Tg~\cite{gunnarsson:91}.  This work was followed by simulations~\cite{palassini:99, ballesteros:00} that supported the conclusion that spin glasses experience a continuous phase transition.

\changes{A recent compendium of new spin-glass materials can be found in the review paper of \citet{mydosh:15}.  He introduced four criteria to distinguish a ``proper'' spin glass. Quoting in part here:
\begin{enumerate}
\item A tiny shift $\Delta T_\text{g}/(T_\text{g} \ln\omega)$ means a strongly interacting cooperative freezing spin glass.
\item For the canonical spin glass, the irreversible temperature, where the first deviation between ZFC and FC occurs, coincides with the $T_\text{g}$ from the cusp (see below for the definition of these two protocols).
\item The magnetic specific heat breaks away from its linear-in-$T$ dependence at $T_\text{g}$ followed by a broad maximum 20\%--40\% above $T_\text{g}$.
\item The waiting-time effect [see Sec.~\ref{sect:off-equilibrium-no-field}] is a unique characteristic of the spin-glass phase.
\end{enumerate}
}

\begin{figure}[tb]
    \centering
    \includegraphics[width=8.5cm]{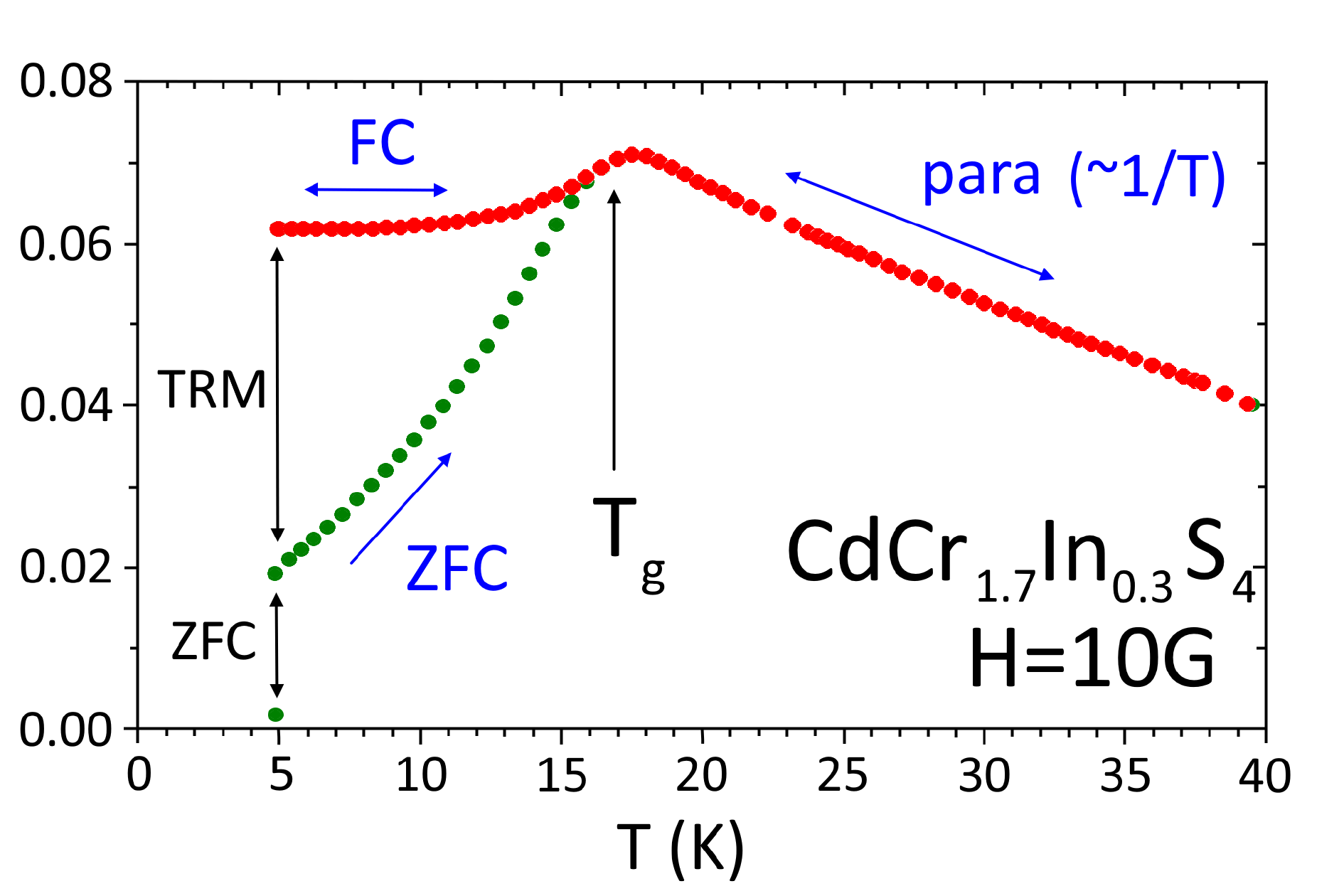}
    \caption{The field-cooled (FC, upper), zero-field-cooled (ZFC, lower), and thermoremanent (TRM, middle) magnetizations against  temperature for CdCr$_{1.7}$In$_{0.3}$$S_4$ in an external magnetic field $H=10$ G.  Reproduced from Fig. 1 of \citet{vincent:24} and \citet{dupuis:02}.}  
    \label{fig:ZFC_TRM_FC_protocols}
\end{figure}

\changes{Unfortunately, the last part of criterion 2 is seldom the case.  The cusp is most often rounded, so that the spin-glass transition temperature is usually defined by the first part of 2, namely} the point at which there is an onset of irreversibility \changes{seen by comparing two measurements of the spin-glass magnetization:} lowering the temperature in the presence of a magnetic field [field-cooled magnetization (FC)]; and lowering the temperature in the absence of a magnetic field and then turning it on [zero-field-cooled magnetization (ZFC)].  Fig.~\ref{fig:ZFC_TRM_FC_protocols} displays the difference, the so-called thermoremanent magnetization (TRM), between both protocols \cite{dupuis:02}.  Experimentally, the FC magnetization has a weak time dependence, while ZFC and TRM measurements are strongly time dependent.  \changes{This irreversibility, the difference between the FC and ZFC magnetization, was first reported by \citet{ferre:81,chamberlin:84,monod:79} and \citet{tholence:74}}.

It is tempting to conclude, because of the weak time dependence of the FC magnetization, that the sum of the time-dependent ZFC + TRM magnetizations should always equal the FC magnetization.  This led the Uppsala group to pose the ``extended principle of superposition'' \cite{nordblad:86,lundgren:86,nordblad:87,djurberg:95},
\begin{equation}\label{eq:superposition_M}
M_{{\text {TRM}}}(t, \tw) + M_{{\text {ZFC}}}(t,\tw)=M_{{\text {FC}}}(0,\tw+t),
\end{equation}
where $\tw$ is the waiting or aging time (see glossary in Sec.~\ref{subsect:glossary}).  However, \citet{orbach-janus:23} showed that Eq.~\eqref{eq:superposition_M} is valid only as $H\rightarrow 0$ (see Sec.~\ref{subsect:off-equilibrium-in-field-fixed-T-4}).

\begin{figure}[tb]
    \centering
    \includegraphics[width=\linewidth]{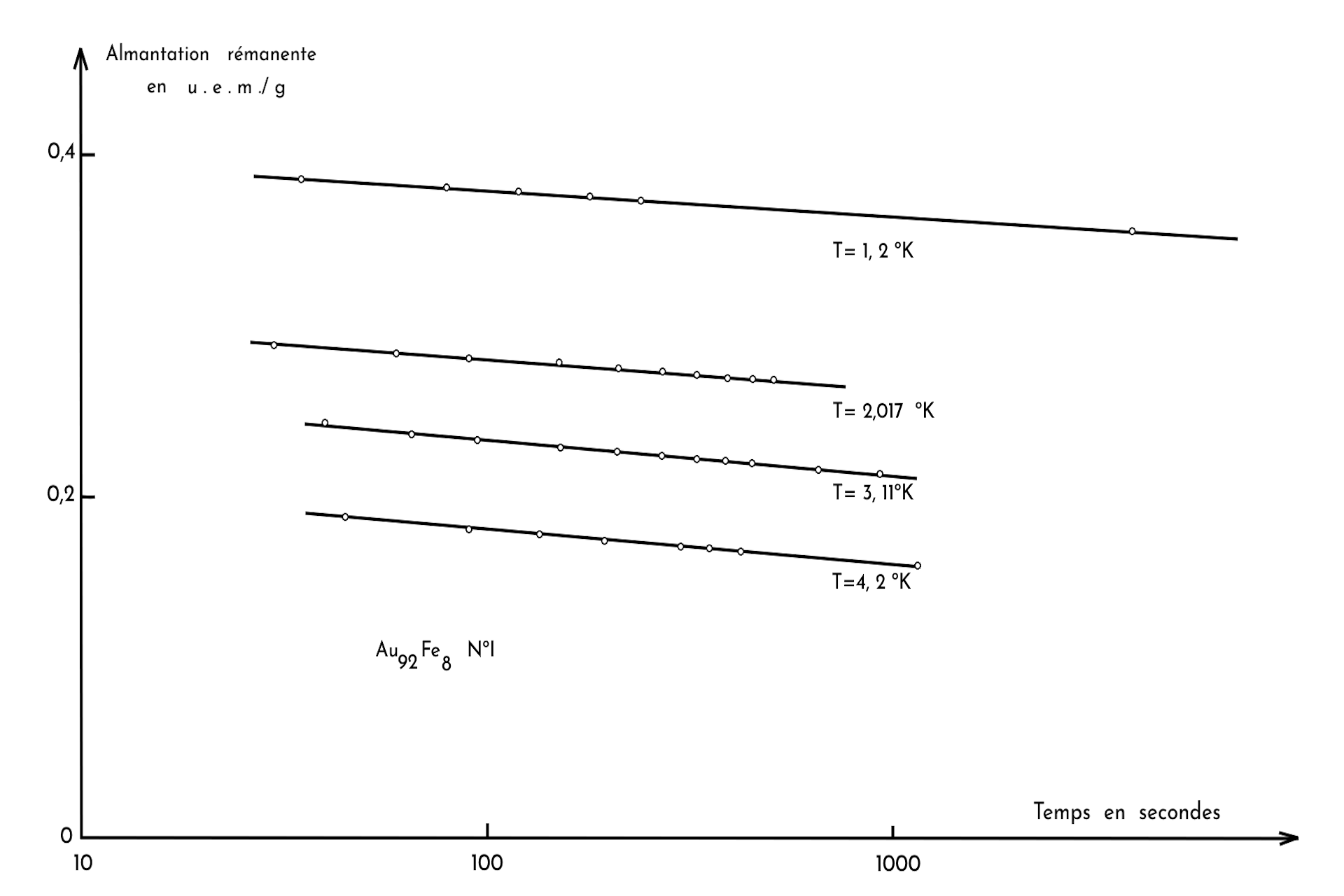}
    \caption{Time decay of the thermoremanent magnetization (TRM) of  Au$_{92}$Fe$_{8}$.  Reproduced from \citet{tournier-1965}.}
    \label{fig:tournier}
\end{figure}

Among the numerous time-dependent studies, there are some key elements that deserve mentioning. The first is the quasi-logarithmic time dependence of the magnetization, first observed in AuFe alloys by \citet{tournier-1965}, see Fig. \ref{fig:tournier}, and  in FeCr by \citet{ishikawa-1965}.

 \begin{figure}[tb]
    \centering
    \includegraphics[width=\linewidth]{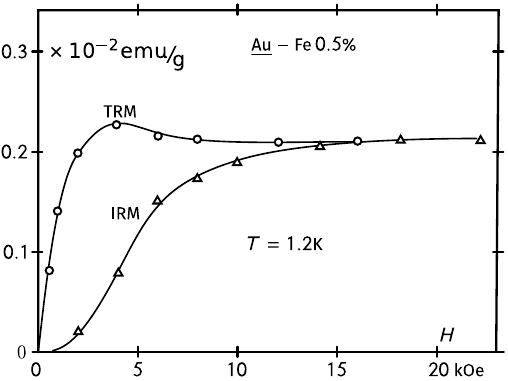}
\caption{Field dependence of the thermoremanent magnetization (TRM) of AuFe 0.5 at.\% obtained after cooling from $T >\Tg$ to $T=1.2$ K in a magnetic field $H$ which is them removed and of the isothermal remanent magnetization (IRM) obtained when cooling in zero magnetic field, turning on the magnetic field $H$ at 1.2 K and then removing it.  Reproduced from Fig. 4 of \citet{tholence:74}.}
    \label{fig:tholence_tournier}
\end{figure}

Early on, the TRM and the isothermal remnant magnetization (IRM) were most commonly measured \cite[for example,][see Fig.~\ref{fig:tholence_tournier}]{tholence:74}. An IRM measurement consists of cooling to a temperature below the glass temperature in zero applied magnetic field. A magnetic field is applied and removed with the TRM being the remaining magnetization.

\begin{figure}[t]
    \centering
    \includegraphics[width=\linewidth]{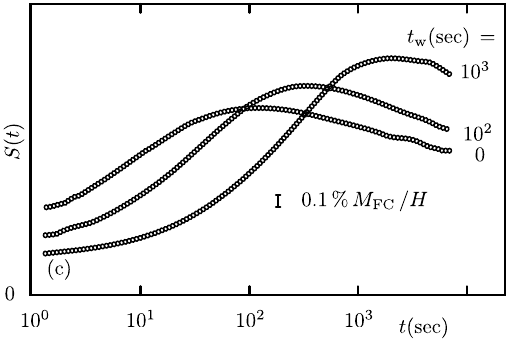}
    \caption{Relaxation rates $S(t)$ of the remanent magnetization of CuMn 5 at.\% ($\Tg=28$ K) at $T=21$ K for different waiting times \tw.  A relaxation rate of 0.1\% of the field-cooled susceptibility value is indicated on the figure. Reproduced from Fig. 2 (c) of \citet{nordblad:86}.}
    \label{fig:peak_St_nordblad}
\end{figure}

A useful tool for quantitative extraction of spin-glass dynamics has been the measurement of $S(t,\tw;H)$,
\begin{equation}\label{eq:St_def}
\changes{S(t, \tw; H) = (\pm ) {\frac {\dd M(t,\tw;H)}{\dd\ln t}}}\,,
\end{equation}
where $+$ ($-$) refers to ZFC (TRM) measurements.  This was first developed by
\citet{nordblad:86}, who showed that the time for $S(t,\tw;H)$ to peak was a useful measure of a spin-glass response time, $\tw^{\text {eff}}$ (see Fig. \ref{fig:peak_St_nordblad}).  Typically, $\tw^{\text {eff}}\approx \tw$, but there are important exceptions \cite{kenning:20} at temperatures close to $\Tg$. 
\begin{figure}[tb]
\centering
\includegraphics[width=\linewidth]{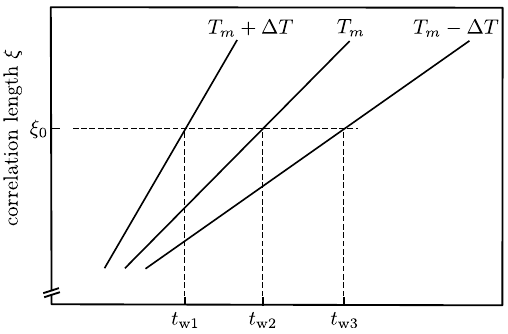}
\caption {The correlation length $\xi$ against the age $t_\text{w}$ (log scale) at different temperatures $\Tm-\Delta T, \Tm,$ and $\Tm+\Delta T$.  The sample was CuMn 10 at.\%, with $\Tg = 45.2$ K, and $\Tm = 41.2$ K.  \citet{granberg:88b} claimed that the different wait times $t_{\text{w}1}, t_{\text{w}2}$, and $t_{\text{w}3}$ yielded the same value $\xi_0$ of the correlation length. Reproduced from Fig. 6 of \citet{granberg:88b}.}
\label{fig:uppsala_fig}
\end{figure}

Later work by the Uppsala group \cite{granberg:88b} used temperature shifts to explore the $S(t)$ peaks as a function of wait times to determine how the growth rate of the correlation length depends upon temperature.  They assumed the correlation length scaled with the time $t$ and argued that at a given temperature, the correlation length (their $\xi_0$) was wait-time independent.  Their data is summarized in Fig. \ref{fig:uppsala_fig}.

\begin{figure}[tb]
\includegraphics[width=.9\linewidth]{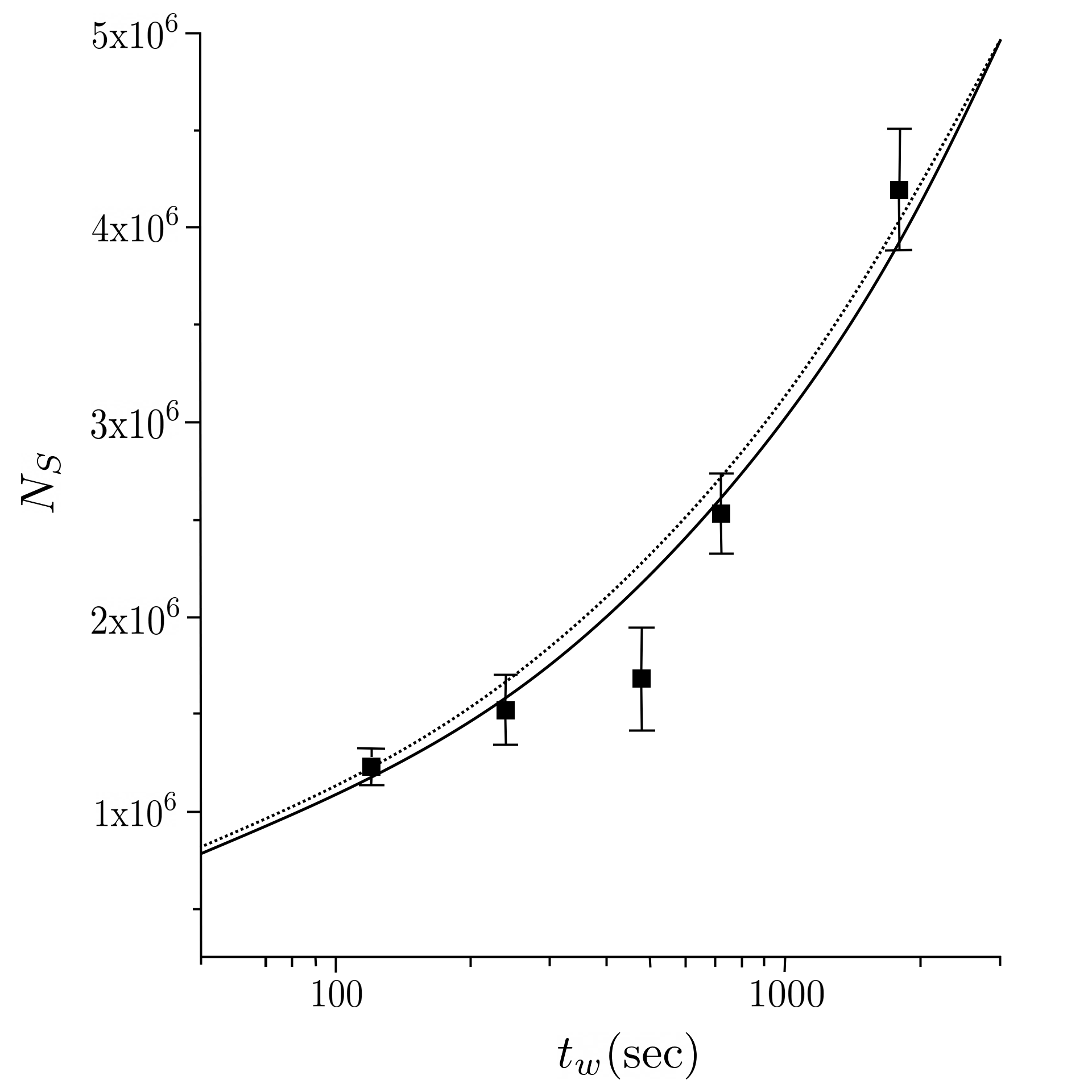}
\caption{A plot of $\Ns$, the number of spins participating in barrier quenching (and barrier hopping), for CuMn 6 at.\% vs $\tw$ on a log scale at fixed  $T=0.89\,\Tg=28$ K.  The solid curve drawn through the points is the prediction for power-law dynamics \cite{koper:88}, while the dashed curve is the prediction for activated dynamics \cite{fisher:88,fisher:88b} with the exchange factor set equal to $\Tg$ (\emph{i.e.}, independent of $T$ and $t$).  The two fits are equally good. Reproduced from Fig. 3 of \citet{joh:99}.}
\label{fig:Ns_vs_tw_Joh}
\end{figure}

\changes{Following this work, \citet{vincent:95} proposed a scaling law for the extraction of a length scale from the field dependence of the $S(t,\tw;H)$ peak as a function of the waiting time $\tw$. \citet{joh:99} made use of this scaling law to extract explicit values of the correlation length of the spin-glass state $\xi(\tw,T)$ ---the $\xi_0$ of Fig. \ref{fig:uppsala_fig}--- and its growth rate as a function of $t$, $\tw$, and $T$ for the metallic spin glass CuMn 6 at.\% and the insulating spin glass CdCr{$_{1.7}$}In{$_{0.3}$}S{$_4$}. 
These authors implicitly assumed a nonfractal geometry for
the bulk of the spin-glass domains by  
setting the number of correlated spins $\Ns\approx \xi(\tw,T)^3$ to measure $\xi(\tw,T)$ as a function of $\tw$.  Their results for CuMn 6 at.\% are exhibited in Fig.~\ref{fig:Ns_vs_tw_Joh}.  Note the different dependence upon $\tw$ between $\xi_0$ and $\xi(\tw,T)^3$ in Figs.~\ref{fig:uppsala_fig} and~\ref{fig:Ns_vs_tw_Joh}, respectively.  
We now know that the number of correlated spins is actually $\Ns = \xi(\tw)^{D-\theta/2}$, where $\theta$ is the so-called replicon exponent ---see, \emph{e.g.}, \citet{marinari:00,marinari:00b} for the meaning 
of the replicon, and Sec.~\ref{subsect:intro-xi_eff} for the current method of extraction of $\xi$ from experiment.
}

The development of the ability to determine an experimental value for the correlation length was critical for connecting experiments to models and simulations. The rationale lies in the dynamics of a spin glass. The defining physical property of glassy materials is determined by the energy landscape and that, in turn, is created by the growth of the correlated domains.

Progress in the experimental state has largely been based on three physical properties intrinsic to the spin-glass state. They are aging, rejuvenation, and memory. The aging phenomenon has been mentioned earlier beginning with the work of \citet{tournier-1965}. The discovery of rejuvenation and memory was first reported by \citet{jonason:98} using low-frequency ac susceptibility measurements. This experiment will be discussed in detail in Sec.~\ref{subsec:memory_and_rejuvenation} as the basis for more recent work on memory and rejuvenation.
\begin{figure}[tb]
    \centering
    \includegraphics[width=\linewidth]{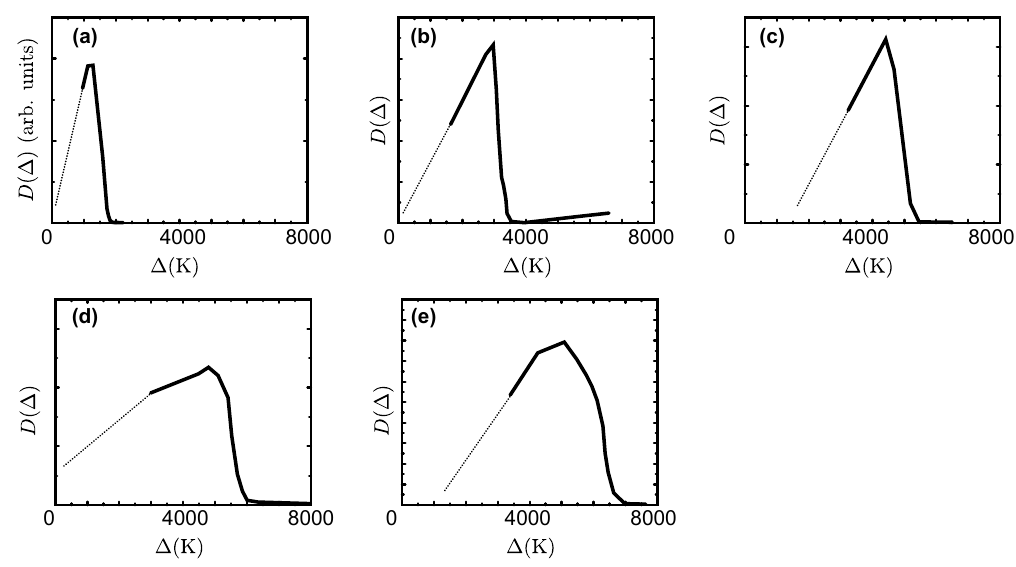}
    \caption{Barrier distributions, $D(\Delta)$, at $T=0$, based on measurements of $S_R(f,T)$ \changes{, the spectral density of the resistance fluctuations as a function of frequency $\it f$ and temperature $T$,} for five film thicknesses: (a) 10 nm, (b) 18 nm, (c) 25 nm, (d) 40 nm, and (e) 80 nm.  The bold portions fall within our measurement bandwidth.  The dashed portions are an extrapolation to guide the eye.  Reproduced from Fig. 3 of \citet{harrison:22}.}
    \label{fig:barrier_distributions}
\end{figure}
Using the development of an experimental quantification of the memory effect, there has been recent progress in understanding its origin in terms of temperature-specific correlations \cite{freedberg:24,paga:23b}. 

Given that the energy barriers, $\Delta$, are critical to the dynamics, there have been attempts to determine their distribution, $D(\Delta)$, for the spin-glass state. Recent $1/f$ electronic-noise measurements have been found to provide a measure of the energy-barrier distribution in thin films \cite{harrison:22,tsai:24}. Although a distribution was obtained, the relevant length scale was that of the electron mean free path and thus may be slanted towards the larger barriers (Fig. \ref{fig:barrier_distributions}).

\subsection{A glossary for aging dynamics}\label{subsect:glossary}
Important concepts used repeatedly in this review:

\begin{enumerate}

\item Two relevant temperatures are $\Tg$ and $\Tm$. The glass transition occurs at $\Tg$, which also is the critical temperature for the phase transition that separates the high-temperature paramagnetic phase from the spin-glass phase that becomes stable below $\Tg$. In a typical experimental setting, the system is quickly cooled from some very high temperature $T>\Tg$ to the \emph{measuring} temperature $\Tm$.

\item Two relevant experimental protocols, explained in Sec.~\ref{subsect:intro-2}, are named zero-field cooling (ZFC) and thermoremanent magnetization (TRM). 

\item Two relevant times are $t$ and \tw. We set our initial time as the time for the temperature quench from $T>\Tg$ to $\Tm<\Tg$. The system is allowed to relax at $\Tm$ for a \emph{waiting} time $\tw$. Then, at time $\tw$ the value of the externally applied magnetic field is varied (in the ZFC protocol the field is switched on at time $\tw$, while in the TRM protocol the field is switched off). The magnetization is measured at a \emph{later} time $t+\tw$, giving us information about the sample's response to the variation of the magnetic field at the earlier time $\tw$.

\item We shall need two kind of averages. The thermal average of a quantity $\mathcal{A}$ will be denoted by $\langle \mathcal{A}\rangle$. \changes{The Monte Carlo method~\cite{sokal:97,landau:05} is ideally suited to estimate $\langle \mathcal{A}\rangle$. Details specific for nonequilibrium simulations can be found in~\citet{janus:09b}; see also \citet{janus:10b,martin-mayor:22} for details about equilibrium simulations and \citet{yllanes:11} for a general discussion.} \changes{To obtain the average over disorder, which is always carried out \emph{after} the thermal average,  one first independently simulates several systems, each with an independent choice of the random couplings, and then takes the average of the different thermal averages.
The average over disorder will} be denoted by $\overline{\langle \mathcal{A}\rangle}$. Although the average over disorder is absolutely crucial for numerical simulations,  experiments are usually carried out for just one sample. Due to self-averaging, this is usually a sound procedure under experimental conditions (see Sec.~\ref{subsect:intro-dedicated_computers}).

\item The coherence length $\xim(\tw)$ is the size of the slowly growing glassy domains. See Sec.~\ref{subsect:off-equilibrium-no-field-3} for details [{\it e.g.}, Eq.~\eqref{eq:C4scaling}].
If the growth of $\xim$ is not isotropic, such as in a film geometry, we consider the perpendicular direction, $\ximT$, and the parallel direction, $\ximP$ (see Sec. \ref{subsect:off-equilibrium-in-field-fixed-T-4}). 
\item The two-time correlation length $\zeta(t,\tw)$ describes the growth of order in the aging regime when the two times $\tw$ and $t+\tw$ are compared \cite{jaubert:07,aron:08,janus:09b}. This quantity, defined in Sec.~\ref{subsect:off-equilibrium-no-field-3}, will be vital for understanding temperature-varying protocols.\footnote{\changes{Our distinction between coherence and correlation lengths is inspired in the coarsening dynamics of a ferromagnetic system with Ising spins. In this analogy, the coherence length is the size of the magnetic domains while the correlation length indicates the length scale to which statistical correlations propagate \emph{inside a domain\/}.}} 
 \item The experimentally accessible $\xiZ$ introduced by \citet{joh:99} will be explained in Sec.~\ref{subsect:intro-5}. This length is related to the lowering of free-energy barriers because of the Zeeman effect and is nonlinear in the magnetic field. 
\end{enumerate}
As explained in Sec.~\ref{sect:off-equilibrium-in-field-fixed-T}, the relationship between the three length scales \xim, \xiZ and $\zeta$ is the key to the study of the nonlinear response of a spin glass in a magnetic field. 
\changes{A central result will be that $\xim$ and $\xiZ$ turn out to behave very similarly whenever they can be compared. In other words, it is possible to use \emph{bulk} measurements 
to determine apparently microscopic quantities (the size of the  spin-glass domains). This correspondence could suggest also referring to $\xiZ$ as a coherence length, but we shall follow the experimental tradition of referring to $\xiZ$ as a correlation length.}

\section{Preliminaries}
\label{sect:preliminaries}
Before getting into detailed analyses of spin-glass dynamics, it is useful to establish the ``ground rules.'' These are the basic tools, both theoretical and experimental, that are used to elucidate the ``amazing cornucopia'' anticipated by \citet{anderson:89}.

\subsection{The Edwards-Anderson and Sherrington-Kirkpatrick models and overlap as an order parameter}\label{sect:models}

The large impact of the \textcite{cannella:72} experimental results, describing a phase transition with novel features, soon made it a relevant goal to establish a theoretical description and, if possible, a complete understanding of the behavior of diluted solutions of, say, Mn in Cu, which just started to be called \textit{spin glasses}.  

By assuming that $\sum_{i,j}J_{ij} \epsilon_{ij}=0$ on all length scales of the system (where $J_{ij}$ is the strength of the interaction between sites $i$ and $j$ and $\epsilon_{ij}$ is the probability of finding two interacting spins at those positions), \citet{edwards:75} introduced a very simplified model of classical spins (which will turn out to embed all relevant features of the problem), the Edwards-Anderson (EA) model. A different approach, based on the presence of potentially magnetized domains, had been proposed before by \textcite{adkins:74}. The fact that the disordered interactions are distributed at random and cannot change in time ---we say that the disorder is \emph{quenched}--- is a crucial element of the model.

The EA model provided an Ising-model-like description of spin glasses. The Hamiltonian is
\begin{equation}
\label{eq:1C:EAHam}
\mathcal H_\text{\tiny EA} \equiv - \sum_{i,j}\phantom{}^{'} \; J_{ij} \sigma_i \sigma_j+ H \sum_i \sigma_i\;,
\end{equation}
where the primed sum runs over the pairs of nearest neighbors in a $D$-dimensional hypercubic lattice. The spins $\sigma_i$ 
can take different forms. For instance, in the Heisenberg spin glass, 
they are unit vectors in space. Unless otherwise specified, however,  in this review, the EA model will always be considered with Ising spins that can only take the values $\pm 1$ \changes{(see Sec.~\ref{sec:chiral} for a discussion 
of the practical differences between these two models)}. The applied magnetic field $H$ will be considered to be zero, for now. 

Notice that this Hamiltonian is written \changes{in units such that $k_\text{B}=1$}. To compare with experimental results, temperatures can be expressed in terms of \Tg (more sophisticated scaling variables
will be introduced later). As for the magnetic field, it can be estimated \cite{janus:17b,aruga_katori:94}, that $H=1$ roughly corresponds to $5\times10^4$~Oe. \changes{The renormalization group \cite{wilson:75,zinn-justin:05,parisi:88} teaches us that the effect of the magnetic field scales with its value in natural units
that depend on the coherence length. This important point will be explored throughout the review.} Finally, the natural unit of time in simulations is the lattice sweep, which is equivalent to $\sim 1$~ps \cite{mydosh:93}.

It is crucial that the quenched couplings, which do not change in the course of the dynamics, can have both signs. 
A possible choice is  
$J_{ij}=\pm 1$ with a probability of one half, 
or that they be normally distributed, with zero mean  and unitary variance, $P\left(J_{ij}\right)
\sim \exp(-{J_{ij}^2}/{2J^2})$. We will not discuss here the possibility of a nonzero expectation value for the couplings.

Because the quenched couplings can be both positive or negative and are distributed at random, some of the products of the couplings on a closed path of consecutive lattice links will be positive and some will be negative. This implies, in turn, that not all interactions can be satisfied at the same time.  The system is frustrated in the sense that all interactions can never be completely satisfied. The energy cannot be as low as it could be in a uniform system. \changes{Typically, frustration leads to degeneracy: the low-energy states of such a system have a higher energy than those states found in a pure system, but  are more  numerous in the disordered system.}

In the EA model, because the interactions between spins oscillate in sign as a function of separation, ferromagnetism does not appear at any relevant length scale. On the contrary, for $T$ lower than a critical \Tg, a \textit{spin-glass phase} appears, with the spins staying close to fixed directions that \textit{appear} to be random. 
This results in overall zero magnetization, but with a nonzero square of the local magnetization leading to the cusp in the susceptibility observed in  experiments (see Sec. \ref{subsect:intro-2}).

\begin{figure}[tb]
    \centering
    \includegraphics[width=\linewidth]{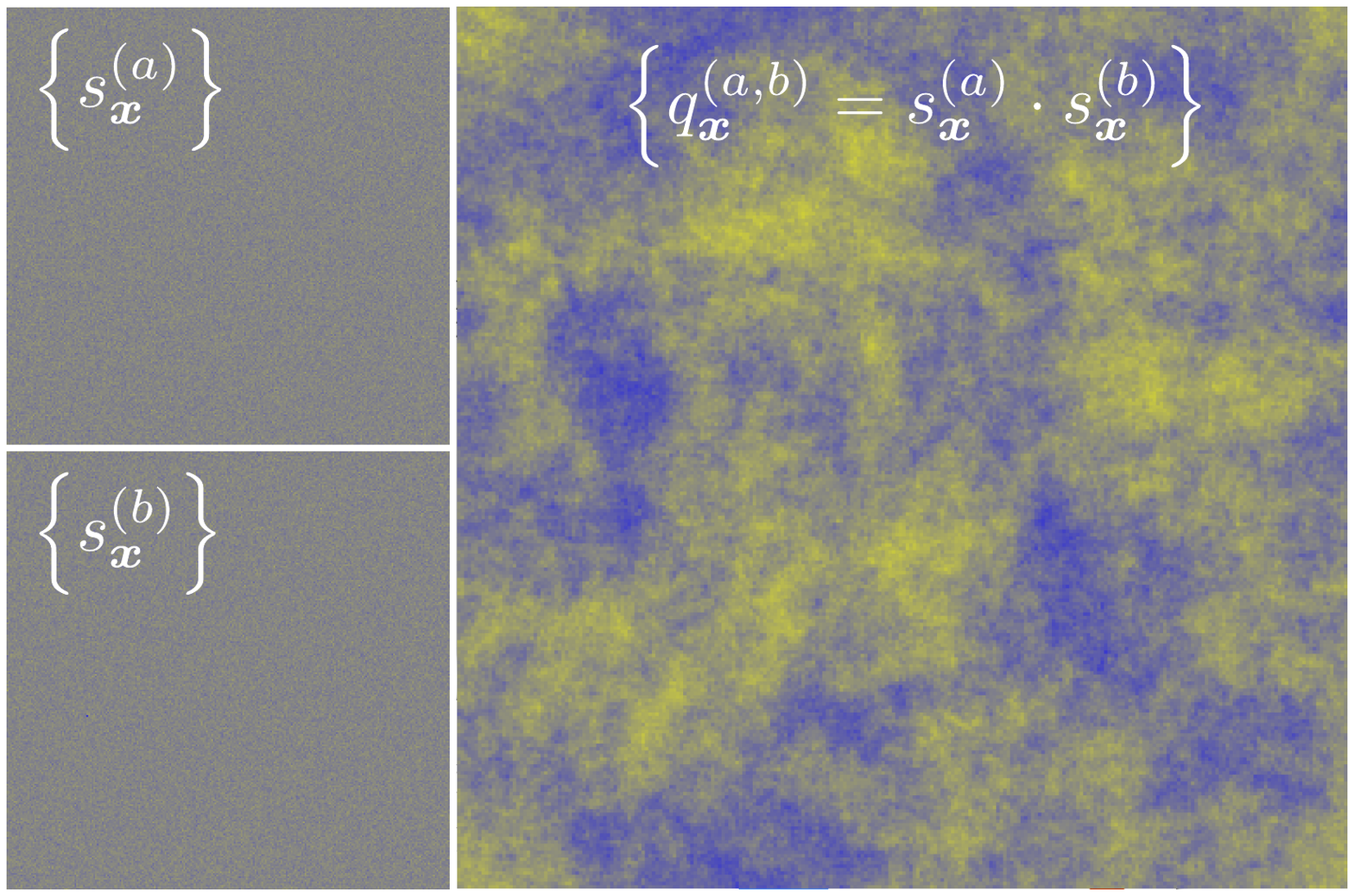}
    \caption{Illustration of the spin overlap. The left panels show configurations of two real replicas of the same spin glass, which have evolved for a long time in the broken-symmetry phase in a Monte Carlo simulation. The color map shows the average magnetization in the XY plane, averaged over the Z axis, which is always zero. In contrast, the right panel shows the spin overlap of Eq.~\eqref{eq:def-overlap}, where a clear pattern is visible.   
    Figure taken from \citet{janus:19}.}
    \label{fig:def-overlap}
\end{figure}
The absence of long-range magnetic order leads to the
concept of spin overlap as an order parameter, crucial to all further developments of the field. One considers the spin $\sigma_i$ at site $i$ and defines:
\begin{equation}
\label{eq:1C:TimeOverlap}
q_i\left(t\right)
\equiv\lim_{\tau\to\infty}
\left\langle
\sigma_i\left(t\right)
\sigma_i\left(t+\tau\right)
\right\rangle,
\end{equation}
which can be different from zero in the low-$T$, broken-symmetry phase.  The brackets $\langle\cdots\rangle$ signify the average over different thermal histories. A more convenient way of measuring the same overlap is based on the use of two replicas: copies of the system with the same quenched disorder (same random couplings in the Hamiltonian)
\begin{equation}\label{eq:def-overlap}
q = \frac1N \sum_{i=1}^N \sigma^{(1)}_i \sigma^{(2)}_i.
\end{equation}
This is the most common observable in numerical simulations. The concept of a nonzero overlap of configurations with zero magnetization is illustrated in Fig.~\ref{fig:def-overlap}.

After averaging $q$ over the disorder realizations, we can also introduce its probability density function, denoted by $P(q)$ (see Sec.~\ref{subsect:Equilibrium-1} for some examples). In addition, we can define the link overlap, which can be written as
\begin{equation}\label{eq:def-linkoverlap}
q_\text{l} = \frac{1}{ND} \sum\phantom{}^{'} \; \sigma^{(1)}_i \sigma^{(2)}_i \sigma^{(1)}_j \sigma^{(2)}_j\,.
\end{equation}
As with the overlap,
we can associate a $P(q_\text{l})$ to this observable. The behavior of the overlap and the link overlap under spin flip is very different: consider two configurations of spins with the same disorder which differ in a domain of size $L$, where the spins have flipped. The link overlap provides an estimate for the surface of the domain, whereas the overlap reflects its volume.

The EA spin glass, an Ising-like finite-dimensional model with quenched, frustrated interactions, cannot be solved exactly and cannot be dissected too deeply by analytical techniques. The experience of the Ising ferromagnet suggested introducing the mean-field, nonlocal Sherrington-Kirkpatrick (SK) model \cite{sherrington:75,kirkpatrick:78}.

The SK Hamiltonian has the same form of Eq. (\ref{eq:1C:EAHam}), but now the sum runs over all pairs of spins. Also if, for example, the couplings have a Gaussian distribution, one requires that $J=\tilde{J}/N^{\frac12}$ in order to make the thermodynamic limit of the model consistent,  where $\tilde{J}$ is an intensive number of order one.

This mean-field  model has the usual paramagnetic phase at high temperatures. The delicate point, which will be the source of trouble and delight, is that, because the couplings are quenched, we have to compute the disorder average
not of the partition function $Z$ (which would anneal the disorder degrees of freedom) but of its logarithm, $\ln Z$. That is, we average physical observables in a given disorder realization (a sample) and then we average these thermal expectation values over the disorder. To do this, EA and SK use the identity (the so-called replica trick)
\begin{equation}
\label{eq:1C:Zn}
\ln Z_J = \lim_{n \to 0} 
\frac{Z_J^n-1}{n}\,
\end{equation}
and apply it by computing $Z^n_J$ for integer $n\ge 1$ and eventually taking the limit for $n \longrightarrow 0$. $Z_J^n$ can be interpreted as the partition function of $n$ copies of the system that are defined in the same disorder landscape (replicas). In principle, this procedure works in the thermodynamic limit if there are no singularities in $n$ when $n\to0$. In the SK model, however, this approach is successful, in spite of the fact that there \emph{is} a singularity.

\citet{kirkpatrick:78} refined their analysis two years after their first, groundbreaking paper \cite{sherrington:75}. They described the same situation, but with an analysis based on the assumption of replica symmetry (RS). \changes{Not all was well, however.
The solution of the RS ansatz not only disagrees with Monte Carlo simulations but also leads to a nonsensical result, namely a negative value at low $T$ for the entropy, which is a semi-definite positive quantity (proving the inadequacy of the replica-symmetric ansatz).}
Before this second SK paper, \citet{thouless:77} had introduced a different approach, avoiding the replica trick, that shed further light on the low-$T$ behavior of the SK model. Their so-called TAP equations, together with their many variations, have been for many years and continue to be important for the analysis of spin-glass physics. When the temperature decreases below a critical value $\Tc$, the correct solution found at high $T$ under the assumption of symmetry between replicas turns out to be undoubtedly wrong.

\citet{dealmeida:78} analyzed the situation and established that the RS solution of the SK model in the  high-$T$ (paramagnetic) region was stable, but was unstable at low $T$, with the instability persisting in a magnetic field. 
\textcite{parisi:79} started from here in his (correct) solution of the mean-field spin glass, beginning with the assumption that \textit{replica symmetry is broken} in the low-temperature phase. Space limits the discussion of his solution. The interested reader is referred to \citet{mezard:87}, which includes some of the original work, or to more recent accounts \cite{dotsenko:01, dedominicis:06, charbonneau:23}.

The above-described mean-field approximation
is only correct above the so-called 
upper critical dimension $D_\text{u}$, which for spin glasses is
$D_\text{u}=6$~\cite{bray:80}.  On the other hand, below the lower critical dimension  \changes{$D_\text{l}$, the low-temperature phase disappears. It is believed that $D_\text{l}=5/2$ and it is clear that $2<D_\text{l}<3$~\cite{franz:94,maiorano:18}; for our purposes what matters is that $D=3$ lies in the nontrivial region}. There has been an intense quest for a successful extension of the Parisi theory to $D=3$ in the renormalization-group style \cite[see for example][]{dedominicis:98,dedominicis:06}. Because of the infinity of states that appear in Parisi's solution for the mean-field theory, renormalizing the system is a daunting task.  Today, after close to 50 years of studies, there still remains much need for clarification. The interested reader can start from \citet{dedominicis:98} and \citet{dedominicis:06}, then continue with the following papers and included references: \textcite{bray:78,bray:79,bray:80, dedominicis:83,dedominicis:84,temesvari:94,dedominicis:97,temesvari:17}.

\subsection{Main theoretical pictures for the low-temperature phase}\label{subsect:intro-4}
As explained above, the EA model can only be solved in mean field, \emph{i.e.}, in the limit of a large spatial dimension. 
In $D=3$, therefore, several competing theoretical pictures coexist. This
section briefly describes them.

\begin{itemize}
\item Droplet. The droplet picture assumes that the low-temperature phase of spin glasses in the absence of an external magnetic field  is composed of only two pure states related by global spin-flip symmetry. The droplet model was developed from a phenomenological theory~\cite{fisher:86} as well as from the Migdal-Kadanoff approximation of the renormalization group~\cite{bray:87}.
The behavior of the low-temperature phase derives from the assumed existence of compact excitations (droplets) in the ground state, with a fractal surface.
It is also assumed that the energy of these droplets of linear dimension $L$ scales as $L^y$. These authors showed that $y \le (D-1)/2 < D-1 < \Ds < D$, where $\Ds$ is the fractal dimension of the surface of the droplets and $D$ is the spatial dimension. Further, the free-energy barriers for the dynamics grow as $L^\psi$, with $D-1 \ge \psi \ge y$.   The presence of an external magnetic field destroys the spin-glass order at low temperatures. 
The Chaotic Pairs (CP) scenario \cite{newman:96c} is similar to the droplet picture with the peculiarity that the two pure states (in the absence of a magnetic field) depend chaotically on the system's size.
\item Trivial-Non-Trivial (TNT). In the droplet model (in the absence of a magnetic field) the probability distribution $P(q)$ of the overlap $q$ between two pure states, and the link overlap, are both trivial (\emph{i.e.}, $P(q)$ is composed of two Dirac delta functions and one Dirac delta function, respectively). In the TNT approach, $P(q)$ exhibits a nontrivial probability distribution, whereas the link overlap exhibits a trivial probability distribution~\cite{krzakala:00,palassini:00}.
\item Replica-Symmetry Breaking (RSB).
In this scenario,  which consists in extending the features of the mean-field solution to finite dimensions \cite{marinari:00,dedominicis:06}, the spin glass has infinite pure states not related by any kind of symmetry.  However, this infinite set of  states can be organized in an ultrametric fashion.   Furthermore, the excitations of the ground state are space-filling. Finally, in this theoretical approach, the spin-glass phase is stable to small magnetic fields (\emph{i.e.}, there exists a phase transition between a paramagnetic phase and a spin-glass phase in the presence of a magnetic field).
\end{itemize}

\begin{figure}[tb]
    \includegraphics[width=\columnwidth]{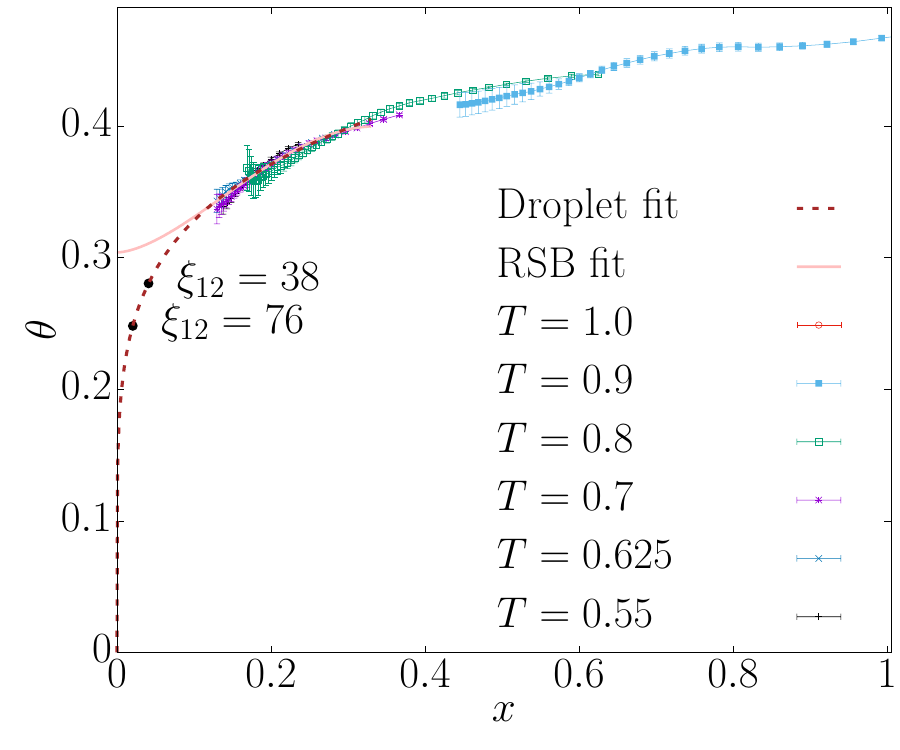}
    \caption{The main quantitative difference between the competing 
    theoretical pictures of spin-glass dynamics is related to the 
    value of an exponent $\theta$, the \emph{replicon}, see Eq.~\eqref{eq:C4scaling} and \citet{marinari:00,marinari:00b}. 
    This exponent is nonzero for RSB and zero in the droplet picture,
    at least in the long-time limit.
    When represented as a function of a scaling variable $x$ \cite[a properly rescaled $1/\xi(\tw, T )$, see][]{janus:18}, the values of $\theta$ obtained for several temperatures fall on a single curve. In addition we show a droplet extrapolation [dashed line, $\theta(x\!\to\! 0)\!\propto\! x^\iota$, with $\iota\!\approx\! 0.15$] and an RSB extrapolation [full line, $\theta(x\!\to\! 0)\! >\! 0$]. Even the droplet extrapolation,
which has $\theta(x)\to 0$ as $x\to 0$, predicts sizable values of $\theta$ for the values of $x$ relevant to experimental length
scales (black dots). Data taken from \textcite{janus:18}.}\label{fig:replicon-scaling}
\end{figure}
One of the main messages of this review is that the controversy discussed above is of no real consequence for understanding the off-equilibrium dynamics of spin glasses at experimentally relevant scales (see Fig.~\ref{fig:replicon-scaling}).

\subsection{Phenomenological dynamical model on an ultrametric tree} \label{subsect:intro-5}
\begin{figure}[tbh]
    \centering
    \includegraphics[width=\columnwidth]{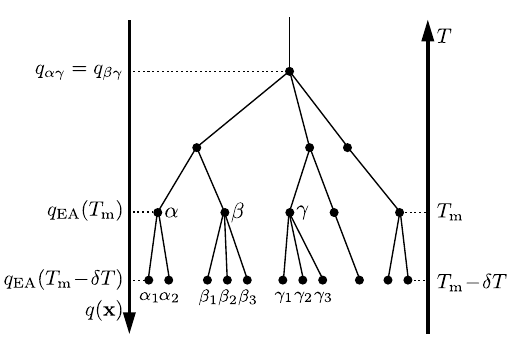}
    \caption{Three representative spin-glass states at a measurement temperature \Tm are labeled $\alpha,~\beta,~{\text {and}}~\gamma$.  Their overlaps $q$, defined in Eq. (\ref{eq:overlap_def}), satisfy the ultrametric topology $q_{\alpha\gamma}=q_{\beta\gamma} \leq q_{\alpha\beta}$.  The self-overlap is independent from the state $q_{\alpha\alpha}=q_{\beta\beta}=q_{\mathrm{EA}}(T)$.  When the temperature is lowered from $\Tm$ to $\Tm-\delta T$, the states ``foliate'' to daughter states, each of which now possesses the self-overlap $q_{\text {EA}}(\Tm-\delta T)$. Figure reproduced from~\citet{orbach-janus:23}.}
    \label{fig:ultrametric_tree}
\end{figure}
The Parisi solution of the Sherrington-Kirkpatrick mean-field model is pictured in Fig.~\ref{fig:ultrametric_tree} in terms of a ``tree'' that is organized in such a way that ultrametric symmetry relates its states (denoted as $\alpha$, $\beta$,...).  Immediately after a temperature quench from above \Tg to a measurement temperature $\Tm$, the spin-glass states have a ``self-overlap'' $q_{\alpha\alpha}(\Tm)\equiv q_{\text {EA}}(\Tm)$, the Edwards-Anderson order parameter \cite{edwards:75},
\begin{equation}\label{eq:qEA_def}
q_{\alpha\alpha}\equiv q_{\text {EA}}(\Tm)={\frac {1}{N}}\sum_i \langle \sigma_i(t=0)\rangle_\alpha^2,
\end{equation}
where $\langle \cdot\cdot\cdot\rangle_\alpha$ represents a thermal average restricted to a single pure state ($\alpha$).

There are no dynamics associated with pure states in that the free-energy barriers between them are infinite.  Experimentally though, time-dependent phenomena are observed.  It takes a large amount of {\it chutzpah} to suggest that {\it between} the pure states there could exist states between which the free-energy barriers were finite and even more to suggest that these states between infinite barriers were self-similar. No matter the temperature, the same ultrametricity holds for metastable states, which are organized by the same tree as the pure states.  The interpretation of dynamical measurements goes even further to suggest that the finite barriers between metastable states are temperature dependent, and apparently diverge into the pure states as the temperature is lowered \cite{hammann:92}, though of course the divergence can not be accessed experimentally.

\begin{figure}[htb]
    \centering
    \includegraphics[width=\linewidth]{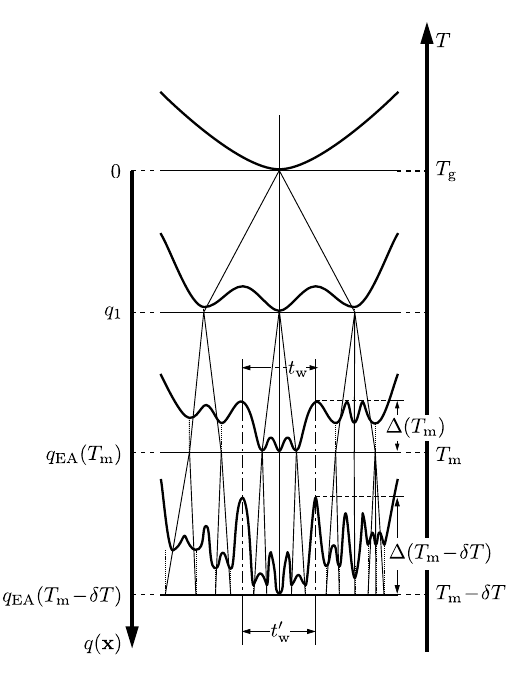}
    \caption{Hierarchical organization of metastable states. The coarse-grained free-energy surface is represented at each level corresponding
to a given temperature.  When temperature is decreased, each valley is subdivided into others. The times $\tw$ and $\tw^\prime$ are indicated that are
necessary to explore, at $\Tm$ and $\Tm - \delta T$, respectively, the region of
phase space bounded by the same barriers. The closest
common ancestor to all states within the space bounded by \changes{$\Delta(T_{\text {m}})$ and $\Delta (T_{\text {m}} - \delta T)$} is the same, and its corresponding value of the overlap functions are 
\changes{$q_{\text {EA}}(T_{\text {m}})$ and $q_{\text {EA}}(T_{\text {m}}-\delta T)$.} The sketch also shows that, as the system explores more of phase space, it encounters ever-increasing barrier heights, and that the free-energy surface has a self-similar structure. Reproduced from Fig. 6  of \citet{lederman:91}.}
    \label{fig:hierarchical_organization}
\end{figure}

There is no theoretical proof that these observations are correct.  Nevertheless, they have proven remarkably prescient experimentally \cite{vincent:09}.  A useful picture with finite barriers between metastable states is exhibited in Fig.~\ref{fig:hierarchical_organization}.

The dynamics associated with the constructs exhibited in Figs.~\ref{fig:ultrametric_tree} and~\ref{fig:hierarchical_organization} follow from their structure.  At $t=0$, the temperature is quenched from above \Tg in the paramagnetic phase to a measuring temperature $\Tm$ in the spin-glass phase. Instantaneously, the spin glass is found in some state $\alpha$ at $\Tm$ with self overlap $q_{\alpha\alpha}=\qEA(\Tm)$ [Eq.~\eqref{eq:qEA_def}].  As time progresses, the quenched initial state organizes from its paramagnetic structure into a progressively growing spin-glass state, expressed through a growing spin-glass coherence length $\xim(t)$, which is a function of the time elapsed since the quench, termed the waiting time $\tw$.  As $\xim(\tw)$ grows, larger and larger free-energy barrier heights are occupied, having been created by the growing coherence length.  The overlap $q_{\alpha\beta}$ with the initial state diminishes and is given by
\begin{equation}\label{eq:overlap_def}
q_{\alpha\beta}(\tw)= {\frac {1}{N}}\sum_i\langle \sigma_i^\alpha(\tw=0)\sigma_i^\beta(\tw)\rangle,
\end{equation}
where the state has evolved from its initial quenched state $\alpha$ to $\beta$.  This is shown in Fig. \ref{fig:ultrametric_tree}  for the ultrametric tree.

\changes{The main consequences of this framework will be considered in Sec.~\ref{subsect:intro-xi_eff}, where the extraction of $\xi$ from experiment is described. An analysis in terms of the phenomenological model discussed here can be found in \textcite{orbach-janus:23}.   The use of Fig.~\ref{fig:hierarchical_organization} to display both the self similarity of the metastable states between the infinite barriers of the Parisi pure states, and their evolution toward the pure states as the temperature is lowered, can be found in \textcite{hammann:92}.}

\subsection{The elusive, but crucial, temperature chaos}\label{subsect:intro-6}

Among the many interesting properties of spin glasses, the contextual measurement of both rejuvenation and memory effects was an astonishing finding \cite{jonason:98}, which in turn asked for theoretical understanding, eventually aided by numerical simulations of the EA model.
These effects occur in the out-of-equilibrium regime and will be mainly covered in Sec.~\ref{subsec:memory_and_rejuvenation}.
Rejuvenation may, however, be related to an equilibrium effect, \emph{temperature chaos}, which we briefly describe in this section. Furthermore, as we explain in Sec.~\ref{subsect:intro-dedicated_computers}, temperature chaos is believed to be the main difficulty that impedes reaching thermal equilibrium even for a very small spin-glass sample.

The term \emph{chaos} refers to the strong sensitivity of the spin-glass state ---the one dominating the Gibbs-Boltzmann (GB) measure for $T<\Tg$--- to the change of some parameter in the GB measure: either the temperature $T$ (temperature chaos), the couplings $\{J_{ij}\}$ (disorder chaos) or the external magnetic field $H$ (field chaos).
These effects require the existence of a spin-glass phase, so they cannot take place in any dimension less than the lower critical dimension $D_\text{l}\approx 2.5$ \cite{franz:00,boettcher:04,maiorano:18}.  When they are present, they have very different intensities, temperature chaos being by far the weakest. Indeed, for quite a long time the very existence of temperature chaos was controversial \cite[see, \emph{e.g.},][]{ney-nifle:97,ney-nifle:98,billoire:00,mulet:01,billoire:02,krzakala:02,rizzo:03,sasaki:05,katzgraber:07,parisi:10,fernandez:13}, while field chaos and disorder chaos were evident.

In order to quantify the intensity of the chaotic effects, one needs to measure the statistical similarity between two typical configurations generated according to two GB measures differing in a small $\Delta T$ (for temperature chaos) or small $\Delta J$ (for disorder chaos). For example, for temperature chaos, one could measure the statistics of the following chaoticity parameter
\begin{equation}
X^J_{T_1,T_2} = \frac{\langle q^2_{T_1,T_2} \rangle_J}
{(\langle q^2_{T_1,T_1} \rangle_J \langle q^2_{T_2,T_2} \rangle_J)^{1/2}}\, ,
\label{eq:chaotic_parameter}
\end{equation}
where the thermal averages are computed for a given sample indicated by the superscript/subscripts $J$.
While the denominator is just a normalization factor to ensure that $X^J_{T,T}=1$, the numerator measures the similarity between replicas at temperatures $T_1$ and $T_2=T_1+\Delta T$.

If the sample under consideration is not chaotic, we expect the statistics (\emph{e.g.}, the variance) of overlap $q_{T_1,T_2}$ between replicas at temperatures relatively close to one another to be very similar to those of $q_{T_1,T_1}$ and $q_{T_2,T_2}$, thus giving $X^J_{T_1,T_2} \simeq 1$.
Under chaotic conditions we expect $q_{T_1,T_2}$ to take the smallest possible value (which is zero if the external field is absent).
In other words, in a chaotic sample, the configurations at temperatures $T_1$ and $T_2$ are typically very different (often maximally different).

This is a very relevant physical phenomenon.  It would imply that such a chaotic spin-glass sample would develop very different long-range order at temperatures $T_1$ and $T_2$. If chaos is present, it would be the most natural explanation for rejuvenation: the long-range order developed in the spin glass at the higher temperature $T_2$ would be replaced by a new long-range-ordered state when the sample is cooled down to the lower temperature $T_1$. How the first ordered state could be recovered when the sample is heated back to $T_2$ is discussed in Sec.~\ref{subsec:memory_and_rejuvenation} and is the basis of the memory effect.

The approaches to studying and quantifying chaotic effects in spin glasses have followed different paths:  a scaling approach based on the droplet model (which is expected to be correct for low-enough dimensionalities); the solution to mean-field models, like the SK model (which is believed to be exact for high-enough spatial dimension); and the approach based on numerical simulations.

According to the droplet theory, the low-temperature properties of a spin glass are controlled by a $T=0$ fixed point, and the large-scale excitations from the ground state (compact droplets with fractal dimension $\Ds$) are controlled by an exponent $y$, which is positive (negative) if a spin-glass phase is present (absent). 
Using scaling arguments, \citet{bray:87} showed that a perturbation of typical intensity $\Delta J$ to the couplings would destabilize the ground state on length scales larger than $\ell_\text{c} \sim (\Delta J)^{-1/[(\Ds/2)-y]}$, provided $\Ds/2-y$ is positive.

Moreover, they extended the argument to a temperature change $\Delta T$ (inside the spin-glass phase), finding again that a typical configuration would be modified on length scales larger than $\ell_\text{c} \sim (\Delta T)^{-1/[(\Ds/2)-y]}$.
Hence, for any temperature in the spin-glass phase, the overlap $q_{T_1,T_2}$ would eventually trend to zero for large sizes.
Within the droplet model, the picture one gets is very simplistic (probably too simple): the spatial correlation function between two replicas at different temperatures decays exponentially with a characteristic length scale $\ell_\text{c}$. Sample-to-sample fluctuations are not discussed, but we shall see that they play an important role in  simulations.

A complementary analytical approach can be developed for the mean-field SK model through the replica method \cite{mezard:87}. In this case, it is more convenient to compute the free-energy cost $\Delta f(q_\text{c})=f(q_\text{c},T_1,T_2)-f(T_1)-f(T_2)$ to keep two replicas at different temperatures $T_1$ and $T_2$, with an overlap $q_\text{c}=q_{T_1,T_2}\neq 0$. The probability distribution function for that overlap is
\begin{equation}
P_N(q_\text{c}) \propto \exp[-N \Delta f(q_\text{c})]\;,
\end{equation}
which becomes a delta function at $q_\text{c}=0$ for $N\to\infty$, unless $\Delta f(q_\text{c})=0$ in a finite range around $q_\text{c}=0$. The free-energy cost $\Delta f(q_\text{c})$ is so small that temperature chaos was not found for a long time until the analytical work by \textcite{rizzo:03}, who carried out a perturbation expansion close to the critical temperature ($T_1=\Tg-\tau_1$ and $T_2=\Tg-\tau_2$) up to the ninth order in $\tau=\Tg-T$ to find
\begin{equation}
\Delta f(q_\text{c}) = \frac{12}{35} |q_c|^7 (\tau_1 - \tau_2)^2\,.
\end{equation}
Close to the critical point, $q_\text{c} \sim (\Tg - T)$. See also the numerical results by \textcite{billoire:02}, consistent with these findings.

This analytical result has convincingly clarified that spin glasses do exhibit temperature chaos (at least in the mean-field approximation), but the intensity of this phenomenon is very weak.  This explains why it had not been observed for a long time in numerical simulation of the SK model \cite{vannimenus:jpa-00209043,sommers:84,nemoto:87,biscari:90}.

The elusiveness of temperature chaos is enhanced when we move away from the realm of mean-field models and study 
short-range systems. Numerical simulations of the EA model tell a complicated story, where sample-to-sample fluctuations
play a key role and which builds on more recent equilibrium studies. Because these 3D results will be directly needed
to understand some aspects of nonequilibrium dynamics, we leave them for Sec.~\ref{subsect:Equilibrium-1}.

\subsection{Advent of spin-glass single crystals}
\label{subsect:intro-7}
Up to recent years, nearly all experiments on spin glasses were done on poly-crystalline samples.  A typical example is CuMn dilute alloys, where often the crystallite size is found to be distributed around a characteristic length of 
approximately $80\,\text{\AA}$ \cite{guchhait:15a,rodriguez:04}.  At first sight, given that poly-crystalline Cu is such a good conductor, one might be inclined to ignore grain boundaries.  However, what drives the spin-glass nature in dilute magnetic alloys is the RKKY interaction.  Its long-range and oscillatory nature is associated with the sharpness of the Fermi surface.
\begin{figure}[tb]
    \centering
    \includegraphics[width=8.5cm]{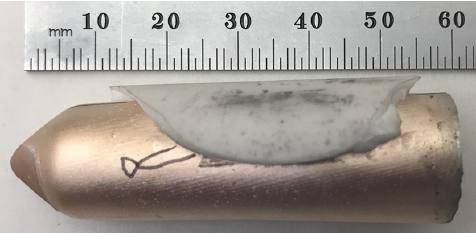}
    \caption{An as-grown crystal with part of the alumina crucible still attached.  A small secondary grain is outlined by a marker.  Later, acid etching of the ingot reduced the size and number of the secondary grains, indicating that they are shallow.  Reproduced from Fig. 5 of \citet{zhai:19}.}
    \label{fig:boule_picture}
\end{figure}

As \citet{kasuya:66} has shown, electron scattering will change the long-range $\cos(2k_\text{F} \,r_{ij})/[k_\text{F} \,r_{ij}]^3$ dependence of the RKKY interaction to $r_0/[(r_0^2-r_{ij}^2)^2+4r_0^2r_{ij}^2]$, where $k_\text{F}$ is the Fermi wave vector, $r_{ij}$ is the distance between spins $i$ and $j$, and $r_0$ is a characteristic scattering length that we take to be of the order of the crystallite size.  The lack of an oscillatory character removes {\it frustration}, a requirement for spin-glass behavior (in addition to randomness), and thus effectively decouples the individual spin-glass grains, limiting the growth of the spin-glass correlation length.  Adding to this difficulty is the distribution of grain sizes that must be taken into account for time-dependent properties.

\begin{figure}[tb]
    \centering
    \includegraphics[width=8.5cm]{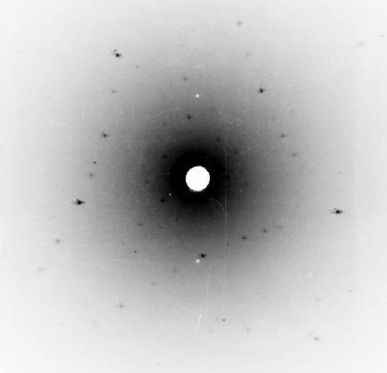}
    \caption{The Laue X-ray diffraction pattern of the sample in Fig.~\ref{fig:boule_picture} confirms it is a single crystal; the Cu$_{94}$Mn$_6$ cube sample was etched in 15\% nitric acid.  The 6 at. \% Mn concentration was estimated from scaling using the observed temperature ($\Tg=$ 31.5 K) at which the remanence disappeared.  Reproduced from Fig. 6 of \citet{zhai:19}.}
    \label{fig:Laue_x_ray}
\end{figure}

All of the complications associated with poly-crystalline materials can be avoided by the use of single crystals.  In that regard, a major step forward took place at the Materials Preparation Center (MPC) of the Ames Laboratory, U.S. Department of Energy, under the direction of D. L. Schlagel.  She was able to fabricate two single-crystal CuMn boules with $\Tg=31.5$~K and 40.6~K, respectively.  The details can be found in Appendix A of \textcite{zhai:19}.  A picture of the $\Tg=$ 31.5 K boule is reproduced in Fig. \ref{fig:boule_picture}.  Laue X-ray diffraction patterns were taken at various points along the length of the body of the same boule.  A representative example is exhibited in Fig.~\ref{fig:Laue_x_ray}, confirming that the majority of the body is a single grain.  The spots make a clear pattern that is a qualitative measure of crystal quality.  There is no evidence of twinning or strain in the crystal.

Use of small samples cut from the 6 at. \% CuMn boule allowed growth of the spin-glass correlation length to as much as 150 nm, far larger than any other glassy correlation length measured to date.  As such, it enabled observation of a phenomena predicted from simulations of the Janus Collaboration: a dynamic slowing down of the growth rate of the correlation length as the correlation length grows \cite{zhai:19}.  Many other observations will be reported in this review that were only possible because of the availability of these single crystals.

\subsection[Experimental extraction of \texorpdfstring{$\xi(t,\tw;H)$}{\$xi(t,t\_w;H)\$}]{Experimental extraction of \boldmath $\xi(t,\tw;H)$}
\label{subsect:intro-xi_eff}

\changes{Experimentally, one needs a finite magnetization to make a measurement.  This requires a magnetic field that can polarize the otherwise nonmagnetic quenched state.  \citet{bouchaud:92} introduced another piece of phenomenology, a Zeeman energy $E_{\text {Z}}$, which reduces each of the barrier heights ---recall Sec.~\ref{subsect:intro-5}--- and is given by \cite{joh:99}
\begin{equation}\label{eq:Zeeman_energy_def}
E_{\text {Z}}=-M_{\text {FC}}H = -\chi_{\text {FC}} \Ns H^2,
\end{equation}
where $\chi_{\text {FC}}$ is the field-cooled magnetization susceptibility {\it per spin} when the spin glass is cooled in a magnetic field to $\Tm$.  $\Ns$ is the number of correlated spins within a volume subtended by $\xi(\tw)$,
\begin{equation}\label{eq:Nc_correlated_spin}
\Ns = \xi(\tw)^{D-\theta/2},
\end{equation}
where $\theta$ is the replicon exponent \cite[][see also Eq.~\ref{eq:C4scaling}]{janus:17b,orbach-janus:23} and $D$ is the spatial dimension.}

\changes{Specifically, one measures}  the time dependence of the magnetization $M(t,\tw; H)$ at a fixed temperature.  \citet{nordblad:86} introduced the derivative $S(t,\tw; H)$ [we rewrite Eq. (\ref{eq:St_def}) for the reader's convenience]
\begin{equation}\label{eq:S_def}
S(t, \tw; H) = \pm {\frac {\dd M(t, \tw; H)}{\dd \ln t}}\,,
\end{equation}
with the $-$ for TRM and $+$ for ZFC experiments. This quantity exhibits a maximum at a characteristic time $\tw^{\text {eff}}$ that scales with the waiting time $\tw$.
Extraction of $\xi(t,\tw;H)$ from measurements of the time at which $S(t,\tw;H)$ has a maximum, yielding $\tw^{\text {eff}}$, derives from \citet{vincent:95} using a free-energy ``trap model'' and \citet{joh:99} using a free-energy barrier model.  These barriers or traps are created by the growth of $\xi(t,\tw;H)$ and they are invoked to account for the dynamical properties of spin glasses.  Further, the deepest trap or the maximum barrier height controls the dynamics because of the exponential growth of occupancy with increasing barrier height.  This can be seen from Fig. \ref{fig:hierarchical_organization}, and is a direct consequence of ultrametricity.

These works note that the presence of a magnetic field reduces the depth of all the traps or, equivalently, the height of all the energy barriers, by the amount of energy associated with the presence of the magnetic field, or the ``Zeeman energy'' $E_\text{Z}$.  They, however, differ in their expressions for $E_\text{Z}$.  For magnetic fields of laboratory strength the approach of \citet{joh:99} is the more appropriate of the two.

The relationship between the maximum barrier height $\Delta_{\text {max}}$ and $E_\text{Z}$ is
\begin{equation}\label{eq:arrahenius_law_Delta_max}
\Delta_{\text{max}} - \Ns\chi_{\text{FC}}H^2=k_{\text {B}}T\,\ln \tw^{\text{eff}}-k_{\text{B}}T\,\ln \tau_0,
\end{equation}
where $E_{\text {Z}},~\Ns$, and $\chi_{\text{FC}}$ are defined below Eq.~\eqref{eq:Zeeman_energy_def}, the relationship between $\Ns$ and the correlation length $\xi(\tw)$ exhibited in Eq.~\eqref{eq:Nc_correlated_spin}, and $\tau_0 \approx \hbar/k_{\text {B}}\Tg$ is an ``exchange time.''  Thus, by measuring $\tw^{\text {eff}}$ as a function of magnetic field $H$, one can extract $\Ns$ and thence $\xi(t,\tw;H)$.  This is the \changes{main} vehicle one has available to determine $\xi(t,\tw;H)$ from experiment \changes{(see Sec.~\ref{sec:alternative_xi} for possible alternatives)}. The usual approaches of neutron diffraction or X-ray scattering measure only two-spin correlations.  Unfortunately, for spin glasses, this averages to zero over sample dimensions.  Instead, $\xi(t,\tw;H)$ derives from a four-spin correlation function \cite[see Sec.~\ref{subsect:off-equilibrium-no-field-3}]{marinari:96}.

\changes {Combining the reduction of the energy barrier, Eq.~\eqref{eq:Zeeman_energy_def}, to the peak of the relaxation function $S(t,\tw ;H)$ of Eq.~\eqref{eq:S_def} one obtains:
\begin{align}\label{eq:xi_zeeman_def}
    \ln \left[ \frac{t^\mathrm{eff}_\mathrm{H}}{t^\mathrm{eff}_\mathrm{H \to 0^+}}\right] \propto  \Ns H^2\propto   \xi^{D-\theta/2} \, H^2  \, .
\end{align}
\changes{Because the replicon exponent $\theta$ depends on $\xi$ (see 
Fig.~\ref{fig:replicon-scaling}), the resolution of Eq.~\eqref{eq:xi_zeeman_def} is self-consistent in $\xi$. \citet{zhai:19} proposed an interpolation function for $\theta$ that holds for \emph{both} the RSB and the droplet approach.\footnote{Due to space constraints, we do not reproduce the full calculation here; it can be found in Appendix B of \citet{zhai:19}.}
}
It is commonplace to age the spin glass over a time $\tw$, then probe its time dependent magnetization at a time $t$ after a magnetic field is turned on at $\tw$, the zero-field-cooled (ZFC) magnetization.  Alternatively, the system can be prepared in the presence of a magnetic field, at a time $t$ after which the field is reduced to zero at time $\tw$, the thermoremanent magnetization (TRM); see Sec. \ref{subsect:intro-2}. The spin-glass correlation length is then denoted by $\xi(t,\tw;H)$.}

\begin{figure}[tb]
    \centering
    \includegraphics[width=8.5cm]{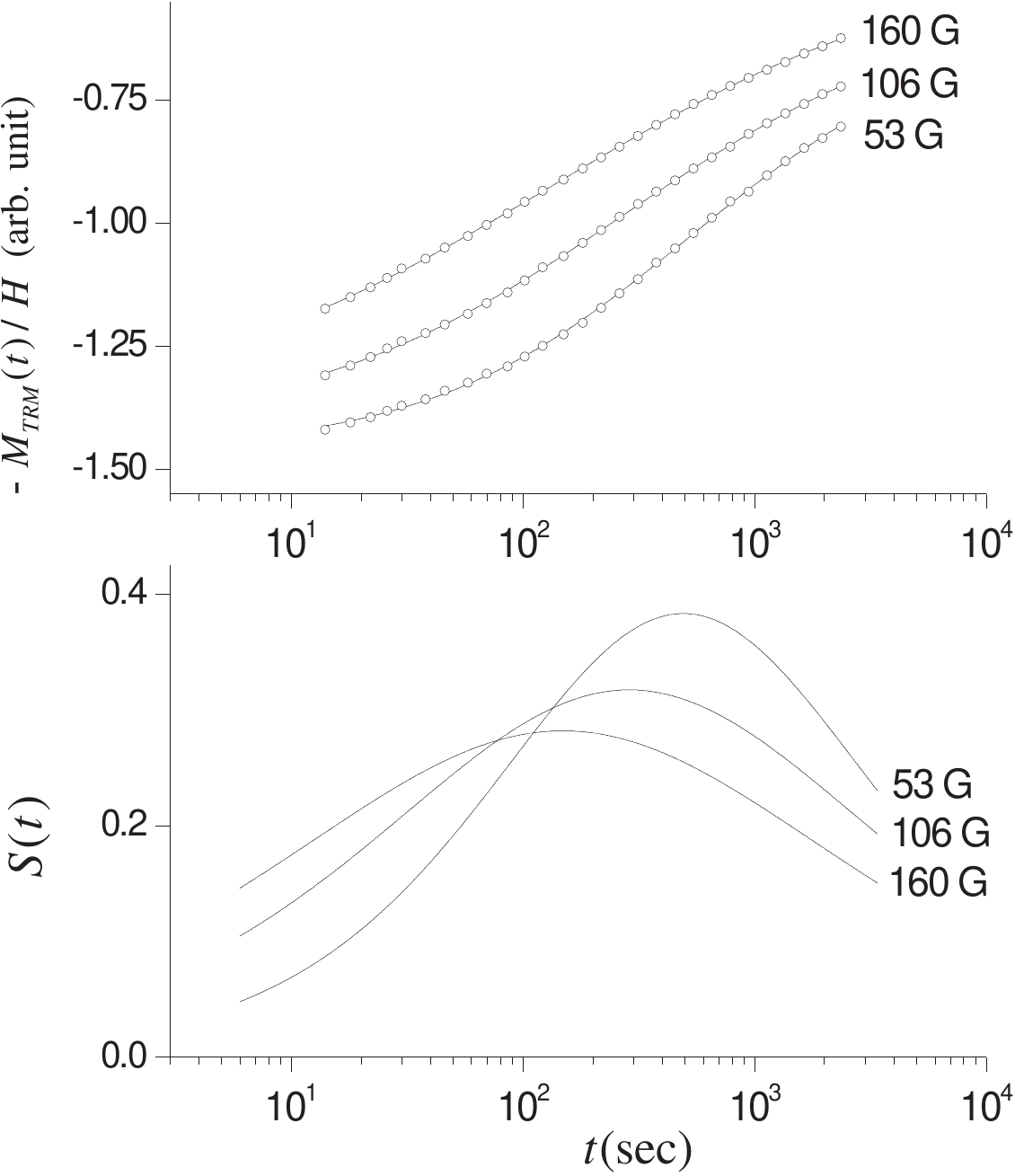}
    \caption{$-M_{\text {TRM}}/H$ against  measurement time $t$ (top), with the solid lines from  the analytic fit, for CuMn 6 at\% for various changes in magnetic field, and for $S(t,\tw; H)$ (bottom) calculated by taking the time derivative of the analytic fit (solid line) of $-M_{\text {TRM}}/H$ (top). Reproduced from Fig.~1 of \citet{joh:99}.}
    \label{fig:St_joh}
\end{figure}

\begin{figure}[tb]
    \centering
    \includegraphics[width=8.5cm]{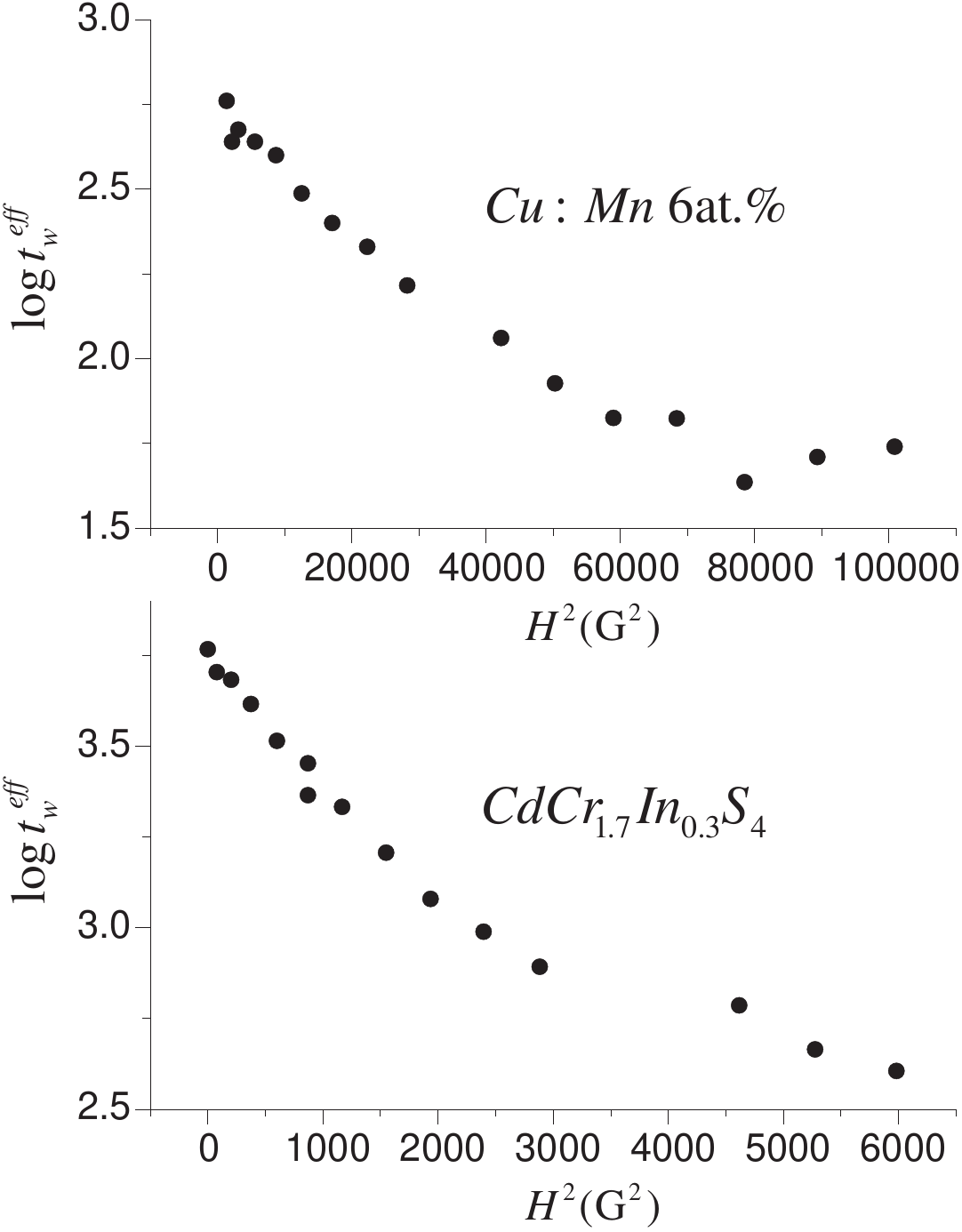}
    \caption{A plot of log\,$\tw^{\text {eff}}$ [equivalently $\Ns\chi_{\text {FC}}H^2$ from Eq. (6)] vs $H^2$ for CuMn 6 at.\% ($T/\Tg=0.83, \tw=480$~s) (top) and for the insulating spin-glass CdCr$_{1.7}$In$_{0.3}$S$_4$ ($T/\Tg=0.21, \tw=3410$~s) at fixed $\tw$ and $T$ (bottom). The dependence is linear for magnetic fields less than 250 and 45 G, respectively.  Reproduced from Fig. 2 of \citet{joh:99}. }  
    \label{fig:teff_vs_H2_joh}
\end{figure}

\changes{\citet{joh:99} used Eq.~\eqref{eq:S_def} to display} the time dependence of $S(t,\tw;H)$ for different values of $H$ as displayed in Fig. \ref{fig:St_joh}. Following Eq.~\eqref{eq:arrahenius_law_Delta_max}, they were able to plot the log\,$\tw^{\text {eff}}$ vs $H^2$, as exhibited in Fig. \ref{fig:teff_vs_H2_joh}.  The break away from linearity with $H^2$ has now been ascribed to nonlinear effects \cite{orbach-janus:23} rather than any breakdown of the relationships in Eq.~\eqref{eq:arrahenius_law_Delta_max}.

The extraction of $\xi(t,\tw;H)$ follows immediately from Eq. \eqref{eq:Nc_correlated_spin} once $\Ns\chi_{\text {FC}}H^2$ has been determined from plots such as Fig.~\ref{fig:teff_vs_H2_joh}.  The combination of the use of this quantity, and the availability of single crystals discussed in Sec.~\ref{subsect:intro-7}, has enabled the first direct comparisons between numerical and experimental measurements of the growth of spin-glass domains. 

This fundamental question of relating the experimental correlation length derived from the Zeeman effect, which will be termed $\xi_\text{Zeeman}(t,\tw;T)$, to the different length scales extracted from simulations (see the glossary in Sec.~\ref{subsect:glossary}) will be the topic of Secs.~\ref{subsect:off-equilibrium-in-field-fixed-T-4} and~\ref{subsect:off-equilibrium-in-field-fixed-T-5}.

\subsection[Possible alternatives for extraction of \texorpdfstring{$\xi(t,\tw;H)$}{\$xi(t,t\_w;H)\$}]{Possible alternatives for extraction of \boldmath $\xi(t,\tw;H)$}  \label{sec:alternative_xi}

\changes{There have been other experimental methods suggested for the existence or extraction of $\xi(t,\tw;H)$ beyond the use of the magnetic-field dependence of the response function $S(t)$.
\subsubsection{Nonlinear susceptibility}\label{sec:non-linear-susc}
The nonlinear susceptibility in spin glasses was discovered by  \citet{miyako:79}.  It has been recently pointed out \cite{biroli:21} that using the nonlinear susceptibility might be a nice way to access $\xi(t,\tw;H)$. In our context, one would study the magnetization $M(t,\tw)$ at several values of the external magnetic field $H$ and expand  $M$ as
\begin{equation}
 M=\chi_1 H +\frac{\chi_3}{3!}H^2+\frac{\chi_5}{5!}H^5+\ldots
\end{equation}
For sure, the study of nonlinear susceptibilities such as $\chi_3$ and $\chi_5$ is a classic topic in the spin-glass literature \cite{binder:86}, particularly in an equilibrium context (\emph{i.e.}, for temperatures above $\Tg$). Furthermore, the experimental finding  of scaling behavior for $\chi_3$ near $\Tg$~\cite{gunnarsson:91} played a major role in convincing the community that a thermodynamic phase transition takes place at $\Tg$, implying that spin-glass phenomenology is not merely a dynamic effect. Perhaps unsurprisingly, a similar approach has been followed in the study of supercooled liquids, a context where the physical origin of the glassy dynamics is still debated. In particular, the analogues of $\chi_3$ and $\chi_5$ have been measured in the dielectric polarizability of a supercooled liquid~\cite{berthier:05,albert:16,albert:20}.

Given these precedents, it would be nice to extend this approach to the experimental study of spin glasses at $T<\Tg$ because it would bridge the study of magnetic disordered systems with that of supercooled liquids. A first step in this sense was taken in \citet{janus:17b}, where $\chi_3(t,\tw)$ was investigated numerically. The following simple ansatz was found to approximately account for the simulation data at long times,
\begin{equation}
\chi_3(t,\tw)= [\xim]^{D-\theta} G(t/\tw)\,,
\end{equation}
where $G$ is a simple scaling function displaying a single maximum. The analogy between the above expression and the quantities employed in the analysis in Secs.~\ref{subsect:off-equilibrium-in-field-fixed-T-4} and
\ref{subsect:off-equilibrium-in-field-fixed-T-5} is striking.
}
\changes{
\subsubsection{Nonresonant hole burning}

A series of papers has established nonresonant spectral hole burning (NSHB) as a vehicle for investigation of the nonexponential relaxation of glass-forming liquids and solids  \cite{chamberlin:18,schiener:96,chamberlin:97}. \citet{chamberlin:99} applied a magnetic analogy of this technique to study the dynamical properties of spin glasses.  His technique utilized a large-amplitude low-frequency pump oscillation, followed by a probe step, with a phase-cycling scheme.  The technique requires that the sample absorb sufficient energy during the pump oscillation to modify its response spectrum, and that this modification persists long enough to be measured on laboratory time scales. Thus, NSHB can investigate linear response and thermal recovery.

\citet{chamberlin:99} measured magnetic NSHB on a AuFe 5~at.\% sample.  He found that the bulk spectrum is heterogeneously broadened.  Individual domains in the sample are heated, independent of one another.  Their relaxation, on the time scale of the experiments, was also found to be independent of the other domains.  As a function of recovery time, the monotonic return to equilibrium of the selected slow degrees of freedom shows that the domains are coupled to a common thermal bath, but with negligible direct coupling to the other slow degrees of freedom.  This independence strongly suggests that the domains represent correlated regions with finite length scales.  Thus, this study underpins the basis for the existence of the correlation length $\xi(t,\tw;H)$.

\subsubsection{Mesoscopic electrical resistance noise measurements}
Weissman and colleagues, in a series of papers \cite{israeloff:89, weissman:88, weissman:90, weissman:93, israeloff:91, Alers:92}, proposed and showed experimentally that noise in the electrical resistance $R$ can provide a probe of spin dynamics in spin glasses because of universal conduction fluctuations (UCF) effects coupled to spin configurations.  Their results for CuMn \cite{israeloff:91} displayed a minimum typical event size of $\approx 3 \times 10^4$ spins at 13 K and $\approx 1.7\times~10^5$ spins at 23 K.  The spectral density of $\delta R/R$ of ``large samples'' (with number of spins, $N_\text{s} > 10^{12}$) rises sharply at a temperature $T_\text{n}$ near the spin-glass magnetic freezing temperature $\Tg$, and has a spectral shape sensitive to the magnetic field. 

These results, when interpreted as the number of spins in a correlated volume, lead directly to the concept of a temperature dependent correlation length.  Subsequent experimental results for CuMn \cite{joh:99} extracted the number of correlated spins as a function of the waiting time $\tw$.  The results for a temperature $T= 28\text{ K} \approx 0.89 \Tg$ varied from $1\times 10^6$ to $4\times 10^6$ (their Fig.~3), quite consistent with the $\delta R/R$ minimum estimates cited above for a typical event.  The important point is that universal conduction fluctuations provided evidence for correlated spins in spin glasses, generating a minimum number that is consistent with later observations utilizing a completely different approach.

Further insight from UCF addressed alternate physical descriptions for spin-glass dynamics in \citet{israeloff:91}.  To quote: ``we find noise statistics in extreme disagreement with a dilute droplet or cluster model.  We then determine a size-frequency scaling parameter which is inconsistent with a basic assumption of all droplet-like models [\emph{i.e.}, no sign is found of a size-frequency correlation, while the hierarchical picture fits in `surprising detail'].  This parameter and others are shown elsewhere \cite{sherrington:75, fert:80} to fit a hierarchical kinetic picture \cite{levy:81} inspired by the Parisi solution \cite{mezard:84}.''  This finding is important because, for example, later extraction of the spin-glass coherence length $\xi(t,\tw;H)$ was unable to exhibit a difference between the prediction of the droplet and the hierarchical model \cite[see Fig. 3 of ][]{joh:99}.
}
\subsection{Spin-glass simulations and dedicated computers}
\label{subsect:intro-dedicated_computers}
Simulating the low-temperature behavior of a spin glass is a challenge common to both physicists and computer scientists.

Interest from computer scientists arises because finding the ground state of a spin glass is an archetypically hard computational problem. Indeed, if the interactions of the spin glass can only be drawn in a nonplanar graph,\footnote{A planar graph
is one that can be drawn on a piece of paper without
edge crossings, such as the square lattice with open boundary conditions. The difference between planar and two-dimensional is important, as discussed below.}
then finding the ground state belongs to the category of NP-complete problems \cite{barahona:82b,istrail:00}. Should a truly efficient algorithm for finding the ground state of a nonplanar spin glass be discovered, efficient algorithms for a large category of hard computational problems (such as the traveling salesman problem) would follow~\cite{papadimitriou:13}.

Physicists are often more interested in the behavior of a system at finite temperature $T$
(one could say that physicists wish to minimize the free energy, rather than
the energy). However, the physicist and the computer-scientist problems are
tightly connected, because the free energy becomes the energy when we approach
$T\!=\!0$.

Difficulties faced by physicists depend strongly on whether they consider
equilibrium or off-equilibrium dynamics:
\begin{itemize}
\item Equilibrium studies are (unfortunately) limited to small systems because
reaching thermal equilibrium may require an inordinately long simulation time
for some \emph{samples} \cite[in cubic lattices, equilibration
times at low temperatures fluctuate from sample to sample by \emph{orders
of magnitude},][]{janus:10,yllanes:11,fernandez:16}. Further computational difficulties are posed
by the fact that a large number of samples need to be simulated in order to
get meaningful results, because quantities of interest wildly fluctuate from
sample to sample.

\item Nonequilibrium studies are conducted on lattices
of large linear size $L$, sufficient to ensure that the coherence length $\xi(t,\tw;H)$
is always much smaller than $L$. This condition is required in order to compare with laboratory correlation lengths, which, in general, are smaller than physical sample sizes (with the exception of thin films, see Sec. \ref{subsect:off-equilibrium-in-field-fixed-T-3}). Indeed, in space dimension
$D$, the system can be approximated as composed of $[L/\xi(t)]^D$ more or less
independent samples of effective size $L_{\text{eff}}\approx \xi(t)$. Hence,
in close analogy to experiments, a simulation of just one very large sample
automatically carries out the average over samples. The main difficulty is
that one is limited to simulation algorithms that mimic natural dynamics, such
as Metropolis \cite[see, \emph{e.g.},][]{sokal:97} and, as already explained in
Sec.~\ref{subsect:intro-xi_eff}, the natural growth
of $\xi(t,\tw; H)$ is slow. This slowness
calls for very long simulation times on few very large lattices.
\end{itemize}

Simulating spin glasses is not always hard.  Efficient
algorithms for finding the ground state are known for planar interaction
graphs. It is also important to note that planar and two-dimensional are not quite the same thing: the bilayer that was shown in \citet{barahona:82b} to pose an NP-hard problem is certainly two-dimensional but nonplanar. The square lattice with periodic
boundary conditions is mildly nonplanar, but still allows one to find ground
states for very large lattices~\cite{khoshbakht:17,caracciolo:22}. For $T>0$,
the problem is also simpler on planar (or quasi-planar) graphs, such as the
square lattice with periodic boundary conditions, for which nonlocal simulation algorithms (named cluster
algorithms) can shorten simulation times by orders of
magnitude \cite{houdayer:01,wang:05}.

Unfortunately, the situation is much worse for nonplanar graphs, such as the
simple-cubic lattice used in most $D=3$ simulations. At $T\!=\!0$,
ground states have been found in cubic lattices with size $L\!\leq\!
14$ with the help of a heuristic algorithm~\cite{marinari:01}.  See
\textcite{caracciolo:22} for information on sure solvers. Many algorithms have been proposed for $T\!>\!0$: simulated tempering~\cite{marinari:92}, umbrella sampling~\cite[see, \emph{e.g.},][]{berg:92,janke:98c}, the Wang-Landau algorithm~\cite{wang:01} or population annealing~\cite[see, \emph{e.g.},][]{machta:10}. Nevertheless, the best-performing
algorithm for $T\!>\!0$ equilibrium simulations appears to be parallel
tempering~\cite{geyer:95,hukushima:96,tesi:96}, which, with the help of a custom-built computer, has been able to equilibrate cubic lattices of size $L\!=\!32$
at $T\!\approx\! 0.64\Tc$~\cite{janus:10b}.
Temperature chaos, Sec.~\ref{subsect:intro-6}, is
the physical mechanism that makes it difficult for algorithms that exploit the correlation at different temperatures (\emph{e.g.}, simulated annealing or parallel tempering) to reach thermal equilibrium at even lower temperatures \cite{fernandez:13,billoire:18}.\footnote{In a very chaotic sample, thermalization achieved at a higher temperature would be almost irrelevant for sampling at a lower temperature.  Thus, the thermalization process would need to restart at the lower temperature.}

Lacking better algorithms, the best we can
do is to increase computing power. Computers specifically designed for
high-energy-physics simulations, such as RTN~\cite{azcoiti:92} and
APE100~\cite{battista:92}, have also been quite useful for spin glasses [see \textcite{badoni:93,ciria:93} in the case of RTN and
\textcite{marinari:96,marinari:98d,marinari:98f} for APE100]. More recently,
graphics processing units (GPUs) have proven to be extremely powerful for simulating \emph{quantum} spin glasses~\cite{bernaschi:24}. Furthermore, a
few computers have been \emph{specifically} designed for the simulation of
spin glasses, such as the Ogielski machine~\cite{ogielski:85},
SUE~\cite{cruz:01}, and the Janus~I \cite{janus:09,janus:12b} and~II supercomputers~\cite{janus:14}.

The great generality of the spin-glass problem uncovered through
the computer-science perspective (Sec.~\ref{subsect:intro-1}) has encouraged people to design and build computers for finding
ground states of spin glasses of various types. This includes D-wave's quantum
annealing chips \cite{johnson:11,mcgeoch:22}, as well as hardware based on a
variety of algorithms and/or physical principles
\cite[see, \emph{e.g.},][]{hayato:19,matsubara:20,takemoto:20,mcgeoch:22,mcmahon:16}.

\section{The equilibrium picture.}\label{sect:Equilibrium}
\subsection{Some important numerical results in \boldmath $D=3$}
\label{subsect:Equilibrium-1}

An overview of analytical and numerical results for the equilibrium spin-glass phase
is beyond the scope of this review \cite[for a recent account see, \emph{e.g.},][and references therein]{martin-mayor:22}. This section discusses some results that will be of use for the interpretation of nonequilibrium data. 

As explained in Sec.~\ref{sect:models}, numerical studies are conducted on an Ising-like system, the Edwards-Anderson model,
whose Hamiltonian is defined in Eq.~\eqref{eq:1C:EAHam}. In $D=3$, this system experiences a continuous phase transition at $\Tg = 1.1019(29)$. See \citet{janus:13} for this value of $\Tg$ and for critical exponents. 

\begin{figure}[tb]
    \centering
    \includegraphics[width=\linewidth]{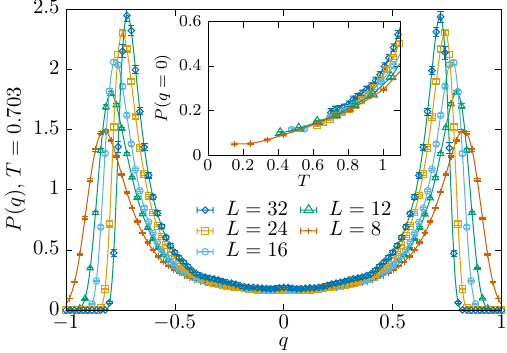}
    \caption{Probability density function $P(q)$ of the spin overlap at $T=0.703$
    for different system sizes. A strictly positive plateau is quickly reached in the $q=0$ sector, even if the position
    of the peaks is still very noticeably shifting with increasing $L$. The inset shows how $P(0)$ grows roughly linearly with temperature. Data from \citet{janus:10}.
    }  
    \label{fig:Pq}
\end{figure}

Perhaps the most straightforward quantitative description of the spin-glass phase comes from the 
probability density function $P(q)$ for the spin overlap $q$ exhibited in Eq.~\eqref{eq:def-overlap}. This is plotted
in Fig.~\ref{fig:Pq} for several system sizes and a temperature of $T=0.703\approx0.64\Tg$.  Notice the two symmetrical peaks at $q=\pm\qEA(L)$
and, crucially, that the probability density does not go to zero at $q=0$. Many other studies conducted over the years \cite{berg:98,marinari:98c,marinari:98d,katzgraber:01,berg:02,wang:20} are in agreement on this point. Should $P(q)$ conserve this shape in the large-$L$ limit, the resulting picture would be 
consistent with RSB ---recall Sec.~\ref{subsect:intro-4}--- but this extrapolation is controversial to this day \cite{moore:21,yucesoy:12,billoire:13,billoire:14b,billoire:17,wang:17}. The dispute is, however, immaterial for our purposes because, in order to model the behavior of nonequilibrium spin-glass dynamics 
at finite times, we shall only need the value of $P(q)$ for finite sizes.

\begin{figure}[tb]
    \centering
    \includegraphics[height=\linewidth, angle=-90]{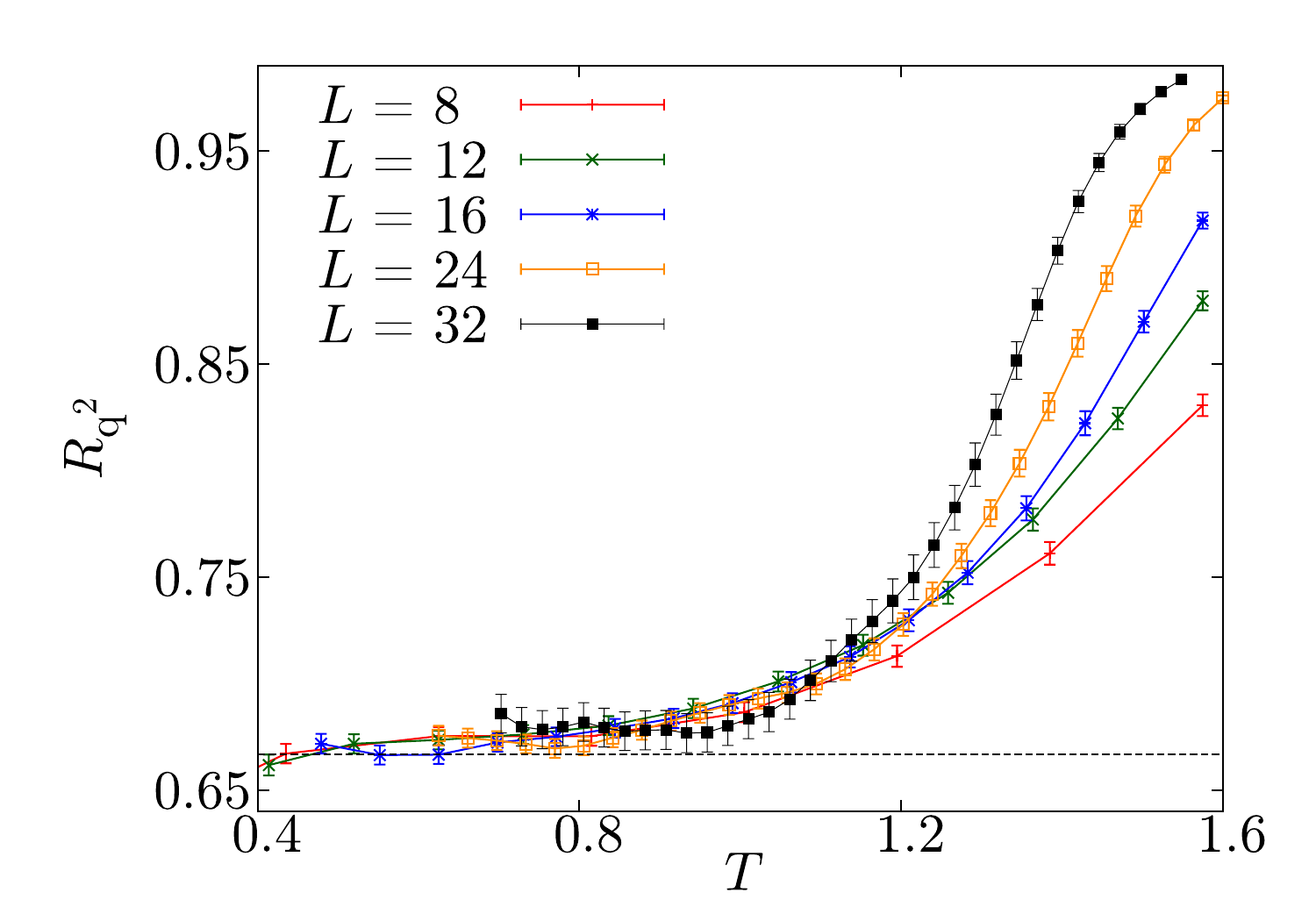}
    \caption{The ratio of Eq.~\eqref{eq:stochastic-stability} in the EA model approaches the expected value $R_{q^2}=2/3$ for stochastically stable 
    systems as $L$ grows. Reproduced from \citet{janus:10}.}  
    \label{fig:stochastic-stability}
    \end{figure}

A second important feature of the equilibrium spin-glass phase is stochastic stability \cite{kushner:74,parisi:01}. In simple terms, if a system with random components is generic, {\it i.e.}, it behaves as a typical random system and is not drastically changed by a small perturbation, then it is stochastically stable. Several papers have conclusively established that the mean-field spin glass has this property~\cite{guerra:97, ghirlanda:98, aizenman:98}. A numerically testable formulation, relying on overlap equivalence \cite{contucci:06,janus:10}, is provided by the following relation
\begin{equation}\label{eq:stochastic-stability}
    R_{q^2} = \frac{\overline{\langle q^4\rangle - \langle q^2\rangle^2}}{ \overline{\langle q^4\rangle} - \overline{\langle q^2 \rangle}^2} = \frac23\, .
\end{equation}
Fig.~\ref{fig:stochastic-stability} plots this ratio and shows that the value $R_{q^2}=\tfrac23$ is reached for large $L$ below the critical temperature. Overlap equivalence and stochastic stability are closely related to the ultrametric property of the RSB phase \cite{parisi:00, iniguez:96,ghirlanda:98, panchenko:13}, which has been more difficult to observe directly \cite{hed:04,janus:11,parisi:15}.
Stochastic stability allows us to establish a direct quantitative relationship between the off-equilibrium dynamics and the equilibrium properties of a spin-glass system (see Sec.~\ref{subsect:off-equilibrium-no-field-4}).

The final key property of the low-temperature phase that we shall need is temperature chaos, introduced in Sec.~\ref{subsect:intro-6}. 
The main numerical results have been reported in \citet{fernandez:13} and are summarized here.
\begin{figure}[tb]
    \centering
    \includegraphics[height=\linewidth,angle=-90]{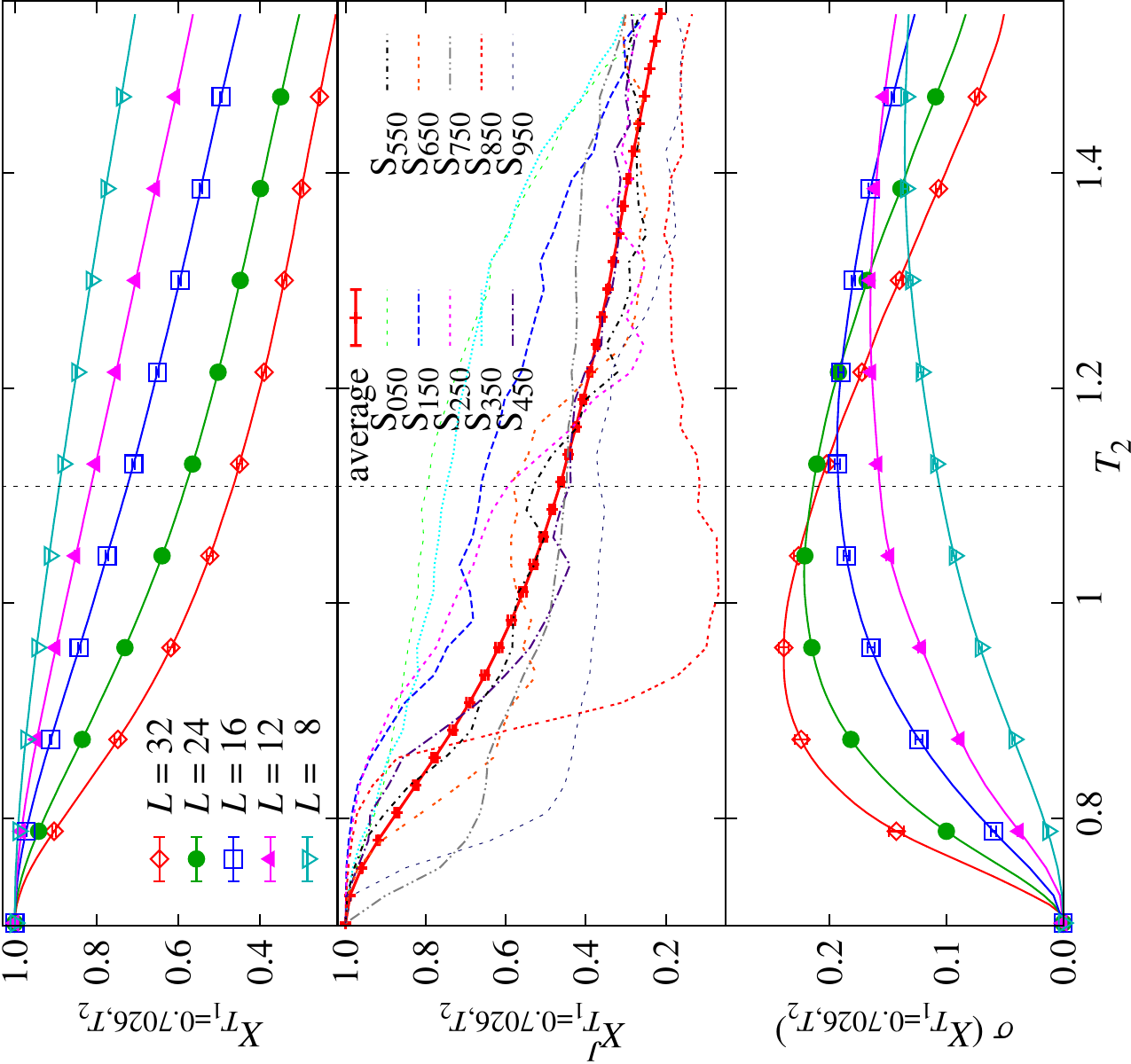}
    \caption{Different views on $X_{T_1,T_2}^J$, Eq.~\eqref{eq:chaotic_parameter}, as a function of $T_2$ ($T_1\approx0.64\Tg$, the vertical line is $T_2=\Tg$).  Top: for all our system sizes, sample-average of $X_{T_1,T_2}^J$.  Center: for $L=32$, we show $X_{T_1,T_2}^J$ for ten samples $S_a$ evenly spaced in a list of growing $\tau_{\text {exp}}$.  Bottom: dispersion ({\it i.e.}, square root of variance over the samples) of $X_{T_1,T_2}^J$ for $8\leq L \leq 32$.  Reproduced from Fig.~2 of \citet{fernandez:13}.}
    \label{fig:chaotic_parameter}
\end{figure}

The authors of~\citet{fernandez:13} have studied the chaoticity parameter $X^J_{T_1,T_2}$, defined by Eq.~\eqref{eq:chaotic_parameter}, with $T_1=0.703\approx 0.64 \,\Tg$ and $T_2$ in a range extending from $T_1$ to above \Tg.  A summary of their results is exhibited in Fig.~\ref{fig:chaotic_parameter}.
The upper panel reports the sample average of the chaoticity parameter, Eq.~\eqref{eq:chaotic_parameter}.  It decreases with system size, but is far from being zero (the predicted value for a model expected to have complete temperature chaos).

The central panel in Fig.~\ref{fig:chaotic_parameter} provides deeper insight.  The chaoticity parameter is shown for ten samples that have different autocorrelation times (a proxy for the difficulty in thermalizing that particular sample).  In some samples, the chaoticity parameter changes very smoothly, while in others sharp changes are observed over a small temperature range. Such large sample-to-sample fluctuations are quantified in the bottom panel, where they are shown to increase with system size in the spin-glass phase.

The overall picture that comes out from the numerical study in \citet{fernandez:13} is that, for the sizes currently accessible, temperature chaos is a phenomenon only visible in rare samples and controlled by the large-deviation principle
\begin{equation}
\text{Probability}[X^J_{T_1,T_2}>\varepsilon] = \exp[-L^D \Omega_{T_1,T_2}(\varepsilon)]\;.
\end{equation}
The large-deviation function is well approximated by $\Omega_{T_1,T_2}(\varepsilon) \propto |T_1-T_2|^b \varepsilon^\beta$ with $\beta \approx 1.7$. 
The exponent $b$ can be related to the chaos exponents of Sec.~\ref{subsect:intro-6}
through a scaling argument: $(\Ds/2-y)=D/b$. In this refined scenario, however, the length scale $L_\text{c} \sim (\Delta T)^{-b/D} \sim (\Delta T)^{-1/(\Ds/2-y)}$ plays a different role with respect to the chaotic length scale $\ell_\text{c}$ in the droplet model: the large-deviation principle above implies that chaotic samples will become typical for $L > L_\text{c}$.

Nonetheless, the experiments measuring temperature chaos are made in the out-of-equilibrium regime. For this reason, numerical simulations that better reproduce such an effect have been reported in~\citet{janus:21} and will be discussed in more detail in Sec.~\ref{subsubsect:off-equilibrium-in-field-several-T-1-numerical}.

\subsection{Brief overview of results in space dimensions larger than three and the spin-glass transition in a field}\label{subsect:Equilibrium-2}
We start by describing the behavior of spin glasses in space dimensions larger than three in the absence of an external magnetic field.

The critical behavior of the EA model is well described by an $n$-component $\phi^3$ field theory with a tensorial coupling in the $n \to 0$ limit \cite{harris:76}. In six dimensions (the upper critical dimension in zero magnetic field, recall Sec.~\ref{sect:models}), the phase transition exhibits mean-field exponents with computable logarithmic corrections~\cite{ruizlorenzo:98,ruizlorenzo:17}. Near and below $D=6$, there should be a phase transition with nontrivial critical exponents~\cite{harris:76,green:85}. This scenario has been tested in numerical simulations in dimensions four and six with very good agreement~\cite{badoni:93,wang:93,parisi:96b,marinari:99,aguilarjanita:24}.
The low-temperature phase in four dimensions follows the RSB predictions.  Both the overlap and the link overlap probability distributions are nontrivial.

The behavior of the EA model in the presence of a magnetic field is not well understood. 
A first attempt is to try the Wilsonian approach~\cite{wilson:74,amit:05}. One begins with the mean-field effective Hamiltonian~\cite{bray:80} 
and introduces fluctuations to determine at which dimension these fluctuations are relevant.  This determines the upper critical dimension in a field, $D^H_\text{u}$.

This upper critical dimension can also be defined as the smallest dimension in which the critical exponents take the mean-field values. However, in contrast with the zero-field case, where the effective Hamiltonian provides $D_\text{u}=6$, no stable fixed point at zero coupling has been found in six dimensions for one-loop computations~\cite{bray:80,pimentel:02}. The two-loop computation finds that the origin (Gaussian fixed point) is unstable, but finds another (nontrivial and maybe nonperturbative) fixed point that could describe the transition in a field
~\cite{charbonneau:17,charbonneau:19}.

Furthermore, it is interesting to note that the transition in six dimensions could be quasi-first-order \cite{holler:20}. $D^H_{\text{u}}\!=\!8$ has been proposed in a recent computation \cite{angelini:22}.  Finally, an analysis based on high-temperature series finds signatures of a phase transition for $D\ge 6$~\cite{singh:17}.

The droplet model finds that there is no stable spin-glass phase in a magnetic field at any finite dimension~\cite{mcmillan:84,fisher:86,bray:87,fisher:88}. A modified version of the droplet~\cite{yeo:15} finds a phase transition only for $D>6$~\cite[see also][]{bray:11,parisi:12}.

Numerically, four- and six-dimensional models display a continuous phase transition~\cite{janus:12,aguilarjanita:24} in a magnetic field. Unfortunately, the 
estimate of the critical exponents in six dimensions did not allow the determination of whether $D^H_\text{u}=6$. In addition, numerical simulations in three dimensions have not found a transition in a field~\cite{janus:14b,janus:14c,zhai-janus:21}.

As in zero field, numerical simulations in finite dimensions can be supplemented by numerical simulations in one-dimensional lattices with couplings decaying algebraically (the power law of the decay can be related to the dimensions of regular lattices). The results are contradictory: some studies claim that no spin-glass phase in a field survives below six dimensions \cite{katzgraber:05b,katzgraber:09,vedula:23,Vedula:24}, while other groups find a phase transition~\cite{leuzzi:09,dilucca:20}.

To summarize the situation, it is widely accepted that the upper critical dimension in the absence of a magnetic field is 6. The behavior below six dimensions agrees with the predictions of the RSB theory, at least for system sizes that one can equilibrate. However, there remain many important open issues regarding the equilibrium behavior in an externally applied magnetic field.

\section{Off-equilibrium dynamics: time evolution at no (or very small) magnetic field.}\label{sect:off-equilibrium-no-field}

\subsection{Aging in experiments} \label{subsect:off-equilibrium-no-field-1}
The concept of aging is not exclusive to spin glasses.  It is a ubiquitous phenomenon found in a wide variety of materials \cite{keim:19,paulsen:25}.  Examples are  polymer mechanics \cite{struik:78}; structural glasses \cite{song:20,richert:24}; strongly correlated electronic systems \cite{wang:08}; gelatin gels \cite{parker:10}; magnetic nanoparticle assemblies \cite{parker:05}; and hysterons, models of interacting binary systems, from crumpled sheets to frustrated metamaterials \cite{vanhecke:21}.

In the ``early days'' (before 1984, see below), it was believed that the field-cooled magnetization of spin glasses was an equilibrium state \cite {vannimenus:jpa-00209043}.  This was based on the observation that the FC magnetization was roughly time independent, while the ZFC magnetization increased with increasing time.  An example of the two is exhibited in Fig. \ref{fig:ZFC_TRM_FC_protocols} for a CdCr$_{1.7}$In$_{0.3}$$S_4$ sample \cite{vincent:24,dupuis:02}.

This interpretation was accepted until the experiment by \citet{chamberlin:84}, who measured the FC magnetization, then cut the field to zero to extract the time dependence of the TRM.  The TRM time decay was found to depend on the time spent in the FC state.  In other words, though the FC magnetization was itself approximately time independent, its subsequent dynamics after a magnetic field change {\it depended} on the time spent in the FC state.

\begin{figure}
    \centering
    \includegraphics[width=\linewidth]{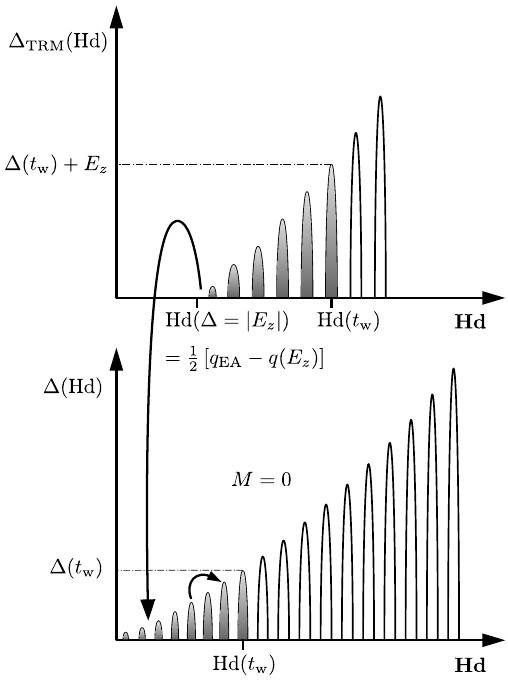}
    \caption{A cartoon illustrating the TRM dynamics associated with the reduction of barrier heights $\Delta(t,\tw;H)$ by the Zeeman energy $E_{\text {Z}}$.  Top:  the system with magnetization $M_{\text {FC}}$, the field-cooled magnetization. Bottom: the system after $H$ was cut to zero with magnetization $M=0$.  The arrow represent the decay of $M_{\text {FC}}$ after $H\rightarrow 0$.  Reproduced from Fig. 11 of \citet{orbach-janus:23}}  
    \label{fig:Hd_barriers}
\end{figure}

This demonstrated that the FC magnetization was anything but an equilibrium state.  \citet{lundgren:83}, using ZFC time-dependent measurements, showed similar dynamics.  The two are complimentary according to the superposition principle first enunciated by the Uppsala group \cite{nordblad:86,lundgren:86,djurberg:95} [we reproduce Eq. (\ref{eq:superposition_M}) for the reader's convenience]:
\begin{equation}\label{eq:superpositionM}
M_{\text {TRM}}(t,\tw) + M_{\text {ZFC}}(t,\tw)=M_{\text {FC}}(0,\tw+t).
\end{equation}
We now know that this relationship is valid only in the $H\rightarrow 0$ limit \cite[][see Sec.~\ref{subsect:off-equilibrium-in-field-fixed-T-5}]{orbach-janus:23}.  That analysis uses the Hamming distance, Hd, the ``distance'' between spin-glass states associated with free-energy barriers $\Delta(t,\tw;H)$.  It is defined in terms of the overlap between two states, $\alpha$ and $\beta$, with $q_{\alpha,\beta}$ as in Eq.~\eqref{eq:overlap_def}, where the state has evolved from its initial quenched state $\alpha(t=0, \tw=0;H)$ to $\beta(t,\tw;H)$ in a magnetic field $H$:
\begin{equation}\label{eq:Hd_def}
\mathrm{Hd}={\frac {1}{2}}(q_{\text {EA}}-q_{\alpha\beta})
\end{equation}
The free-energy barrier separating states \changes{has been found to increase more rapidly than linear with Hd \cite{orbach-janus:23}} as exhibited in Fig.~\ref{fig:Hd_barriers}.  The specific relationship between Hd and barrier heights was first developed by \citet{vertechi:89}.

The decay of $M_{\text {FC}}$ after the magnetic field is cut ({\it e.g.}, a TRM protocol) is controlled by the states that lie at the highest occupied barrier as displayed in Fig. \ref{fig:Hd_barriers}.  This is a direct consequence of an exponentially increasing occupancy of states because of ultrametric symmetry, as can be inferred from Fig. \ref{fig:ultrametric_tree}.

The significance of Fig. \ref{fig:Hd_barriers} is that, in the presence of a magnetic field $H$, the barriers are quenched for values of Hd such that  $\Delta(\text{Hd})\leq |E_\text{Z}|\equiv \Delta_{E_\text{Z}}(\text{Hd})$ with $E_\text{Z}$  defined by Eq. \eqref{eq:Zeeman_energy_def}.  Equation \eqref{eq:Hd_def} can be written alternatively to take into account this cancellation.  Take $q_{\alpha\beta}(E_{\text {Z}})=q_{\text {EA}}-q(E_\text{Z})$ to be the value of $q_{\alpha\beta}$ at the value of $q$ where $\Delta(q)=|E_\text {Z}|$.  The probability of finding a value of $q$ larger than $q(E_\text{Z})$ is zero.  That is, the states between $q_{\text{EA}}$ and $q(E_{\text{Z}})$ have  been immediately emptied after the magnetic field $H$ was cut to zero in a TRM protocol.

The spin-glass coherence length $\xi(t=0,\tw;H)$ therefore grows not from $q=0$ but from $q=q(E_{\text {Z}})$ in a TRM experiment.  If $\Delta(\tw)$ would depend linearly on Hd, equivalently on $q_{\alpha\beta}$ from Eq.~\eqref{eq:Hd_def}, there would be no difference in $\xi(t=0,\tw;H)$ between a TRM and a ZFC experiment.  However, if instead $\Delta(\tw)$ behaves as drawn in Fig.~\ref{fig:Hd_barriers}, that is, exhibits an upward curvature as Hd increases, a TRM protocol would begin at a larger value of $q$ than the ZFC protocol which begins at $q=0$.  The more rapid increase of $\Delta(\tw)$ as Hd increases from Hd($\Delta=|E_Z|)$, as compared to increasing from Hd($\Delta = 0)$, would imply a slower growth of $\xi(t,\tw;H)$ for a TRM protocol than for an ZFC protocol.  Hence, a breakdown of the superposition principle.

\citet{lederman:91} found explicit evidence for an upward curvature of $\Delta$ as a function of Hd in a AgMn 2.56 at.\% spin glass.  The calculation of \citet{vertechi:89} found a similar behavior.  It is for this reason that \citet{orbach-janus:23} concluded that the principle of superposition only holds in the  $H\rightarrow 0$ limit.

\subsection{Correlation functions for nonequilibrium dynamics and the coherence length}\label{subsect:off-equilibrium-no-field-3}

The purpose of this section is to introduce the correlation functions that will be instrumental in showing that off-equilibrium dynamics conveys information about equilibrium structures.

\begin{figure}[t]
    \centering \includegraphics[width=\linewidth,angle=0]{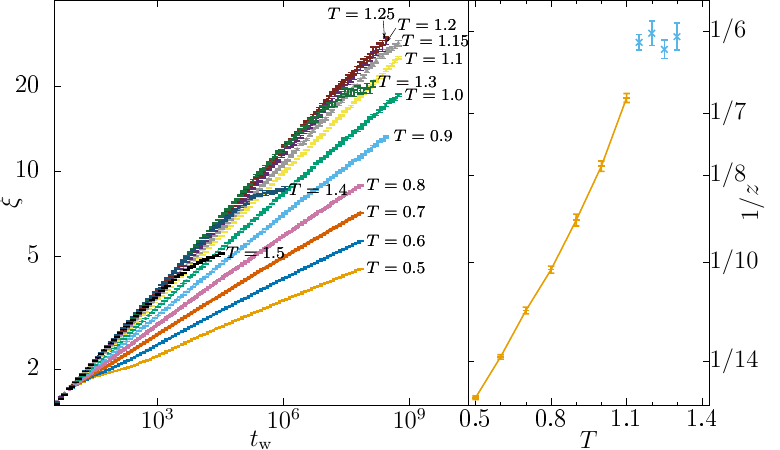}
    \caption{Temperature-dependent growth of the coherence length. {Left:} coherence length $\xi_{12}$($\tw$) vs Monte Carlo time, see Eq.~\eqref{eq:xi_micro_def}, as computed for size $L = 256$, for several temperatures [$\Tg = 1.1019(29)$], data taken from ~\textcite{fernandez:15}. One Monte Carlo step corresponds to 1 ps in physical time. For $T \geq 1.3$, we reach equilibrium. {Right:}
    To a first approximation, see Sec.~\ref{subsect:off-equilibrium-in-field-fixed-T-4} for more, in the nonequilibrium regime $\xi$ grows with time as a power law
    $\xi(\tw) \sim \tw^{1/z(T)}$. Joined points stand for $T \leq \Tg$. Note the constant value $z(T >\Tg) \approx 6$ (cyan circles).}
    \label{fig:xi_many_T}
\end{figure}

In general, we shall consider a spin glass that has been quickly cooled from some very high temperature to the working temperature $\Tm<\Tg$. In computer simulations, the cooling rate can be infinite, so that the starting configuration can be completely disordered. An infinite cooling rate is not experimentally achievable. As explained already several times, the system is then allowed to relax for a time $\tw$. The properties of the system can be studied at the time $\tw$, or at some later time $\tw+t$. Different correlation functions are introduced according to whether we consider only one time, or two.

As the waiting time increases, glassy magnetic domains grow.
This growth is monitored using the correlation function \cite{marinari:96}
\begin{equation}\label{eq:C4_def}
C_4(\boldsymbol{r},\tw)= \disorderav{\frac{1}{N}\sum_{\boldsymbol{x}} \langle q_{\boldsymbol{x}}(\tw)q_{\boldsymbol{x}+\boldsymbol{r}}(\tw)\rangle},
\end{equation}
\changes{where $q_{\boldsymbol{x}}(\tw)=\sigma^{(1)}_{\boldsymbol{x}}(\tw) \sigma^{(2)}_{\boldsymbol{x}}(\tw)$. $C_4(\boldsymbol{r},\tw)$}
 is self-averaging, hence for a sample of macroscopic size the disorder average can be omitted. The expected behavior at long distance is 
\begin{equation}\label{eq:C4scaling}
C_4(r\gg 1)= \frac{1}{r^\theta}f\bigl[r/\xim(\tw)\bigr]\,,
\end{equation}
where $\xim$ is the size of the coherent regions (the glassy domains), $\theta$ is the replicon exponent \cite{marinari:00,marinari:00b}, and $f(x)$ is a scaling function that decreases faster than exponentially for large values of its argument. Typically~\cite[see][]{janus:08b,janus:09b} $\xim$ is taken as the $k=1$ value of 
\begin{equation}\label{eq:xi_micro_def}
\xi_{k,k+1}(\tw)={\frac {I_{k+1}(\tw)}{I_k(\tw)}}\,,\ I_k(\tw)=\int_0^\infty \mathrm{d} r\,r^k C_4(r,\tw)\,.
\end{equation} 
When applying Eq.~\eqref{eq:xi_micro_def}, care must be taken to check for
finite-size effects and to deal with the noisy tails of $C_4$. This can be 
done in a self-consistent manner \cite[see][]{janus:09b,janus:18}.
Fig.~\ref{fig:xi_many_T} exhibits the sluggish growth of $\xim(\tw)$.

\begin{figure}[t]
    \centering    
    \includegraphics[width=\linewidth,angle=0]{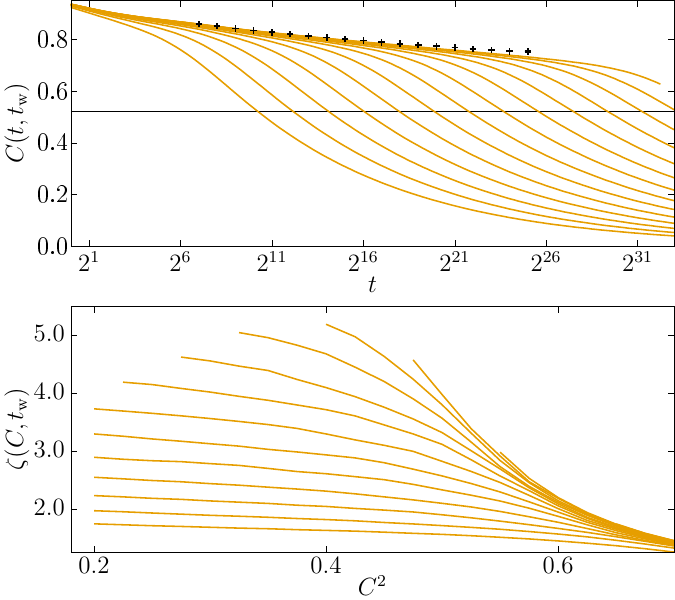}
    \caption{Top: $C(t,\tw)$, see Eq. \eqref{eq:Cttw-def}, as a function of $t$ for $\tw = 4^i$ , $i = 3; \dots ; 16$ (lines, $\tw$ grows from bottom to top). We also plot $\lim_{\tw\to\infty}{C}(t,\tw)$ (points) which generates an enveloping curve that ends at $q_\text{EA}$ (horizontal line), recall Eq.~\eqref{eq:largeTimeLimits}. 
    Bottom: The correlation length $\zeta(C,\tw)$, extracted from the correlation function in Eq.~\eqref{eq:C2+2-def}, as a function of $C^2$ (same values of $\tw$ as in the top panel). Note from the top panel that, for fixed \tw, $C(t,\tw)$ is a monotonically decreasing function of $t$. This length, $\zeta$, indicates the spatial range of the correlations between the changes when the spin configurations at times $t$ and $\tw$ are compared. Figure adapted from \citet{janus:10b}.}
    \label{fig:Ct-qEA}
\end{figure}

\begin{figure}[th]
    \centering    
    \includegraphics[width=\linewidth,angle=0]{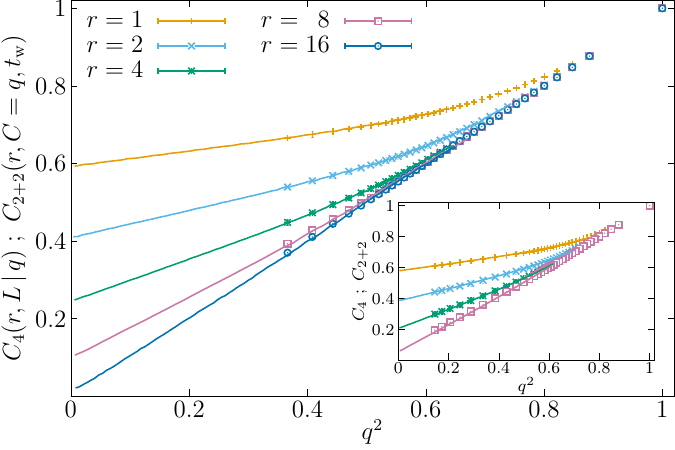}
    \caption{\changes{The comparison of correlation functions evinces a statics-dynamics equivalence between equilibrium quantities computed on small systems (dots) and nonequilibrium quantities obtained from large systems (continuous lines). Specifically, the continuous lines are $C_{2+2}(r,C,\tw)$, see Eq.~\eqref{eq:C2+2-def}, for several values of $r$, and a fixed \tw. As in Fig.~\ref{fig:Ct-qEA}---bottom, we change variables from $t$ to the monotonically \emph{decreasing}
    $C^2(t,\tw)=q^2$ ($q$ is parameter ranging from 0 to 1), hence data at \emph{smaller} $q^2$ correspond to \emph{later} times. This dynamical function coincides with the equilibrium spatial correlation function $C_4(r;L|q)$ defined in Eq.~\eqref{eq:def-C4r-equilibrio} and plotted with points, if the proper equivalence is made between \tw and $L$.
    In particular, the figure, taken from \citet{janus:10}, uses a simple statics-dynamics dictionary where nonequilibrium data at time \tw is 
    compared to equilibrium averages for a value of $L$ such that $L=3.7\xi_{12}(\tw)$. The main plot shows this comparison for $L=32$, which corresponds to $\tw=2^{31}$, and the inset shows the same for $L=24$, corresponding to $\tw=2^{26}$. The errors in both datasets are comparable and smaller than the point size.}}
    \label{fig:statics-dynamics}
\end{figure}

Another correlation function globally checks how much memory is left at time $\tw+t$ of the spin configuration found at time \tw:
\begin{equation}\label{eq:Cttw-def}
    C(t,\tw)=
    \disorderav{\frac{1}{N}\sum_{\boldsymbol{x}} \langle \sigma_{\boldsymbol{x}}(\tw)\sigma_{\boldsymbol{x}}(\tw+t)\rangle}\,,
\end{equation}
It provides, therefore, a direct check of aging \cite{kisker:96,rieger:93,jimenez:03}. 
The order of limits is particularly tricky:
\begin{equation}\label{eq:largeTimeLimits}
    0=\lim_{\tw\to\infty}\lim_{t\to\infty} C(t,\tw)\,,\quad \qEA=\lim_{t\to\infty}\lim_{\tw\to\infty} C(t,\tw),
\end{equation}
as exemplified in Fig.~\ref{fig:Ct-qEA}.  It is worth stressing that the first limit in Eq.~\eqref{eq:largeTimeLimits} is zero under the assumption of weak ergodicity breaking \cite{cugliandolo:93}, but a nonzero limit has been recently measured in numerical simulations of spin glasses on Bethe lattices \cite{bernaschi2020strong}.

\changes{A related quantity is the two-time link correlation function 
\begin{eqnarray}
C_\text{link}(t,\tw)&=&\overline{\frac{1}{D N}\sum_{\Vert\boldsymbol{x}-\boldsymbol{y}\Vert=1}\,\sigma_{\boldsymbol{x}}(\tw) \sigma_{\boldsymbol{y}}(\tw)}\\\nonumber
&&\quad\times\ \overline{\sigma_{\boldsymbol{x}}(t+\tw) \sigma_{\boldsymbol{y}}(t+\tw)}
\end{eqnarray}
where
$D$ is the space dimension and the sum is restricted to lattice nearest neighbors. $C_\text{link}$ conveys information about the density of domain walls in the system. It is very important to  change the variable from $t$ to $C(t,\tw)$. This is always possible because $C$ monotonically decreases with $t$ at fixed \tw. The derivative $\mathrm{d} C_\text{link}(t,\tw)/\mathrm{d} C^2(t,\tw)$ is able to distinguish between the RSB theory, which predicts a nonvanishing value for the derivative at long $\tw$, and the droplets view of coarsening where the derivative should vanish~\cite{janus:08b,janus:09b}. 
}

Finally, one can construct a two-time, two-site correlation function that provides spatial information about how the new ordering at later times $\tw+t$ builds by eroding the old ordering pattern at time $t$:
\begin{equation}\label{eq:C2+2-def}
    C_{2+2}(\boldsymbol{r},t,\tw)= \disorderav{\frac{1}{N}\sum_{\boldsymbol{x}} \langle c_{\boldsymbol{x}}(t,\tw) c_{\boldsymbol{x}+\boldsymbol{r}}(t,\tw)- C^2(t,\tw)\rangle}\,,
\end{equation}
where the shorthand $c_{\boldsymbol{x}}(t,\tw)=\sigma_{\boldsymbol{x}}(\tw)\sigma_{\boldsymbol{x}}(\tw+t)$ is used. Note that $c_{\boldsymbol{x}}(t,\tw)=-1$ if the spin at site $\boldsymbol{x}$ has flipped at time $\tw+t$ as compared to time $\tw$.  We name this a \changes{rearrangement}, and Eq.~\eqref{eq:C2+2-def} allows the extraction of the correlation length between \changes{rearrangements}, $\zeta(t,\tw)$, which is shown in the lower panel of Fig.~\ref{fig:Ct-qEA} by a direct analogy with Eq.~\eqref{eq:xi_micro_def}, see \citet{janus:09b}. By construction, $\zeta(t,\tw)$ is upper-bounded by $\xim(\tw)$.

\changes{The correlation length $\zeta(t,\tw)$, measuring the average size of the regions that rearrange between times $\tw$ and $\tw+t$, is connected to the dynamic heterogeneities that have been measured in spin-glass models with great accuracy \cite{castillo:02,castillo:03,montanari:03a,montanari:03b,jaubert:07,roma:16}. However, a precise connection would require an analysis of local quantities, similar to the one discussed in Sec.~\ref{subsubsect:off-equilibrium-in-field-several-T-1-numerical} for temperature chaos. Such an analysis is not presently available.}

An important result that we shall use throughout the rest of this review is that these off-equilibrium quantities can be related quantitatively to their equilibrium counterparts. To do so, one needs to establish a \emph{statics-dynamics dictionary}, which defines a correspondence between a given finite time \tw (off-equilibrium) and a finite size $L$ (equilibrium). One of the first such dictionaries, based on a single length scale, $\xi_{12}$, is illustrated in  Fig.~\ref{fig:statics-dynamics}\changes{, which employs an equilibrium version of $C_4(r,\tw)$:
\begin{equation}\label{eq:def-C4r-equilibrio}
C_4(r;L|q)=\overline{\frac{1}{N}\sum_{\boldsymbol{x}}\, \langle q_{\boldsymbol{x}} q_{\boldsymbol{x}+\boldsymbol{r}}\rangle_q}\,,
\end{equation}
where the thermal average is restricted only to pairs
of replicas of an equilibrated system conditioned to have overlap $q$, recall Eq.~\eqref{eq:def-overlap}. Mind that $q$ here is a numerical parameter that we are free to vary between 0 and 1.}

As we shall see below in Sec.~\ref{subsect:off-equilibrium-no-field-4}, it is necessary to use 
several length scales to define a comprehensive statics-dynamics equivalence.

\subsection{Connecting aging with the equilibrium picture: the fluctuation-dissipation relations}\label{subsect:off-equilibrium-no-field-2} 

To better quantify spin glasses' aging behavior, we describe the connection between the off-equilibrium aging dynamics and the equilibrium picture with many states and a nontrivial $P(q)$.

The out-of-equilibrium dynamics of strongly disordered systems, like spin glasses, can be solved in a closed form only for models with variables interacting via a fully connected topology. The pioneering work by \textcite{cugliandolo:93} considered the fully connected spherical $p$-spin model
\begin{equation}
\mathcal H = -\sum_{i_1<\ldots<i_p}^N J_{i_1\ldots i_p} \sigma_{i_1} \ldots \sigma_{i_p} - \sum_{i=1}^N h_i \sigma_i\;.
\end{equation}
This model is quite different from realistic spin glasses, as the $\sigma_i$  are unbounded real numbers subject to the spherical constraint $\sum_{i=1}^N \sigma_i^2=N$, and interact through all possible groups of $p>2$ spins with very weak couplings in the large-$N$ limit.\footnote{The most common choice is to consider Gaussian couplings of zero mean and variance $\overline{J_{i_1\ldots i_p}^2}=p!/(2N^{p-1})$.}

The following Langevin equation describes the relaxation dynamics at temperature $T$
\begin{equation}
\partial_t \sigma_i(t) = -\mu(t) \sigma_i(t) - \frac{\partial \mathcal H}{\partial \sigma_i}(t) + \eta_i(t)\;,
\end{equation}
where the first term on the r.h.s. enforces the spherical constraint, and the noise is white: $\langle \eta_i(t) \eta_j(t')\rangle = 2T \delta_{ij} \delta(t-t')$. 

In addition to the correlation function $C(t,\tw)$, Eq.~\eqref{eq:Cttw-def}, the response function is a key observable:
\begin{eqnarray}
\label{eq:4B:C}
R(t,\tw) &=& \frac1N \sum_{i=1}^N \frac{\partial\overline{\langle \sigma_i(t+\tw)\rangle}}{\partial h_i(\tw)}\, .
\end{eqnarray}
Self-consistency equations for $C(t,\tw)$ and $R(t,\tw)$ can be derived through standard functional methods \cite{cugliandolo:93}. Space limitations prohibits writing out their equations here, but we can briefly discuss their solution.
Following a quench from a high temperature ($T>T_\text{d}$) to a low temperature ($T<T_\text{d}$),\footnote{The dynamical critical temperature $T_\text{d}$ is defined by the divergence of the equilibrium correlation time.} the relaxation dynamics do not reach an equilibrium nor a stationary state on any finite time scale.  They remain trapped in an \emph{aging} regime, where functions of two times depend on both times.  In the simplest scenario, we have $C(t,\tw)=\mathcal{C}[\tw/(t+\tw)]$ and $R(t,\tw)= (t+\tw)^{-1} \mathcal{R}[\tw/(t+\tw)]$, so that relaxation becomes slower while time increases.

Perhaps the most striking result in the Cugliandolo-Kurchan (CK) solution \changes{to the spherical $p$-spin model} \cite{cugliandolo:93} is that, in the aging regime, response and correlation functions are connected by a generalized fluctuation-dissipation relation (GFDR)
\begin{equation}
\mathcal{R}(\lambda) = \beta x\; \mathcal{C}'(\lambda)\;,
\end{equation}
where $\lambda=\tw/(t+\tw)$ and $x$ is the Parisi parameter in the equilibrium solution to the model. More specifically, \changes{for free energies $f$} close to the equilibrium free energy $f_0$, the number of metastable states grows like $\mathcal{N}(f) \sim \exp[N \beta x (f-f_0)]$, which in turn generates a nontrivial overlap distribution
\begin{equation}
P(q) = x\, \delta(q) + (1-x)\, \delta(q-\qEA)\;.
\end{equation}
Therefore, $x$ can be regarded as the probability that two replicas at equilibrium are in different states.
The connection between the off-equilibrium GFDR and the equilibrium $P(q)$ is foundational to the comprehension of disordered systems from the study of their aging dynamics.

\changes{In models of structural glasses, like the spherical $p$-spin, the quantity $\beta x$ is (mostly) constant in the entire aging regime. For this reason, it is useful to introduce the concept of an effective temperature $T_\text{eff}=1/(\beta x)$ in the out-of-equilibrium regime \cite{cugliandolo:97}.
However, for spin-glass models having multiple and differently diverging time scales in the aging regime, $T_\text{eff}$ changes with time, and its interpretation as a temperature requires further assumptions, thoroughly discussed by~\textcite{montanari:03b}.}

The CK ansatz for the solution to the out-of-equilibrium dynamical equations has been extended by \textcite{cugliandolo:94b} to the SK model (actually a version of it with soft spins).
Even in this more complicated model possessing many diverging time scales \cite{sompolinsky:82}, the GFDR provides the sought-after connection between off-equilibrium dynamics and equilibrium states. Indeed, assuming
\begin{equation}
\mathcal{R}(\lambda) = \beta X[\mathcal{C}(\lambda)]\, \mathcal{C}'(\lambda)
\end{equation}
in the aging regime, \textcite{cugliandolo:94b} solve the dynamical equations and connect the dynamical quantity $X[C]$ to the Parisi distribution of overlaps between states at equilibrium through the relation
\begin{equation}
X[\mathcal{C}] = \int_0^\mathcal{C}\dd q\ P(q)\;.
\label{eq:XvsPq}
\end{equation}

Nevertheless, it is fair to remember that the CK solution is based on some crucial hypotheses: weak long-term memory and weak ergodicity breaking. Both are required to decouple the aging dynamics at long times from the initial condition and the fast relaxation at short times. Recently, it has been found that weak ergodicity breaking ---which corresponds to the fact that any finite-time configuration is eventually forgotten in the long-time limit--- is not always true and must be sometimes replaced by \emph{strong} ergodicity breaking, meaning that even at very long times the aging dynamics takes place in a restricted space having a positive overlap with the initial condition. Strong ergodicity breaking in off-equilibrium dynamics has been observed in mixed $p$-spin spherical models \cite{folena2020rethinking,folena2023weak} and in spin-glass models on sparse random graphs \cite{bernaschi2020strong}.

All the above results were obtained for mean-field models. In finite-dimensional models, analytical results are much scarcer. An important one was developed by \citet{franz:98,franz:99}. These authors considered a generic model with short-range interactions in finite dimensions.  The only condition was  \emph{stochastic stability}, \emph{i.e.}, robustness to small random perturbations (see also Sec.~\ref{subsect:Equilibrium-1}).
By studying linear response to the application of long-range perturbations, and assuming that the dynamical averages of one-time quantities (\emph{e.g.}, the energy) converge to their equilibrium values, one can again obtain Eq.~(\ref{eq:XvsPq}) relating the equilibrium overlap distribution $P(q)$ to the fluctuation-dissipation ratio
\begin{equation}
\label{subs-4c-xq}
\lim_{\substack{t,\tw\to\infty\\C(t,\tw)\to q}} \frac{T\,R(t,\tw)}{\partial_{\tw} C(t,\tw)} = x(q) \equiv \int_0^q \dd q'\ P(q')\,.
\end{equation}
In numerical simulations, measuring the instantaneous response $R(t,\tw)$ is not advisable, so integrated responses $\chi(t,\tw) = \int_{\tw}^t R(t,t') dt'$ have been used instead \cite{marinari:98f,parisi:99b}, relating to the equilibrium $x(q)$ as follows
\begin{equation}
T \lim_{\substack{t,\tw\to\infty\\C(t,\tw)\to q}} \chi(t,\tw) = \int_q^1 x(u) \dd u \equiv \mathcal{S}(q)\,.
\label{eq:intFDR}
\end{equation}
Usually, the integrated response $\chi(t,\tw)$ is measured as the ratio $m/h$ where $m$ is the magnetization at time $\tw+t$ induced by a field $h$ switched on at time $\tw$. To avoid nonlinear effects in the response, methods have been invented that measure $\chi(t,\tw)$ without switching on the magnetic field \cite{ricci-tersenghi:03,chatelain2003far}.

\begin{figure}
    \centering
    \includegraphics[width=\columnwidth]{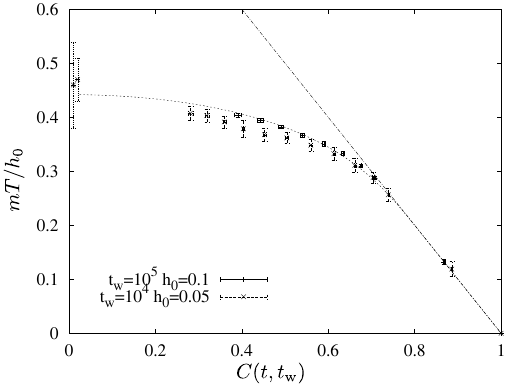}
    \caption{The integrated response $\chi(t,\tw)=m/h_0$ (multiplied by the temperature) as a function of the correlation function $C(t,\tw)$, measured in the off-equilibrium aging dynamics of the $D=3$ EA model, for different values of $\tw$ and $h_0$. It follows closely the dashed curve $S(C)$ obtained via Eqs.~(\ref{subs-4c-xq}) and (\ref{eq:intFDR}) from the equilibrium $P(q)$. This plot, reproduced from \textcite{marinari:98f}, was the first piece of evidence for the validity of statics-dynamics equivalence in spin glasses.}
    \label{fig:Marinari_1998b_fig1}
\end{figure}

Fig.~\ref{fig:Marinari_1998b_fig1} reproduces the result by \textcite{marinari:98f}, which was the first convincing evidence that the connection between dynamical measurements in the out-of-equilibrium regime and the equilibrium properties, as expressed by Eq.~(\ref{eq:intFDR}), holds in the $D=3$ EA model.
Further details on the use of the static-dynamics connection expressed by Eq.~(\ref{eq:intFDR}), and how such a relation needs to be modified in the presence of finite-size effects and growing length scales, will be discussed in Sec.~\ref{subsect:off-equilibrium-no-field-4}.

\begin{figure}
    \centering
    \includegraphics[width=\columnwidth]{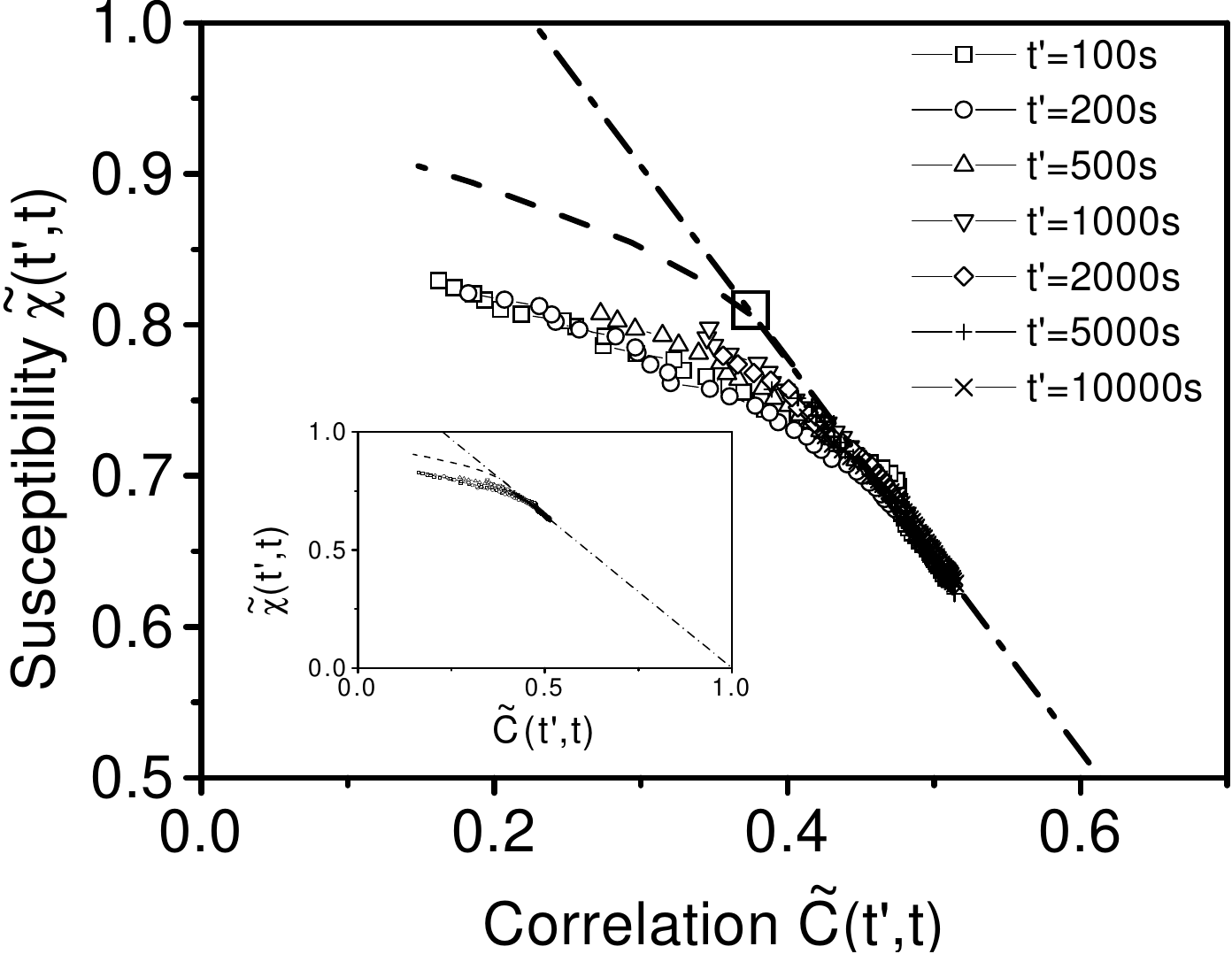}
    \caption{The plot, reproduced from \citet{herisson:02}, shows the relation between the integrated response and the correlation measured in experimental spin glasses. It is worth stressing the similarity with the data reported in Fig.~\ref{fig:Marinari_1998b_fig1}.}
    \label{fig:Herisson_2002_fig3}
\end{figure}

We end this section by presenting the first and most important measurement of the fluctuation-dissipation ratio in experimental spin glasses. \textcite{herisson:02}  obtained an impressive result by measuring, on the same insulating spin glass at a temperature $T=0.8\,\Tg$, both the integrated response $\chi(t,\tw)$ and the magnetization autocorrelation $C(t,\tw)$. By plotting both quantities parametrically in $t$ for several values of $\tw$ (see Fig.~\ref{fig:Herisson_2002_fig3}), \textcite{herisson:02} reported a nontrivial fluctuation-dissipation ratio in experimental spin glasses very similar to the one that was already exhibited in numerical simulations of spin-glass models \cite[][see, \emph{e.g.}, Fig.~\ref{fig:Marinari_1998b_fig1}]{marinari:00b}.

The reader will note, however, that the experimental results in Fig.~\ref{fig:Herisson_2002_fig3} display a clear time dependence. In other words, the experiment of \textcite{herisson:02} was unable to reach the infinite-time limit of Eq.~\eqref{eq:intFDR}.\footnote{Herisson and Ocio were well aware of this fact and attempted to \emph{extrapolate} their results to infinite time.} Because very large times mean very large coherence lengths, the sluggish growth of $\xim$ at low temperatures depicted in Fig.~\ref{fig:xi_many_T} suggests that it is highly unlikely that any near-term experiments will be able to reach the limit required in Eq.~\eqref{eq:intFDR}. This raises the question: how should Eq.~\eqref{eq:intFDR} be used when
the limit of long times is unreachable? This is, precisely, the question that we shall address in the next section.

\subsection{The statics-dynamics dictionary
\label{subsect:off-equilibrium-no-field-4}}

We discuss here the important issue of the quantitative connection between the nonequilibrium evolution of a spin glass (dynamics) and its equilibrium behavior (statics), following \citet{janus:17}. Accurate numerical computation of the nonequilibrium fluctuation-dissipation ratio for the  EA model has been the basis for establishing a \textit{statics--dynamics dictionary}.  This is of great help for computing controlled extrapolations.

A framework has been emerging in the past few years for theoretical explanations of what can be, and is, observed under actual experimental conditions.  One needs to reach a theoretical understanding of what can be measured under the ``best'' experimental conditions. For example, it was found in the seminal experiments of \textcite{joh:99} that the largest spin-glass coherence length one could reach in the laboratory was of order $50~a_0$, with $a_0$ the mean distance between magnetic moments. This was in 1998.  Today, because of the availability of single crystals, one can do better, reaching $234~a_0$ \cite[][recall Sec.~\ref{subsect:intro-7}]{zhai:19}. The two settings, numerical simulations and experiments, are plagued by different limitations.  If they can be overcome, a joint understanding becomes possible.

The root for the theoretical analysis lies in the work of \textcite{cugliandolo:93}, see also \textcite{franz:94b,franz:94c}, which has been discussed in detail in Sec.~\ref{subsect:off-equilibrium-no-field-2}. Interesting work attempting to extend their analysis to finite times can also be found in \textcite{barrat:01,berthier:01,janus:08b,janus:10,janus:10b}.

The basis for our discussion is the very slow growth of the coherence length in the low-temperature phase of spin glasses (already discussed in detail in Sec.~\ref{subsect:off-equilibrium-no-field-3}).
Another crucial ingredient for this analysis is the \textit{stochastic stability}
discussed in Sec.~\ref{subsect:Equilibrium-1} and Sec.~\ref{subsect:off-equilibrium-no-field-2}, because it allows the establishment of a quantitative relationship between dynamics and statics, a major issue. As discussed in Sec.~\ref{subsect:off-equilibrium-no-field-2}, one can use stochastic stability to derive the GFDR \cite{franz:98,marinari:98f}.

We shall consider, instead, quantities that can be computed on a lattice of size $L$, like the equilibrium (static)
\begin{equation}
x(C,L) \equiv \int_{0}^{C} \dd q \;P_L\left(q\right) \equiv -\frac{\dd \mathcal{S}\left(C,L\right) }{\dd C}\, ,
\label{eq:3D:GFDR}
\end{equation}
recall Sec.~\ref{subsect:Equilibrium-1} and the finite-size $P_L(q)$ shown in Fig.~\ref{fig:Pq}; the dynamic linear response function measured in off-equilibrium experiments by switching on a field $\delta H$ at time $\tw$
\begin{equation}
\label{eq:3D:S}
\chi_L\left(t,\tw\right)
\equiv \lim_{\delta H \rightarrow 0}\;
\frac{\delta m(t+\tw)}{\delta H}\;;
\end{equation}
and the corresponding correlation function $C_L(t,\tw)$ defined in Eq.~\eqref{eq:Cttw-def}.

We now go to the core issue: how an effective and well-functioning statics-dynamics dictionary (SDD), based on a numerical GFDR, can be generated \cite{janus:17}. It turns out to be crucial that the SDD be based on matching observables, both at the waiting time $\tw$ (where the experimental conditions are set) and at the probing time $\tw+t$ (where experimental measurements are taken).

Numerical simulations of a $3D$ EA spin glass on a system of linear size $L$ and volume $V=L^3$ can produce data that follows a relevant experimental protocol. One cools the simulated system from high temperatures $T$ down to, say, $T=0.64\, T_{g}$ in zero magnetic field, $H=0$. The system then evolves during the waiting time $\tw$, generating a growing coherence length  $\xi(t,\tw;H=0)$. At $\tw$, one switches on the magnetic field $H$ and evaluates the linear response function, $\chi_L(t,\tw)$ for further times,
together with the correlation function $C_L(t,\tw)$. We denote the infinite-volume limit of different observables by omitting the subscript $L$.  For $T<\Tg$ the GFDR in Eq.~(\ref{eq:intFDR}) has the form
\begin{equation}
\lim_{t,\tw\to\infty} \lim_{L\to\infty} T \chi_L\left(t,\tw\right)=
\lim_{t,\tw\to\infty} \lim_{L\to\infty} \mathcal{S}\left(C_L(t,\tw),L\right)\,,
\label{eq:3D:naive_gfdr}
\end{equation}
where one must be careful with the order of the limits. 
The measurement time $t$ has to be scaled appropriately for increasing $\tw$ as in Eq.~(\ref{eq:intFDR}).

\begin{figure}[t]
    \centering    \includegraphics[width=\linewidth,angle=0]{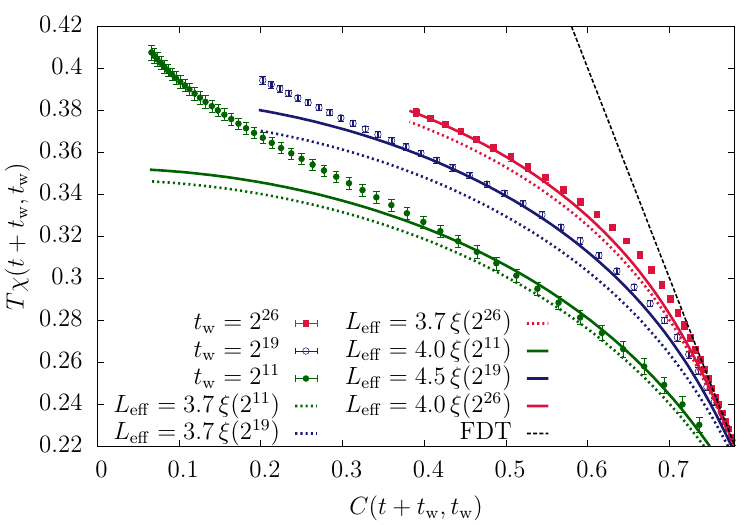}
    \caption{The ``naive'' scaling of Eq.~\eqref{eq:3D:SDD-simple} does not work when the correlation function becomes too small. It also does not work for short times when close to \qEA. Reproduced from Fig. 2 of \textcite{janus:17}.}
    \label{fig:3D:ko}
\end{figure}

Taking the limit of large $L$ in the l.h.s.\ of  Eq.~\eqref{eq:3D:naive_gfdr} means keeping $L\gg \xim$ for the whole time of the experiment (or the simulation).
However, in a numerical simulation, the order of the limits in the r.h.s.\ of Eq.~\eqref{eq:3D:naive_gfdr}, must be reversed.  We reach equilibrium (\emph{i.e.}, infinite time) by using some \emph{unphysical dynamics} (recall Sec.~\ref{subsect:intro-dedicated_computers}) in a relatively small system. The function $\mathcal{S}\left[q=C_L(t,\tw),L\right]$ is computed numerically by integrating the finite size probability distribution for the overlap $P_L(q)$.
In other words, one uses numerical simulations to mimic experiments by relating the nonequilibrium response of a very large system at finite times to the equilibrium overlap for a much smaller system. In particular, one may find the  effective equilibrium size $L_{\text{eff}}(t,\tw)$ as the value of $L$ such that $T\chi(t,\tw)=\mathcal{S}\left[q=C(t,\tw),L\right]$.\footnote{Recall that, if we keep $t$ and $\tw$ fixed, the limit of infinite size can be reached for both $C$ and $\chi$ without much pain in the off-equilibrium experiment (or simulation). Unfortunately, the limit of large $L$ for $\mathcal{S}(q,L)$ is considerably more difficult to obtain.}

\begin{figure}[t]
    \centering    
    \includegraphics[width=\linewidth,angle=0]{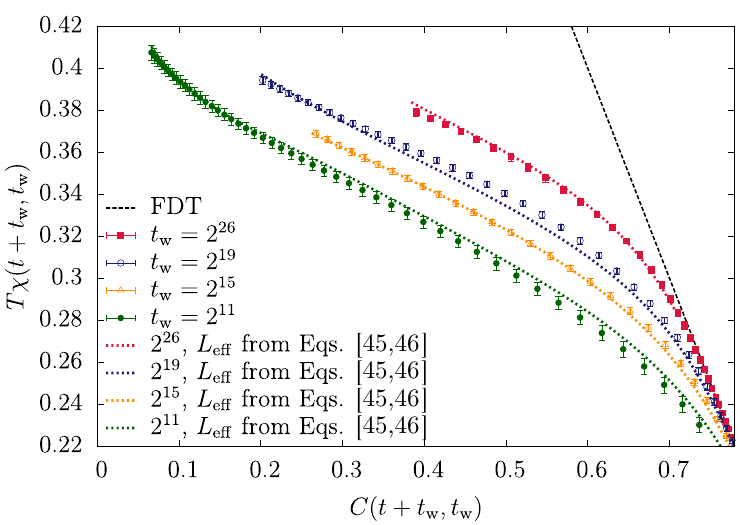}
    \caption{The improved scaling, which depends on two coherence lengths, see Eq.~\eqref{eq:3D:ultrametric-ansatz}, works well in all measured regimes. Reproduced from Fig. 4 of \textcite{janus:17}.}
    \label{fig:3D:ok}
\end{figure}

In a first approach, one tries a single-time ansatz for
the effective equilibrium size $L_{\text{eff}}(t,\tw)$. For instance,
\begin{equation}
\label{eq:3D:SDD-simple}
L_{\text{eff}}(t, \tw) = k\, \xi(\tw)\,,
\end{equation}
This is the method illustrated in Fig.~\ref{fig:statics-dynamics}, where $k=3.7$
was shown to work to scale the $C_{2+2}$ correlation function. When studying the response function, however, this simple statics-dynamics dictionary does not work, as shown in Fig.~\ref{fig:3D:ko}, where it is represented by the dotted lines. Even allowing a different value of $k$ for each value of \tw (solid lines) does not work. Again, we do not get a good description of the data, either for short or for long times. Our scaling hypothesis was, somehow, too naive.

It is not difficult to understand the discrepancy at large $t$. We expect $\xi$ to grow without bounds in the spin-glass phase, resulting in $\xi(t+\tw)$ becoming eventually far larger and very different from  $\xi(\tw)$.

It is clear at this point that the effective length ${L}_{\text{eff}}$ must be analyzed by considering as relevant both coherence lengths $\xi(\tw)$ and $\xi(t+\tw)$. The numerical data suggests that, to good accuracy, $L_{\text{eff}}/\xi(t+\tw)$ is a function of $\xi(t+\tw)/\xi(\tw)$. So a new scaling law, depending on two coherence lengths, can be analyzed by assuming
\begin{equation}
\label{eq:3D:ultrametric-ansatz}
L_\text{eff}(t,\tw)=\xi(t+\tw)\,
h\left[\xi(t+\tw)/\xi(\tw)\right]\,,
\end{equation}
with a scaling function of the form $h(x) = k_1 + k_2 x^{-c}$.
Using numerical data to determine plausible best values for the fitting parameters, one finds that it is legitimate and very sensible to rewrite $h(x)$ in the simpler form
\begin{equation}
h\left[\xi(t+\tw)/\xi(\tw)\right] = k_1 \left(1 + \sqrt{\frac{\tw}{t+\tw}}\right)\;,
\label{eq:3D:simpler-ansatz}
\end{equation}
which depends on a single free parameter.  The success of this assumption is clear from Fig. \ref{fig:3D:ok}, where one sees that the scaling (\ref{eq:3D:ultrametric-ansatz}) works nicely over all time regimes.

The line of reasoning presented here is based on the idea that the experimental aging response function contains information about the Parisi order parameter $P(q)$, and is motivated by the fact that the extrapolation to infinite waiting times is cumbersome and sometimes mysterious. The scaling hypotheses  (\ref{eq:3D:ultrametric-ansatz}) and (\ref{eq:3D:simpler-ansatz}), based on assuming that two different length and time scales have to be incorporated in the SDD, allow an orderly, systematic and fully successful extrapolation. When $\xi(t+\tw)\gg \xi(\tw)$ this dictionary disagrees with the simple SDD base on a single length scale, and correctly reports that, under these conditions,  $L_{\text{eff}} \sim \xi(t+\tw)$.

\section{Off-equilibrium dynamics: nonlinear response at a single temperature}\label{sect:off-equilibrium-in-field-fixed-T}

\subsection{Experiments on films and the lower critical dimension} 
\label{subsect:off-equilibrium-in-field-fixed-T-3}
 The use of thin-film spin glasses has generated some remarkable dynamical properties.  At first glance, it would seem improbable that a sufficient number of spins are contained in a thin film of thickness, say, 2 or 3 nm, for measurement with even the most sensitive of instruments.  The situation changed with the introduction of thin-film multilayers comprised of a thin spin-glass layer separated by a nonmagnetic layer sufficiently thick to eliminate any communication between the spin-glass layers.  These were pioneered by  \citet{kenning:87,kenning:90}, who used thin layers of CuMn separated by thick Cu layers.  The use of the same metal for solvent and separator eliminated the problem of strains at the interfaces.

\begin{figure}
    \centering
    \includegraphics[width=8.5cm]{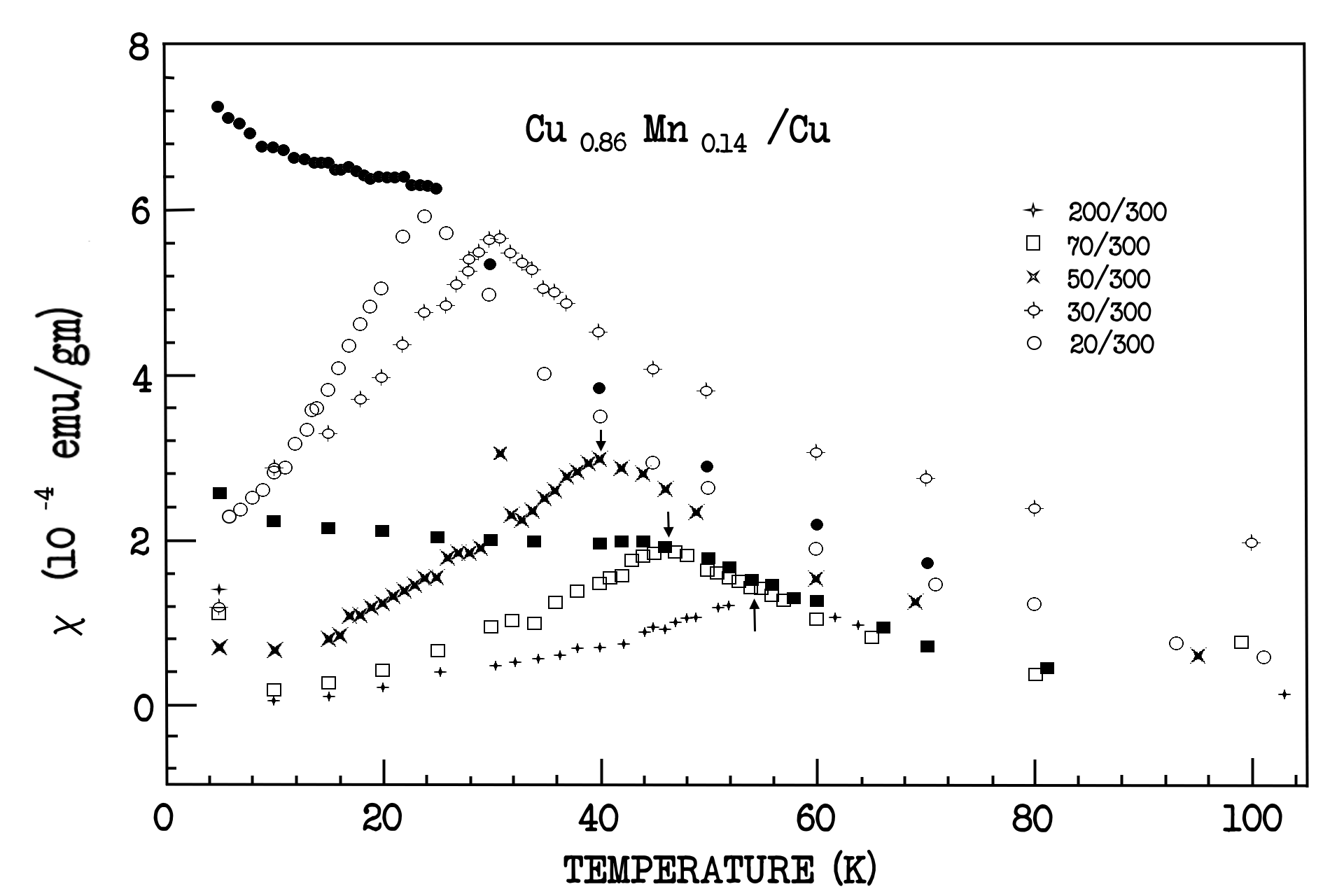}
    \caption{Magnetic susceptibility as a function of temperature for Cu$_{0.86}$Mn$_{0.14}$/Cu multilayer systems. The open symbols are for ZFC data and the solid symbols are for FC data.  The arrows indicated the transition temperatures.  The thicknesses are reported in angstroms. Reproduced from Fig. 14 of \citet{kenning:90}.}  
    \label{fig:chi_vs_T_multilayers}
\end{figure}

\begin{figure}
    \centering
    \includegraphics[width=\linewidth]{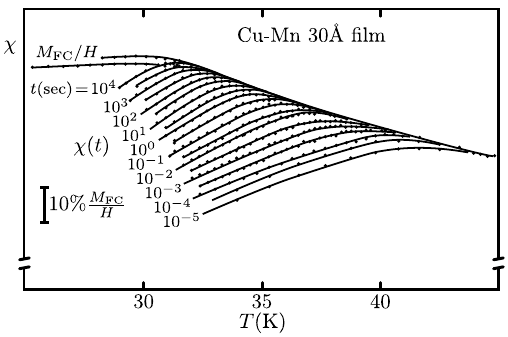}
    \caption{Temperature dependence of the dynamic susceptibility for a CuMn 13.5 at.\% multilayer at different observation times.  Reproduced from Fig. 1 of \citet{sandlund:89}.}  
    \label{fig:chi3_vs_T_film}
\end{figure}

An example of their data is exhibited in Fig. \ref{fig:chi_vs_T_multilayers}.  The peak in the ZFC susceptibility, and its separation from the FC magnetization, moves to lower temperatures as the thickness of the CuMn films is reduced.  This ``freezing temperature'', $T_\text{f}$, is distinguished from the bulk spin-glass condensation temperature $\Tg$.  It arises from the limitation of the growth of the correlation length by the thickness of the film, thereby never reaching the value required for a true phase transition.  Instead, the maximum barrier height, $\Delta_{\text {max}}({\mathcal {L}})$, set by the thin-film parameters, causes an apparent freezing when the laboratory time scale, $\tau_{\text{expt}}$, satisfies
\begin{equation}
{\frac {1}{\tau_{\text{expt}}}}\approx {\frac {1}{\tau_0}}{\text {exp}}[-\Delta_{\text {max}}({\mathcal {L}})/k_\text{B}T_\text{f}].
\end{equation}
The dynamics of the CuMn multilayers were first considered by \citet{sandlund:89}.  They measured the temperature dependence of the dynamic susceptibility at different observation times from ZFC data for a 3 nm film, exhibited in Fig. \ref{fig:chi3_vs_T_film}.  Their data were analyzed by \citet{guchhait:15a} in terms of a temperature-dependent maximum barrier height \cite{hammann:92}. This assumed a hierarchical structure of free-energy states as displayed in Fig.~\ref{fig:hierarchical_organization}. Work by \citet{guchhait:15a} also assumed that the coherent volume reached a steady-state maximum when $\xi(t,T)=\mathcal {L}$, the thickness of the CuMn layer.

In order to account for the thin-film geometry, \citet{guchhait:15a} described the coherent volume as ``pancake-like,'' with a perpendicular length scale $\xi^{\perp}(t,T)$ and a parallel length scale $\xi^{\parallel}(t,T)$.  They defined an effective correlation length,
\begin{equation}
\xi^{\text {eff}}(t,T)=\{\xi^{\perp}(t,T)[\xi^{\parallel}(t,T)]^2\}^{1/3}.
\end{equation}
They assumed that, upon nucleation, the correlation length grew as in a bulk sample until it reached the film thickness, then saturated at $\mathcal{L}$, defining a crossover time,
\begin{equation}\label{eq:xi_parallel}
\xi^{\perp}(t_{\text {co}},T)=\mathcal{L}.
\end{equation}
The parallel component was more troublesome.  After crossover, it was assumed that $\xi^{\parallel}(t,T)$ reflected the behavior of a $D = 2$ spin glass for which $\Tg=0$, the lower critical dimension for an Ising spin glass being $D_\text{l} = 2.5$ (Sec.~\ref{sect:models}).  This was supported by an experimental observation on a thin film of an amorphous GeMn alloy \cite{guchhait:14} that found $2\,< D_\text{l}<3$.  The authors further assumed that, after $t_{\text {co}}$, $\xi^{\parallel}(t\geq t_{\text {co}},T)$ rapidly equilibrated.  Thus, they took
\begin{equation}\label{eq:xi_perp_saturation}
\xi^{\parallel}(t>t_{\text {co}},T)=\xi_\text{eq}^{\parallel}= {\mathcal {L}}\bigg({\frac {T_\text{q}}{T}}\bigg)^{\nu},
\end{equation}
where $T_\text{q}$ is the quenched temperature of Fig.~\ref{fig:chi3_vs_T_film} and $\nu$ \changesbis{is the correlation-length critical exponent for the $\changes{2D}$ Ising Edwards-Anderson model.}


Putting all this together, \citet{guchhait:15a} hypothesized an effective correlation length in equilibrium after crossover to be given by,
\begin{equation}\label{eq:xieff_film}
\xi^{\text {eff}}(t>t_{\text {co}}, T)=\mathcal{L}\bigg({\frac {T_\text{q}}{T}}\bigg)^{2\nu/3}.
\end{equation}
Simulations (see Sec.~\ref{subsect:off-equilibrium-in-field-fixed-T-4}), however, have found that the parallel correlation length continues to grow after $t_{\text {co}}$.  Measurements at times greater than $t_{\text {co}}$ are an opportunity for future experiments (see Sec.~\ref{subsect:conclussions-2}).  
\subsection{Relating the numerical coherence length \boldmath $\xim$ and the experimental  $\xiZ$}
\label{subsect:off-equilibrium-in-field-fixed-T-4}
\begin{figure}
    \centering
    \includegraphics[width=8.5cm]{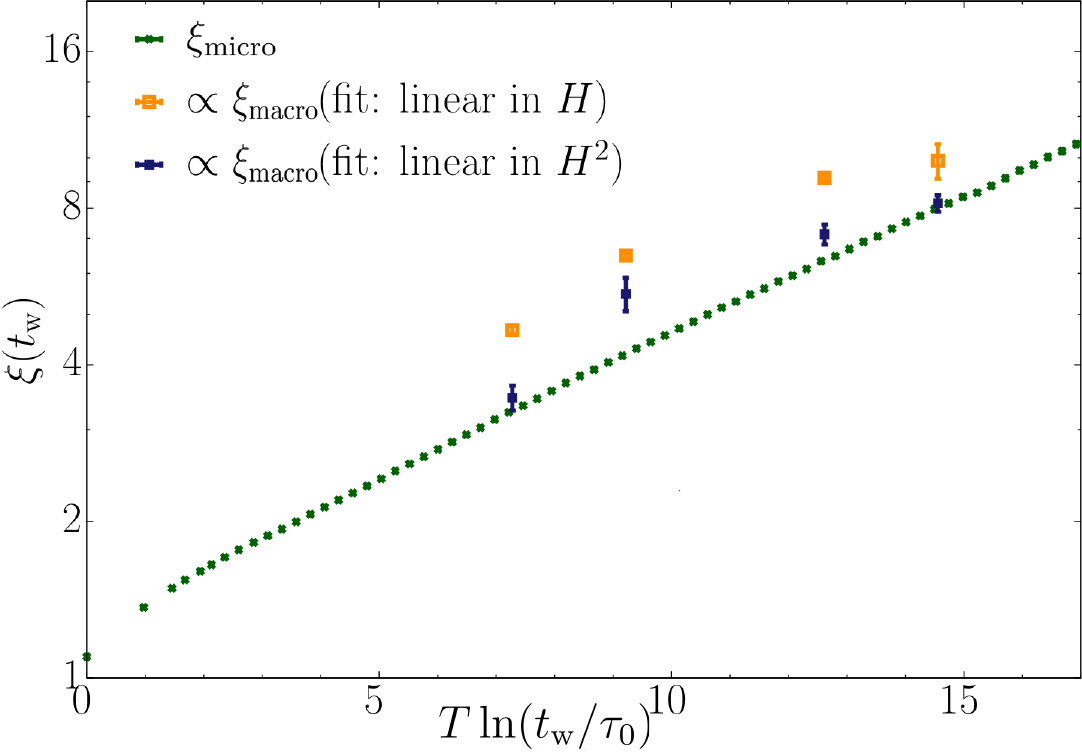}
\caption{The growth of the coherence length $\xim$, computed using 
the microscopic correlation function
$C_4(r,\tw)$, see Sec.~\ref{subsect:off-equilibrium-no-field-3}, is compared to the
  macroscopic $\xi_\text{macro}$, which is 
  calculated by reproducing in simulations the experimental recipe to measure 
  $\xiZ$. The figure shows two alternatives, 
  either the version of Eq.~\eqref{eq:xi_zeeman_def}, based on a quadratic dependence of the Zeeman energy on $H$, or that of Eq.~\eqref{eq:xi_macro_linear}, based on a linear dependence.
  The microscopic time scale $\tau_0=1$
  corresponds to a single lattice sweep.  The  scaling variable $T \ln(\tw/\tau_0)$ is
  standard in the experimental literature~\cite{nakamae:12}. Figure taken from \citet{janus:17b}.}
    \label{fig:ximicmac}
\end{figure}

Throughout this review, recall the glossary in Sec.~\ref{subsect:glossary}, we have considered several length scales in parallel. On the one hand, from theory and simulations, we have the coherence 
length \xim extracted from the $C_4$ correlation function as explained in Sec.~\ref{subsect:off-equilibrium-no-field-3}. Such microscopic detail is not accessible to experiments, where instead a $\xiZ$ based on the Zeeman energy is measured, 
as described in Secs.~\ref{subsect:intro-5} and~\ref{subsect:intro-xi_eff}. On the other hand, while simulations have reached longer times 
with the advent of special-purpose computers, they
are still not quite at the experimental scale.
This section will show how to relate experimental and numerical
length scales and, hence, move toward the joint
quantitative analysis of experiment and simulation
that will be treated in the following sections.

The first major step in this direction was
taken by \citet{janus:17} who measured the 
response function $S$ of Eq.~\eqref{eq:St_def}, the basic tool of experiments,
in simulations of the EA model. This procedure
allowed the calculation of a numerical
$\xi_\text{macro}$ that reproduced the experiments from which \xiZ was extracted.

The key formula is Eq.~(\ref{eq:xi_zeeman_def}),
where a quadratic dependence of the Zeeman energy on the magnetic field was assumed. One could also compute $\xi_\text{macro}$ assuming, instead, a linear dependence \cite{bert:04}. Under that assumption, Eq.~\eqref{eq:xi_zeeman_def} would be modified to
\begin{equation}\label{eq:xi_macro_linear}
   \ln \left[ \frac{t^\mathrm{eff}_\mathrm{H}}{t^\mathrm{eff}_\mathrm{H \to 0^+}}\right] =A^\prime \sqrt{N_f(\tw) H^2}  \, .
\end{equation}
\changes{with $N_f(\tw)$ the number of unpaired moments arising from fluctuations.  Then, $\xi_\text{macro}(\tw)=[N_f(\tw)]^{1/(D-\theta/2)}$.  The linear dependence of $E_Z$ upon $H$ would be expected at very small magnetic fields, while the quadratic dependence is expected for values of $H$ used in normal laboratory experiments.}

The comparison between $\xi_\text{micro}$ and the two determinations of $\xi_\text{macro}$ using quadratic and linear fits to the magnetic field is exhibited in Fig.~\ref{fig:ximicmac}. The behavior of these three lengths is quite similar.

The results of \citet{janus:17} suggested that the numerical \xim would have the same physics as the experimental \xiZ. A major quantitative discrepancy, however, still remained: the aging rates measured in simulations and experiments
greatly differed.

The growth of the coherence length $\xim(\tw,T)$ was usually expressed through a power law that defined a temperature-dependent dynamic critical exponent $z(T)$, 
\begin{equation}\label{eq:z}
\xim(\tw,T) \simeq A(T)\, \tw^{1/z(T)}\,,
\end{equation}
with $z(T)\simeq z_\text{c} \Tg/T$ and $z_\text{c}\equiv z(\Tg)$. According 
to the previous discussion, $z(T)$ should be 
the same for the experimental $\xiZ$.

\begin{figure}[t]
    \centering
    \includegraphics[width=8.5cm]{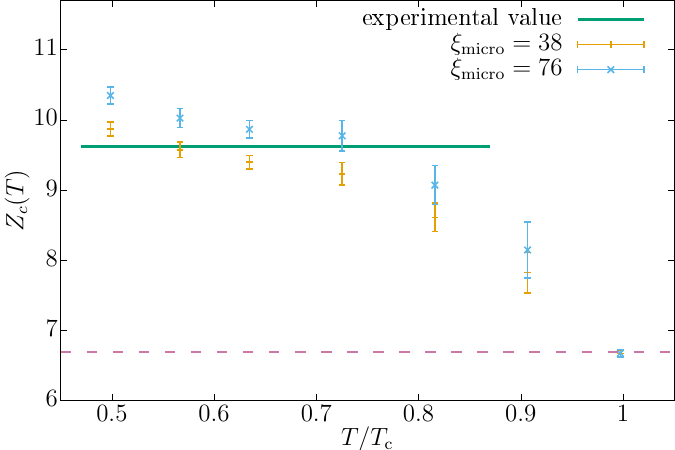}
\caption{Extrapolation of the $\xim$-dependent 
aging rate $z(T,\xim)$ of Eqs.~\eqref{eq:convergente} and~\eqref{eq:aging_rate} to the experimental
range of values of the coherence length
in thin CuMn films. The resulting ratio $Z_\text{c}(T) = z(T,\xim) T/\Tg$ exhibits a very 
good agreement to the experimental value 
$Z_\text{c}\approx9.62$~\cite{zhai:17}, 
indicated by the straight line running along the experimental
temperature range.
Critical effects can only be detected  for $T>0.7$.
Data from \citet{janus:18}.}    \label{fig:film_simulation_z}
\end{figure}

Unfortunately, experiments on CuMn films \cite{zhai:17} reported an almost constant value of the ratio $z(T) T/\Tc\approx 9.62$~\cite{zhai:17} whereas numerical simulations found $z_\text{c}=6.86(16)$~\cite{janus:08b} and $6.80(15)$ \cite{lulli:15}.

Taking advantage of much more precise measurements of $C_4$, the authors of \citet{janus:18} were able to find the missing
piece of the puzzle by observing 
that the dynamical critical exponent 
is actually dependent on \xim. Eq.~\eqref{eq:z}
is only an approximation that works for a limited time range and, to make a connection to experimental time scales, one has to consider
\begin{equation}\label{eq:convergente}
\ln \tw = D(T)  + z_\infty(T)\, \ln\xim + E(T)\, \xim^{-\omega}\,,
\end{equation}
\changes{where $z_{\infty}(T)$ is the value of $z(T)$ in the limit that $\xi_{\text {micro}}\rightarrow \infty$, and the finite $\xim$ dependence of $z$ is controlled by the exponent $\omega$.} At the critical point, $\omega$ should be the exponent
of the first correction to scaling \cite[see][]{janus:13}.  In the spin-glass phase, it would correspond to the replicon exponent, $\omega=\theta$.

Using Eq.~(\ref{eq:convergente}) we can define
an effective dynamic critical exponent (or aging rate) for each value of $\xim$:
\begin{equation}\label{eq:aging_rate}
    z(T,\xim) = \frac{\dd \ln t_\text{w}}{\dd\ln \xim}\,,
\end{equation}
Eq.~\eqref{eq:aging_rate} can be extrapolated 
for typical experimental values in thin films to find $Z_\text{c}(T)=z(T,\xi_\text{film}) T/\Tc$. These extrapolations are shown in Fig.~\ref{fig:film_simulation_z}. The agreement with the experimental value is very good. A modification of Eq.~(\ref{eq:z}) based on activated dynamics~\cite{bouchaud:01b,berthier:02}  has less success in reproducing the nearly constant $Z_\text{c}(T)$ seen in \changes{single-crystal} experiments~\cite{janus:18}.

We are now in a position to compare simulations
with experiments, beginning with the analyses
of thin-film dynamics of Sec.~\ref{subsect:off-equilibrium-no-field-1}. This step was taken in \citet{fernandez:19b}, finding a scenario with some similarities, but also some important differences with the experimental observations.

\begin{figure}
    \centering
    \includegraphics[width=8.5cm]{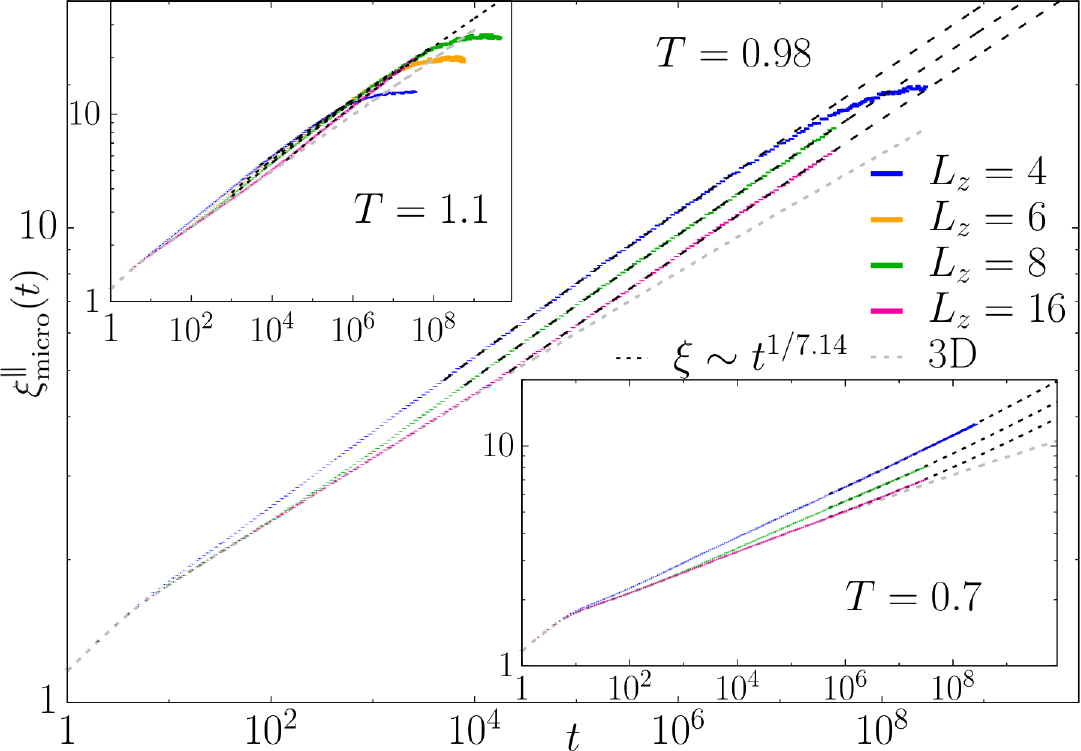}
\caption{The longitudinal coherence length $\xim^{\parallel}(T,t)$, as computed in films of thickness $L_z$, against the waiting time $t$ after a quench to temperature $T$, for $T=0.98$ (main), $T=1.1$ (upper insert), and $T=0.7$ (lower insert).  The critical temperature is $\Tg=1.102(3)$ \cite{janus:13}.  Also shown, as a reference, purely 3D dynamics~\cite{baityjesi:18} and fits to 2D dynamics $\xim^{\parallel}(L_z,T,t)\approx b(L_z,T)+a(L_z,T) t^{1/z_{2D}}$, with $z_{2D}=7.14$ \cite{fernandez:19}, being $b(L_z,T)$ and $a(L_z,T)$ the fit parameters.}  
    \label{fig:film_simulation}
\end{figure}

Fig.~\ref{fig:film_simulation} exhibits the general behavior of $\xim^{\parallel}$ [computed, as usual, with the $k=1$ value of Eq.~\eqref{eq:xi_micro_def} as applied to the parallel correlation function] as a function of time, for three different temperatures (the critical temperature of bulk systems, and two temperatures below) and for four values of the perpendicular size of the sample $L_z$, much smaller than the parallel size.  
Instead of the two time regimes invoked by experiment \cite{guchhait:15a}, simulations find four.

In the first regime, for small times, the growth of $\xim^{\parallel}$ is indistinguishable from what happens in 3D, consistent with the experimental interpretation for $\xi^{\perp}(t,T)$.  However, eventually the growth rate changes (for example for $T=0.98$ and $L_z=16$ at a time larger than $10^4$) and the system enters a second regime where $\xim^{\parallel}$ grows {\it faster} than in $D=3$.  After a transient period, in a third regime, $\xim^{\parallel}$ grows as in 2D which, for $T<\Tg$, is faster than 3D growth.  Finally, the fourth regime corresponds to the saturation of $\xim^{\parallel}$ to its equilibrium value. The fourth regime is completed in the numerical data for $L_z=4$ at $T=0.98$, and for all simulated $L_z$ at $T=1.1$.

A major difference between the experimental interpretation and the findings of simulations is the rate of growth of the correlation length after crossover.  The interpretation of experiments is that $\xi^{\parallel}(t,T)$ is in $D=2$ equilibrium after crossover.  This is in opposition to the findings of simulations that $\xim^{\parallel}(\tw,T)$ increases even more rapidly than in 3D after crossover.  There is clearly need for new experiments to probe thin-film dynamics in this unexplored regime (see Sec.~\ref{subsect:conclussions-2}).

\subsection{Direct comparison of CuMn experiments with Janus~II simulations}\label{subsect:off-equilibrium-in-field-fixed-T-5}

The advent of dedicated supercomputers, such as Janus II, and of single crystals of CuMn \cite{zhai:19} opened up the possibility of combining numerical and experimental results for the first time. Given this opportunity, the initial challenge was to define observables accessible to both theory and laboratory. On the one hand, one has direct access to microscopic configurations through simulations. On the other hand, experiments measure the system's response to external magnetic field perturbations, either switching it on (ZFC) or off (TRM). The previous 
section showed how these two approaches can be reconciled, by
relating the experimental correlation length $\xi_\mathrm{Zeeman}$
to the microscopic coherence length \xim measured in simulations. The key
was the attempt by \citet{janus:17b} to measure experimental response functions in simulations. That approach, however, relied on an approximate
scaling law, valid only for some range of values of the external field and
which broke down close to \Tg (Fig.~\ref{fig:film_simulation_z}), which 
is precisely the most interesting experimental regime for glassy systems
(because it allows the observation of larger coherence lengths).

\begin{figure}
    \centering
    \includegraphics[width=\linewidth]{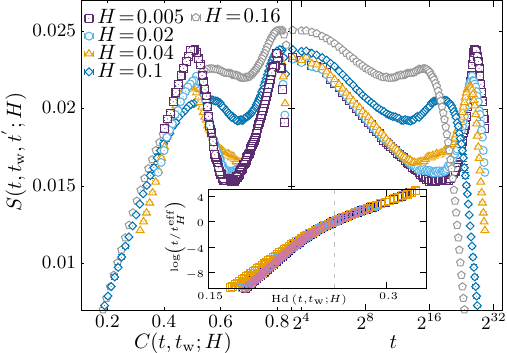}
\caption{Behavior of the relaxation function, $S(t,\tw;H)$ as a function of the correlation function, $C(t,\tw;H)$ ({left}) and as a function of time $t$  ({right}) for several magnetic fields. The peak's position occurs at a constant value $C_\mathrm{peak}$. The insert shows the scaling law that emerges if time is rescaled by the extracted effective time $t^\text{eff}_H$ as a function of the Hamming distance $\mathrm{Hd}$. The dashed line refers to the value of $C_\mathrm{peak}$. See \citet{zhai-janus:20a,zhai-janus:21} for more details.} 
    \label{fig:PRL_scaling_tricks}
\end{figure}

A more sophisticated treatment was clearly needed. In 2020, a
collaboration between experimental and theoretical groups \cite{zhai-janus:20a, zhai-janus:21} took the next step, beginning with a 
more precise replication of the experimental ZFC protocol in simulation.  Two tricks were required:
\begin{itemize}
    \item  Because $H=1$ for the EA model with binary couplings corresponds to $\sim 5 \times 10^4\,\mathrm{Oe}$~\cite{janus:17b,aruga_katori:94}, to match the experimental scales a dimensional analysis \cite{fisher:85} was used to relate $H$ and the reduced temperature $\hat{t}=(\Tg -T)/\Tg$:
\begin{equation}
\label{eq:match_H_between_exp_and_num}
\hat{t}_\mathrm{num} \approx \hat{t}_\mathrm{exp} \bigg( \frac{H_\mathrm{num}}{H_\mathrm{exp}} \bigg)^{\frac{4}{\nu(5-\eta)}},
\end{equation}
where $\nu=2.56(4)$ and $\eta=-0.390(4)$ are the critical exponents at $H=0$ \cite{janus:13}.
\item  Experiments define an effective response time $t^\text{eff}_H$ from the the time at which the relaxation function, $S(t,\tw;H)$, peaks. 
A different strategy was proposed for simulations, where the time at which $S(t,\tw;H)$ peaks was found to occur at a constant 
value of the correlation function $C(t,\tw;H)$ of Eq.~\eqref{eq:Cttw-def}.
This allowed a huge simplification for the extraction of $t^\mathrm{eff}_H$ because
\begin{equation}
\label{eq:Cpeak_vs_teff}
C(t^\mathrm{eff}_H, \tw; H) = C_\mathrm{peak} \, ,
\end{equation}
and could be solved at $H=0$ as well.
Moreover,  the scaling law that appears if one displays $\ln (t/t^\mathrm{eff}_H)$ as a function of the Hamming distance $\mathrm{Hd}$, defined in Eq.~\eqref{eq:Hd_def}, guarantees that the extraction of $t^\mathrm{eff}_H$ is not sensitive to the precise value of $C_\mathrm{peak}$ \cite{orbach-janus:23}:
\begin{equation}
\label{eq:scaling_teff_vs_Cpeak}
\ln \{ t/ t^\mathrm{eff}_H[C_\mathrm{peak}(\tw)] \} = \mathcal{F} [ C(t,\tw;H), \tw ] \; .
\end{equation}
See Fig. \ref{fig:PRL_scaling_tricks} for details.
\end{itemize}

 Because of these observations and tricks, \citet{zhai-janus:20a} and \citet{zhai-janus:21} were able to introduce a scaling law that describes the system’s response over the entire natural range of variation, allowing a bridge to be built between simulations and experiments:
\begin{equation}
\label{eq:scaling_law_ZFC}
\ln \frac{t^\mathrm{eff}_H}{t^\mathrm{eff}_{H \to 0^+}} = \frac{\hat{S}}{T} \xi^{D-\theta/2} H^2 + \xi^{-\theta/2} \mathcal{G} (\xi^{D-\theta/2} H^2 ; T) \, .
\end{equation}
Here $\xi$ stands for $\xim(\tw,T)$ [which it is equivalent to $\xi_\mathrm{Zeeman}(\tw,T)$, recall Fig.~\ref{fig:ximicmac}], $\hat{S}$ is a constant, $D=3$ is the spatial dimension and $\theta$ stands for the replicon exponent $\theta(\tilde{x})$, where $\tilde{x}= \ell_\text{J} (T) /\xi(\tw,T)$ and $\ell_\text{J}(T)$ is the Josephson length.

The derivation of this law follows from:

(i) The assumption that off-equilibrium, and at least for large $\xi$, the equilibrium scaling theory that describes the magnetic response to an external field $H$ holds:
\begin{equation}
\label{eq:scaling_law_M_vsH}
\begin{split}
&M(t,\tw;H) = \left[  \xi(t+\tw) \right]^{- \frac{D}{2}- \frac{\theta}{4} }\\
&\qquad\qquad \times  \mathcal{F} \left(  H \left[  \xi(t+\tw) \right]^{- \frac{D}{2}- \frac{\theta}{4}} , \frac{\xi(t+\tw)}{\xi(\tw)} \right).
\end{split}    
\end{equation}
The full-aging spin-glass dynamics allows the ratio $\xi(t+\tw)/\xi(\tw)$ to be considered as approximately constant.

(ii) However, the Taylor expansion
\begin{equation}\label{eq:M_taylor_expansion_H}
   {\frac {M(t,\tw;H)}{H}} = \chi_1 + \frac{\chi_3}{3!} H^2 + \frac{\chi_5}{5!} H^4 + \mathcal{O}(H^6)\, ,
\end{equation}
where
\begin{equation} \label{eq:chi_vs_xi_dependence}
\chi_{2n-1}     \propto b_{2n} (T) [ \xi(\tw)] ^{(n-1)D -n\frac{\theta({\bar x})}{2}},
\end{equation}
predicts the paradoxical result that $\chi_1\to0$ when $\xi \to \infty$.  In fact, Eq. (\ref{eq:chi_vs_xi_dependence}) neglects the contribution of the regular part of the free energy.  A better description is
\begin{equation}
\chi_1={\frac {{\hat S}(C_{\text {peak}})}{T}}+{\frac {b_2(T)}{\xi^{\theta({\tilde x})/2}}},
\end{equation}
where ${\hat S}(C_{\text {peak}})$ is the function appearing in the fluctuation-dissipation relations \cite[][see Secs.~\ref{subsect:off-equilibrium-no-field-2} and~\ref{subsect:off-equilibrium-no-field-4}]{cugliandolo:93,marinari:98,franz:98,janus:17}. From now on, we shall use the shorthand $\hat S$ for ${\hat S}(C_{\text {peak}})$.
 
\begin{figure}
    \centering
    \includegraphics[width=8.5cm]{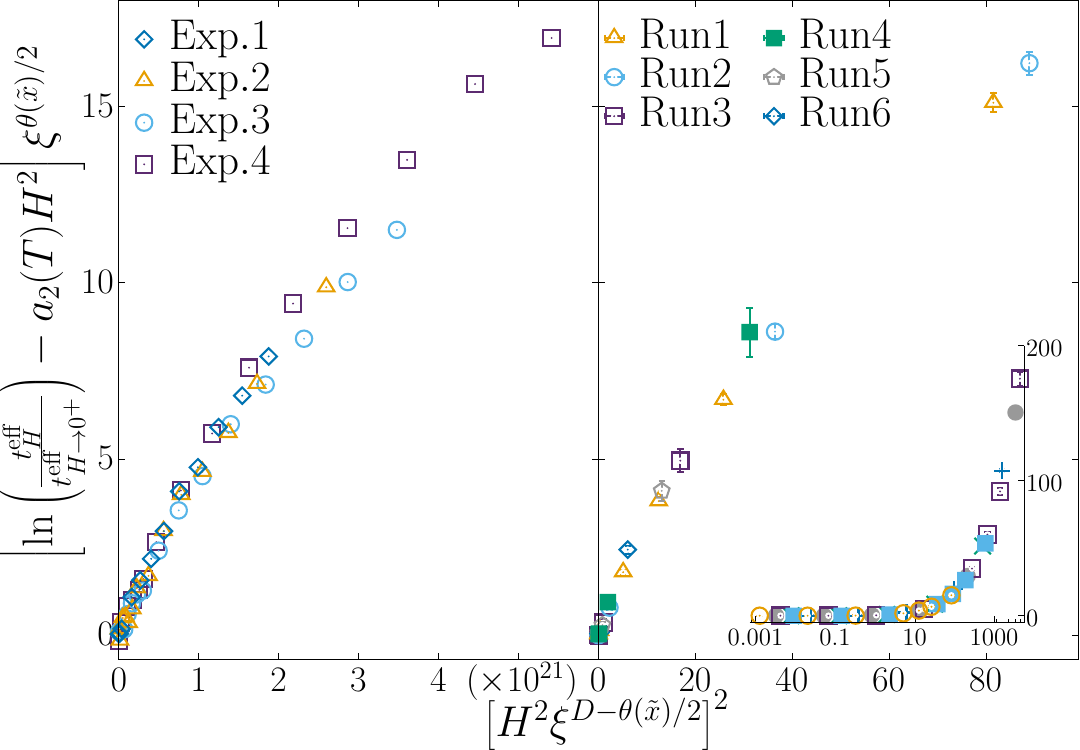}
\caption{The nonlinear part of the response-time data, $[\ln(t^\mathrm{eff}_H/ t^\mathrm{eff}_{H \to 0^+} ) -a_2(T) H^2] \xi^{\theta(\tilde{x})/2}$, plotted against the scaling variable $[H^2 \xi^{D -\theta(\tilde{x})/2}]^2$, see Eq.~\eqref{eq:scaling_law_ZFC}.  On the right, numerical data; on the left the experimental one. See Tables~\ref{tab:exp_details_zhai_janus20a}-\ref{tab:num_detail_native_zhai_janus20}  for parameter details. Figure taken from \citet{zhai-janus:20a}.} 
    \label{fig:scaling_law_ZFC}
\end{figure}

Figure~\ref{fig:scaling_law_ZFC} demonstrates the remarkable collapse of both experimental and numerical data through the scaling law Eq.~\eqref{eq:scaling_law_ZFC}, and solves a thirty-year problem concerning the Zeeman energy.  It creates a road map for further comparison between experiments and simulations, such as the investigation of the exotic phenomena of rejuvenation and memory, see Sec.~\ref{subsec:memory_and_rejuvenation}.

\begin{table}[t]
\begin{ruledtabular}
	\begin{tabular}{c  c c    c    c    c}
		& & \Tm (K) & $\tw$ (s) & $\xi(\tw)/a$ & $\theta(\tilde x)$ \\
                \hline
		\textbf{Exp. 1} && 28.50 & 10\,000 &320.36 &0.337\\
		\textbf{Exp. 2} && 28.75 & 10\,000 &341.76 &0.344\\
		\textbf{Exp. 3} && 28.75 & 20\,000 &359.18 &0.342\\
		\textbf{Exp. 4} && 29.00 & 10\,000 &391.27 &0.349\\
	\end{tabular}
\end{ruledtabular}
        \caption{Main parameters for the four experiments reported in Fig. \ref{fig:scaling_law_ZFC}. Table taken from \citet{zhai-janus:20a}.}
        	\label{tab:exp_details_zhai_janus20a}
\end{table}

\begin{table}[!h]
\begin{centering}
\begin{ruledtabular}
\begin{tabular}{c c c c c c c}
&$\Tm$&$\tw$ & $\xim(\tw)$ & $t_{\text {max}}$ & $\theta(\tilde {x})$&$C_\mathrm{peak}(\tw)$\\ 
\hline\\[-5pt]
${\textbf{Run 1}}$&0.9&$2^{22}$&8.294(7)&$2^{30}$&0.455&0.533(3)\\
${\textbf{Run 2}}$&0.9&$2^{26.5}$&11.72(2)&$2^{30}$&0.436&0.515(2)\\
${\textbf{Run 3}}$&0.9&$2^{31.25}$&16.63(5)&$2^{32}$&0.415&0.493(3)\\
${\textbf{Run 4}}$&1.0&$2^{23.75}$&11.79(2)&$2^{28}$&0.512&0.422(2)\\
${\textbf{Run 5}}$&1.0&$2^{27.625}$&16.56(5)&$2^{32}$&0.498&0.400(1)\\
${\textbf{Run 6}}$&1.0&$2^{31.75}$&23.63(14)&$2^{34}$&0.484&0.386(4)\\
${\textbf{Run 7}}$&0.9&$2^{34}$&20.34(6)&$2^{34}$&0.401&0.481(3)\\
\end{tabular}
\caption{Parameters for the \emph{native} 
 numerical simulations reported in Figs.~\ref{fig:scaling_law_ZFC}--\ref{fig:decay_diff_teff_scaling}:
$\Tm,~\tw,~\xi(\tw)$, the longest simulation time $t_{\text {max}}$, the
replicon exponent $\theta$, and the value of $C_\mathrm{peak}(\tw)$. Table taken from \citet{zhai-janus:20a,zhai-janus:21}.}
\label{tab:num_detail_native_zhai_janus20}
\end{ruledtabular}
\end{centering}
\end{table}

\begin{figure}[t]
    \centering
    \includegraphics[width=8.5cm]{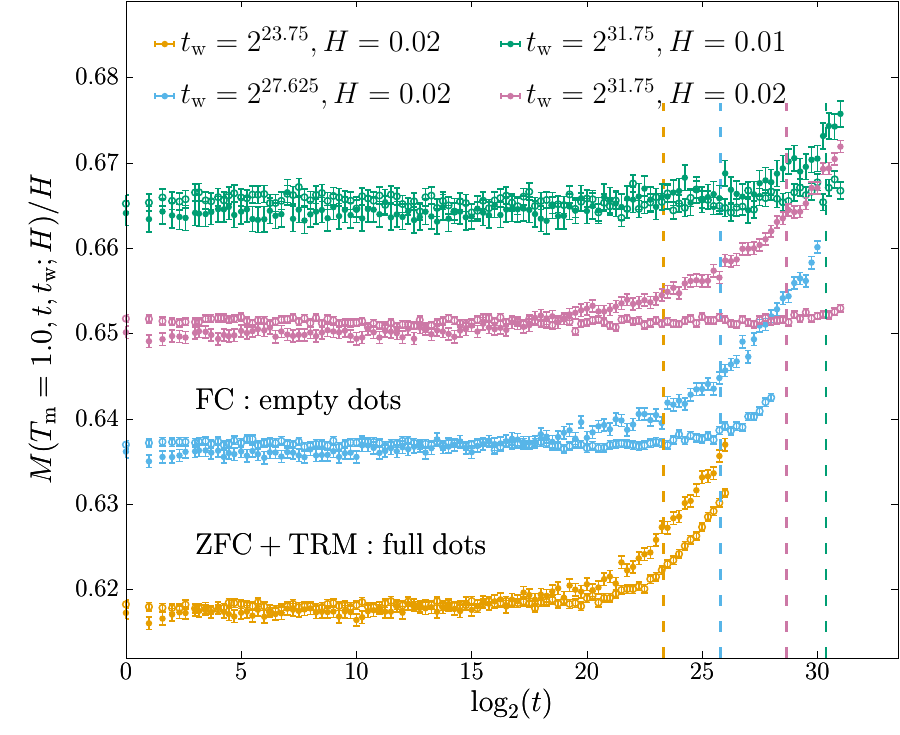}
\caption{Growth of the rescaled magnetization for different experimental protocols: $[M_\mathrm{ZFC}(t,\tw) + M_\mathrm{TRM}(t,\tw)]/H$, and  $M_\mathrm{FC}(0,\tw + t)/H$. Empty symbols are used for the FC case and filled symbols for the quantity ZFC + TRM. The vertical dashed lines indicate the effective times, $t^\text{eff}_H$ , associated with each case. The violation of Eq. \eqref{eq:superposition_M} is evident. Figure taken from \citet{orbach-janus:23}.} 
    \label{fig:superpositionM}
\end{figure}

The scaling law also provides a vehicle to examine the superposition principle, see Eq.~\eqref{eq:superposition_M} and the discussion in Sec.~\ref{subsect:off-equilibrium-no-field-1}. As was briefly explained in the Introduction, this concept was an experimental
milestone, describing the connection between the TRM, ZFC, and FC protocols. Exploiting the same techniques developed for the ZFC protocol, \citet{orbach-janus:23} focused on the TRM protocol to demonstrate, through the use of correlation lengths, that the superposition principle is valid only in the limit that the magnetic field $H\rightarrow 0^+$, see Fig. \ref{fig:superpositionM}.

The argument is as follows.
In the out-of-equilibrium regime, we can connect the growth of $\xim$ to time as in Eq.~\eqref{eq:z}. Additionally, we can relate $\xim$ to an external magnetic field:
\begin{equation}
\label{eq:scaling_xim_vsH}
\xim(\tw) \propto H^{1/y_\mathrm{H}} , \; \;  2 \, y_\mathrm{H}= D-\theta/2     
\end{equation}
where $D$ is the spatial dimension and $\theta$ is the replicon.

\begin{figure}
    \centering
    \includegraphics[width=\linewidth]{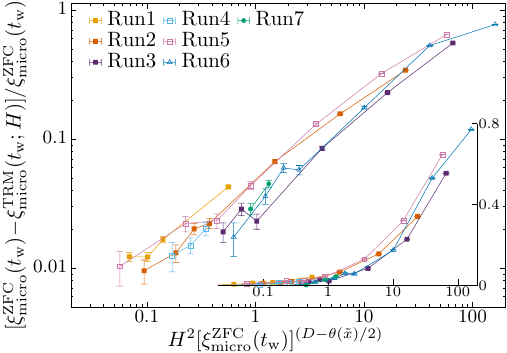}
\caption{Scaling relation according to Eq.~\eqref{eq:diff_xiTRM_xiZFC}. The main plot is in log-log scale; the insert has a linear scale for the ordinate. Figure taken from \citet{orbach-janus:23}. } 
    \label{fig:diff_xiTRM_ZFC_scaling}
\end{figure}

Whether in the presence (TRM) or absence (ZFC) of an external magnetic field, the study begins with the calculation of $\xim$ using Eq.~\eqref{eq:xi_micro_def}. The presence of an external magnetic field slows down the growth of $\xim$. Because $\ximZFC$ and $\ximTRM$ approach the $H^2 \to 0^+$ limit with a linear slope, their difference can be expressed as
\begin{equation}\label{eq:diff_xiTRM_xiZFC}
    1- \frac{\ximTRM(\tw,H)}{\ximZFC(\tw)} = A(\tw,T) [\ximZFC(\tw)]^{D-\theta/2} H^2 \, .
\end{equation} 
A scaling behavior emerges, see Fig.~\ref{fig:diff_xiTRM_ZFC_scaling}.

\begin{figure}
    \centering
    \includegraphics[width=8.5cm]{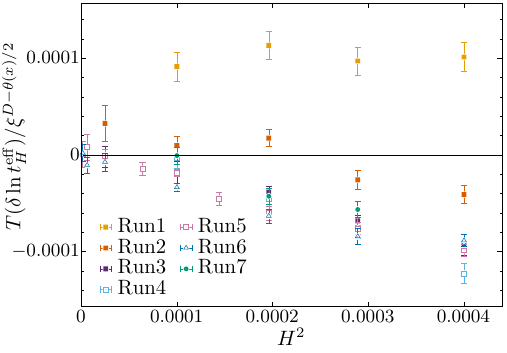}
\caption{The rescaled  quantity $T ( \delta \ln t^\text{eff}_H)/[ \xim(\tw)]^{D-\theta/2}$ as a function of $H^2$. Figure taken from \citet{orbach-janus:23}.} 
    \label{fig:decay_diff_teff_scaling}
\end{figure}

 The same result can be obtained following a different strategy. Because of Eq.~\eqref{eq:scaling_law_ZFC}, the decay of the effective time $t^\text{eff}_H$ can be connected to $\xim$. Thus, the difference in the decay of $t^\text{eff}_H$ between the TRM and ZFC protocols can be written as
 \begin{equation}\label{eq:difference_teff_scaling}
\begin{split}
\frac{\ln t^\text{eff,ZFC}_H-\ln t^\text{eff,TRM}_H}{\xim(\tw)^{D-\theta/2}}= & -[ K^\mathrm{ZFC}\!-\!K^\mathrm{TRM}] H^2\\& + \mathcal{O}(H^4) \, .
\end{split}     
 \end{equation}
The results in Fig.~\ref{fig:decay_diff_teff_scaling} show that scaling behavior emerges.  By combining this with Eq.~\eqref{eq:difference_teff_scaling}, the failure of the superposition principle, Eq.~\eqref{eq:superposition_M}, in finite magnetic fields is shown to be a result of the lowering of the free-energy barrier in the presence of an external magnetic field.

\subsection{Properties in the vicinity of the critical temperature}\label{subsect:off-equilibrium-in-field-fixed-T-6}
A scaling law for the magnetic field valid for large fields and close to $\Tg$ was developed in \citet{zhai:22}. It was based on the assumption that a scaling theory that describes the magnetic response to an external field $H$ in equilibrium \cite{parisi:88,amit:05} also holds in the nonequilibrium regime.  Defining an ``effective'' response time, $t^\text{eff}_H$ in the same sense as Eq.~\eqref{eq:arrahenius_law_Delta_max} above, Eqs.~(\ref{eq:scaling_law_ZFC}--\ref{eq:chi_vs_xi_dependence}) can be written as
\begin{equation}\label{eq:logteff_taylor_exp}
\ln \, t_H^{\text {eff}}=a_0+a_2H^2+a_4H^4+a_6H^6 + {\mathcal {O}}(H^8),
\end{equation}
where
\begin{equation}\label{eq:an_coff_vs_xim_exp}
a_n\propto b_{2n}(T)\xi^{|nD-(n+1)\theta(\bar x)|/2}.
\end{equation}
The terms in Eq. ~\eqref{eq:an_coff_vs_xim_exp} are defined in the text following Eq. (\ref{eq:scaling_law_ZFC}).  Fits of Eqs. (\ref{eq:M_taylor_expansion_H}) and (\ref{eq:chi_vs_xi_dependence}) to the data exhibit nonlinearities in $H^2$, both from experiments and simulations, and are exhibited in \citet{zhai:22}.  
\begin{figure}[t]
    \centering
    \includegraphics[width=8.5cm]{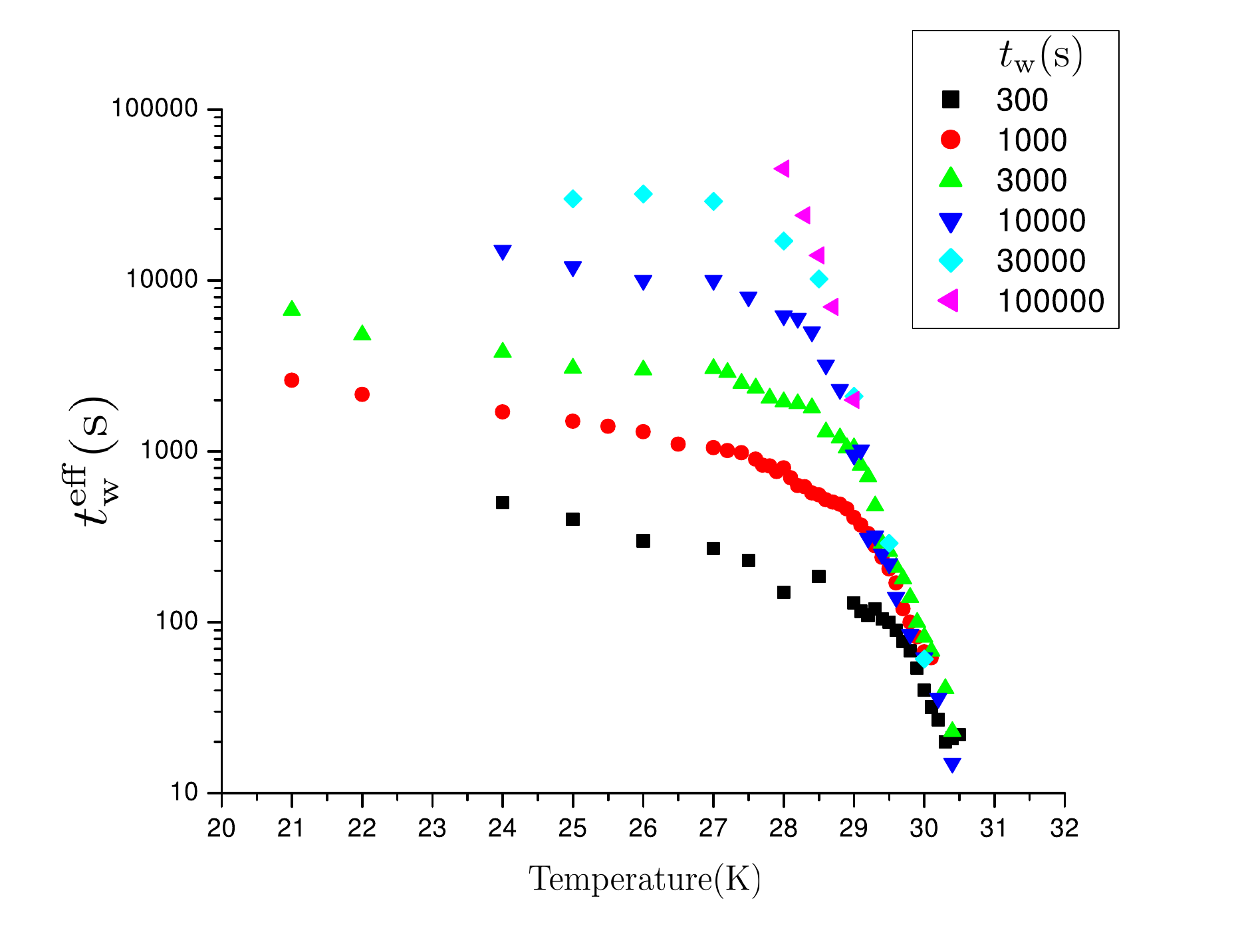}
\caption{Effective time $\tw^{\text {eff}}$ extracted from the time at which $S(t)$ peaks vs temperature for waiting times $\tw$ from 300 to 100\,000 s for a 6 at.\% CuMn single-crystal sample ($\Tg=31.5$~K).  Reproduced from \citet{Kenning:24}.} 
 \label{fig:s(t)_peaks}
\end{figure}

 A spectacular example of the need for the scaling law is the collapse of $t^{\text {eff}}_H$ near $\Tg$ \cite{Kenning:24}. Fig. \ref{fig:s(t)_peaks} exhibits his extraction of $t^{\text {eff}}_H$ for a CuMn 6 at.\% ($\Tg=31.5$ K) as $T\rightarrow \Tg$ from below.

This collapse is evidence of the importance of the higher-order terms in the scaling law [Eqs. (\ref{eq:M_taylor_expansion_H}) and (\ref{eq:chi_vs_xi_dependence})]. 
The values of $a_2$ and $a_4$ were found by \citet{zhai-janus:20a} to be of opposite sign, so that the $H^4$ term is of opposite sign to the $H^2$ term in the expansion of Eq.~(\ref{eq:logteff_taylor_exp}).  The power of the coherence length in the $H^4$ term is roughly twice the power of the coherence length in the $H^2$ term.  Hence, as one approaches $\Tg$ \cite[and $\xim$ increases,][]{janus:18}, the $H^4$ term rapidly grows compared to the $H^2$ term, sharply diminishing $t^{\text {eff}}_H$.

A quantitative fit is underway at the time of this review \cite{Kenning:24}.  For the moment, we believe the scaling law exhibited in Eqs. (\ref{eq:logteff_taylor_exp}) and (\ref{eq:an_coff_vs_xim_exp}) is sufficient to explain the collapse of $t^{\text {eff}}_H$ as $T\rightarrow \Tg$ from below.

\subsection{Brief summary of dynamics in a field}\label{subsec:dynamics_in_a_field}
 The equilibrium and out-of-equilibrium dynamics of the 3D Ising spin glass in the presence of an external magnetic field were addressed in \citet{janus:14b} using  numerical simulations with the Janus~I dedicated supercomputer. The next paragraphs briefly review the main findings of this study.
 
The dynamics appear to be activated, in agreement with the droplet-model predictions that  no phase transition occurs in a field. However, the detailed dependence of the observables with time differs from that predicted by the droplet model.  In particular, the data follows a super-Arrhenius law at temperatures as low as $T=0.36 \Tg^{(H=0)}$. The large value of the correlation length at low temperatures indicates that the lower critical dimension is at most slightly greater than  three.

Furthermore, equilibrium and out-of-equilibrium approaches allow the determination of a kind of transition line.  However, the correlation length does not diverge on this line, against replica-symmetry breaking (RSB) predictions. In addition, if the transition is driven by a $T=0$ fixed point (as in the droplet model) one should expect some kind of activated dynamics also in RSB. Thus, in this scenario, the identified transition line could be interpreted as a dynamic glass transition with the correlation length diverging at lower temperatures.

Finally, \citet{janus:14b} discussed a scenario motivated by the study of supercooled liquids.  On a qualitative level, the dynamic transition could be interpreted as a mode-coupling temperature, and, at lower temperatures, a crossover to activated dynamics.

In summary, the dynamical behavior in and out of equilibrium of the three-dimensional spin glass in a magnetic field exhibits large values of the correlation length. But this behavior is not compatible with the onset of a continuous phase transition (a de Almeida Thouless line). 
These facts, together with the results of equilibrium numerical simulations~\cite{janus:14c,janus:12}, suggest that the lower critical dimension of the Ising spin-glass model in a field is greater than three but less than four \cite[see also][]{zhai-janus:21}.

\section{Off-equilibrium dynamics: nonlinear response when varying the temperature}
\label{sect:off-equilibrium-in-field-several-T}

Rejuvenation and memory experiments \cite{jonason:00} are among the most profound in the field of spin glasses. These experiments suggested that the study of magnetic responses in temperature-varying protocols could demonstrate some of the most interesting features of a complex free-energy landscape (most notably, the multiplicity of equilibrium states and temperature chaos). Once again, it is necessary
to relate equilibrium landscapes to nonequilibrium experiments. This section reports on a continuing effort, originating in \citet{bouchaud:01} and still ongoing  \cite{janus:23,freedberg:24,paga:23b}, to learn about the geometry of the free-energy landscape from temperature-varying experiments.

This section does not take a chronological view. Rather, it intertwines experimental and theoretical progress on different aspects of the problem. Sec.~\ref{subsec:mpemba} is somewhat different, because it deals with a memory of a different kind (the Mpemba effect), which has been theoretically predicted but not yet seen experimentally in spin glasses.

\subsection{Temperature chaos in the off-equilibrium context}\label{subsect:T-chaos-off-eq}
Temperature chaos was introduced as an equilibrium phenomenon in Secs.~\ref{subsect:intro-6} and~\ref{sect:Equilibrium}, while 
experiments are always conducted out of equilibrium. This section will explore
how one can translate the concept of temperature chaos to the experimental setting
and, then, how it can be investigated in nonequilibrium simulations.

\subsubsection{The experimental point of view}
\begin{figure*}[t]
    \centering
    \includegraphics[width=\columnwidth]{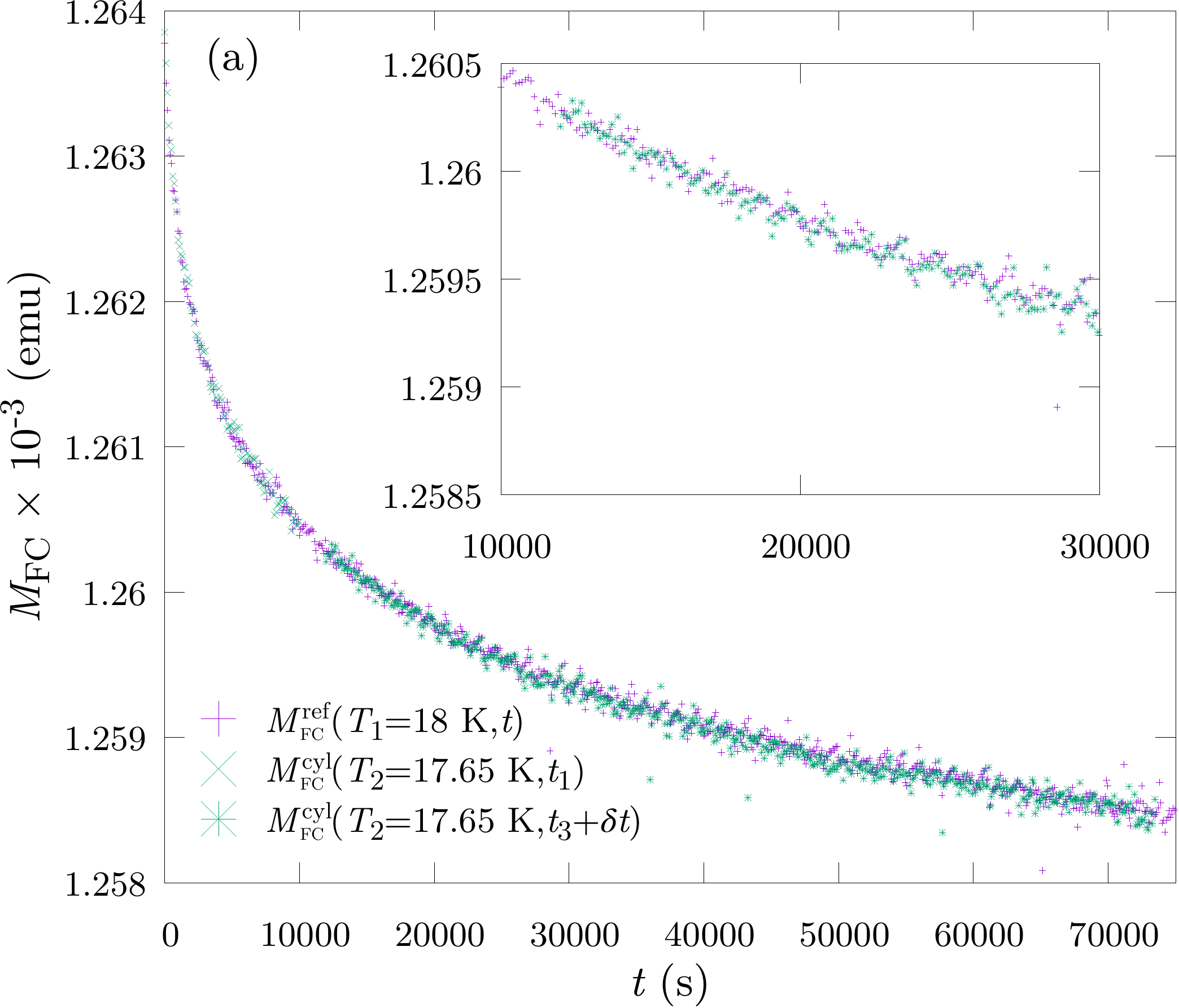}
    \includegraphics[width=\columnwidth]{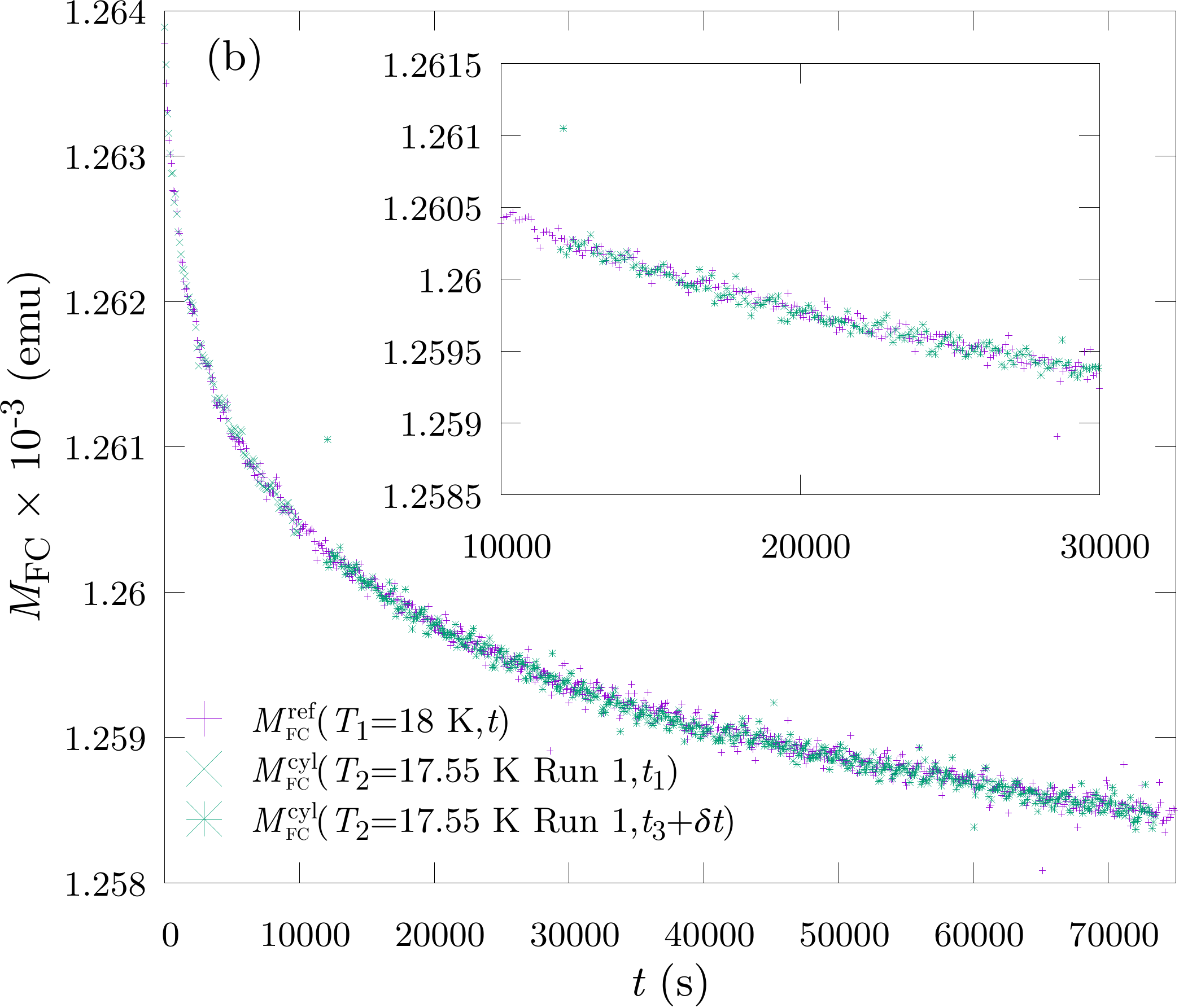}
    \includegraphics[width=\columnwidth]{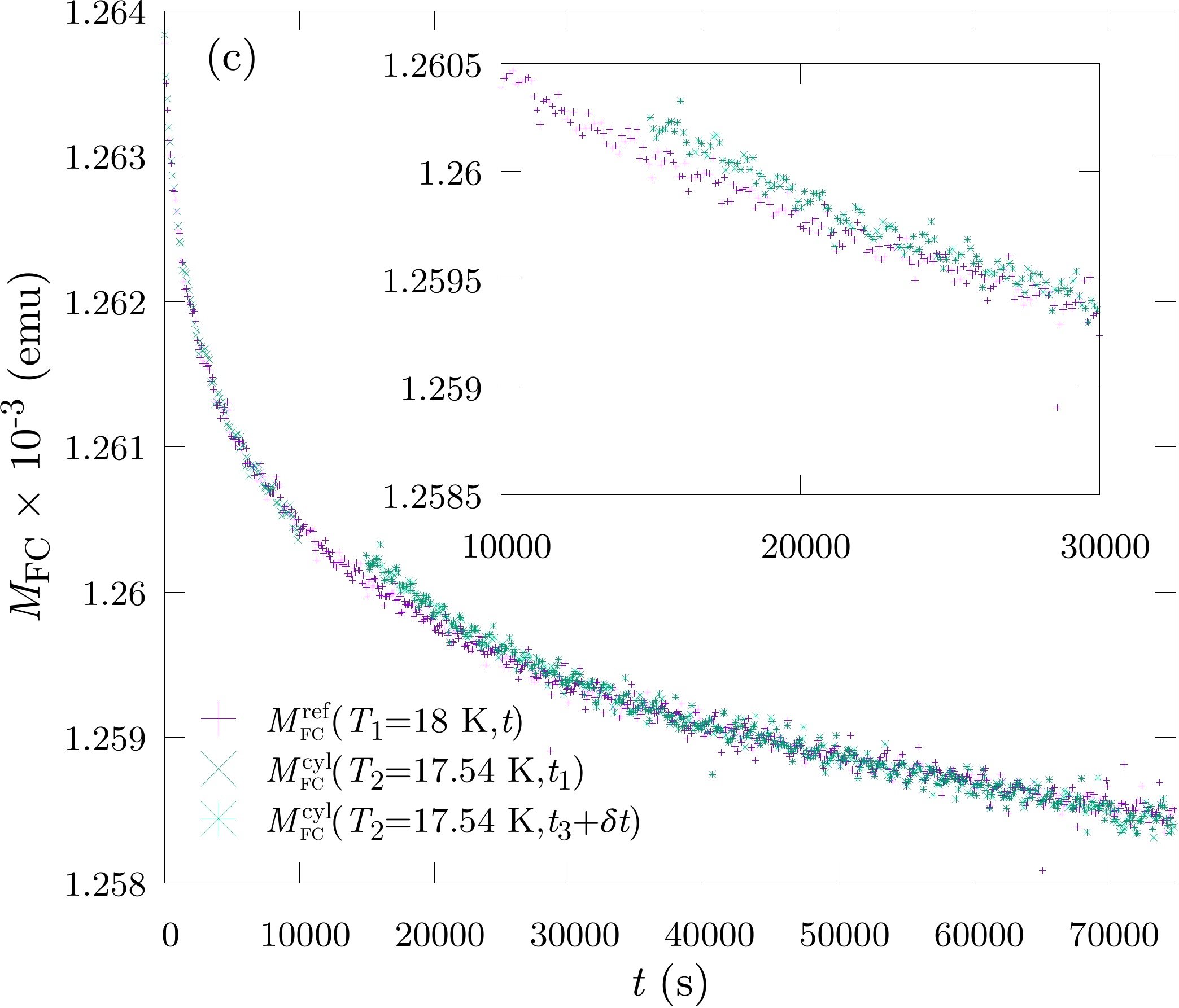}
    \includegraphics[width=\columnwidth]{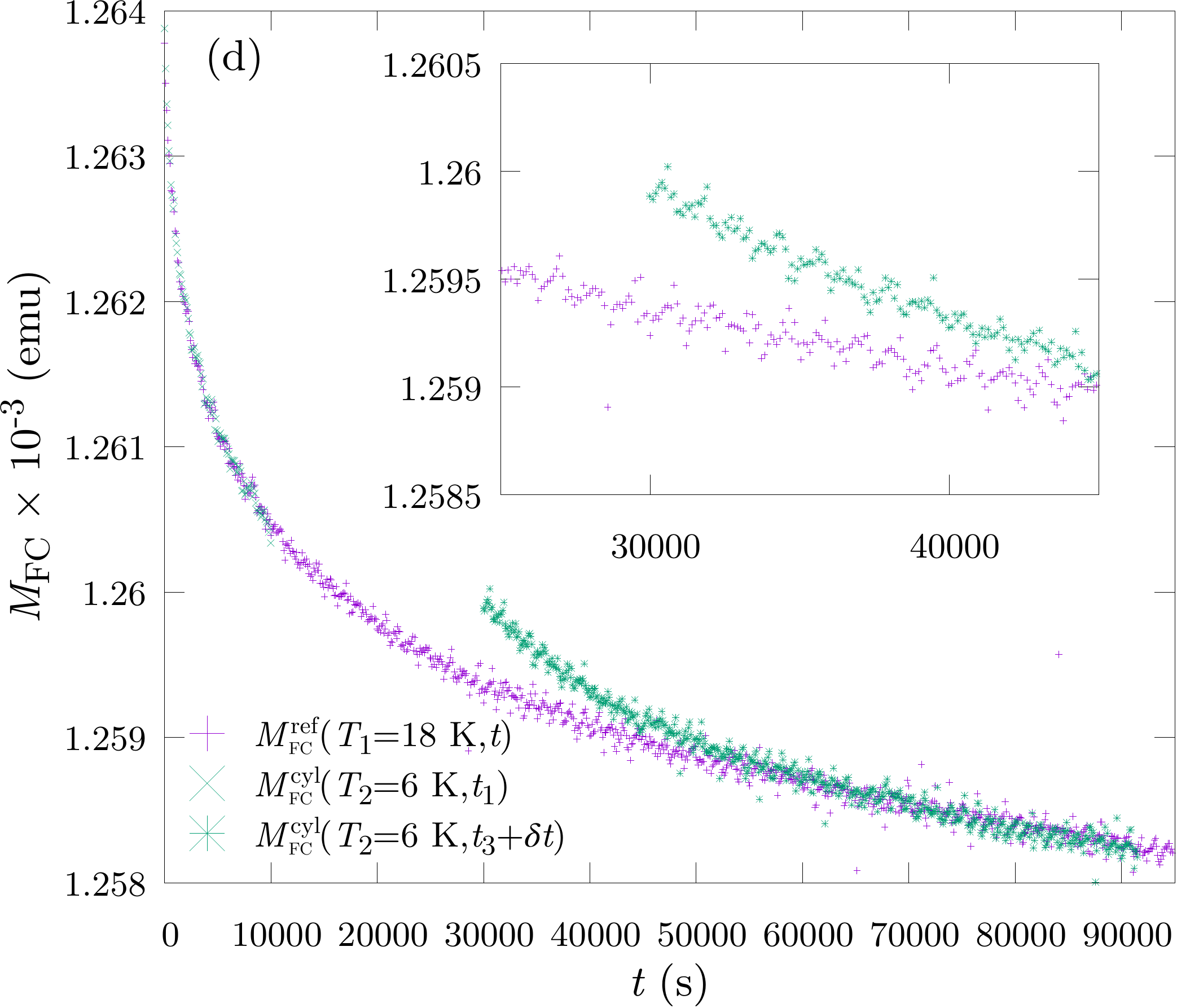}
\caption{The reference and the temperature-cycling curves for $M_{\text {FC}}(t)$ of a CuMn 6 at.\% sample for $T_1=18$ K, and $\twone=10^4$~s. The temperature-cycling curve is shifted by $\delta t$ to overlap the reference curve.  In the reversible range, (a) and (b), the cycling curve can be overlapped with the reference curve over the whole period of measurement, $\approx 7\times 10^4$~s.  In the chaotic range, (c) and (d), the cycling curve can only be partially overlapped.  Hence, temperature chaos sets in for $T_1 = 18$ K, $\twone=10^4$~s at $\Delta T \geq 450$ mK.  Reproduced from Fig. 1 of \citet{zhai:22}.}  
    \label{fig:MFC_chaos_evidence}
\end{figure*}

\label{subsect:off-equilibrium-in-field-several-T-1}
\citet{bray:87b} introduced a length scale for temperature chaos upon a temperature change $\Delta T$:
\begin{equation}\label{eq:chaotic_length_def}
\ell_\text{c}=a_0\bigg[{\frac {T_1}{|\Delta T|}}\bigg]^{1/(\Ds/2-y)}
\end{equation}
where $a_0$ is the average distance between magnetic ions, $\Delta T=T_1-T_2$ is the temperature change, $T_1$ and $T_2$ being the initial and final temperatures, respectively, and $(\Ds/2-y)$ is the exponent combination for temperature chaos (Sec.~\ref{subsect:intro-6}).  Thus, at a given coherence length $\xi(\tw,T)$, the condition for the onset of  chaos is
\begin{equation}\label{eq:chaotic_length_def_vs_xi}
\ell_\text{c}=\xi(\tw,T),
\end{equation}
such that, for $\ell_\text{c}\geq\xi(\tw,T)$ the system is reversible, but for $\ell_\text{c}\leq\xi(\tw,T)$, the system has entered a chaotic state.

The search for temperature chaos in spin glasses has had a checkered history.  The first systematic report \cite{jonsson:02} utilized both positive and negative temperature shifts for $\Delta T$.   \citet{arai:07} were able to change bond strengths to access the chaotic state.  \citet{jonsson:02} found for small $\Delta T$ that aging was ``accumulative'' but for larger $\Delta T$ a transition to ``nonaccumulative'' aging was observed.

We associate accumulative to reversible behavior, and nonaccumulative to chaos.  \citet{jonsson:02} used ZFC measurements to extract the temperature ranges for accumulative and nonaccumulative aging.  They estimated the chaos exponent $(\Ds/2-y)$ as
\begin{equation}\label{eq:zetaJ_value}
1/(\Ds/2-y)_\text{J} = 2.65\pm 0.5 
\end{equation}
where the subscript J refers to their value.  Unfortunately, the accepted value for $1/(\Ds/2-y)$ is near unity (recall Sec.~\ref{subsect:intro-6}).  The origin of their error has been speculated \cite{zhai:22} to arise from the presence of chaos induced by the change of magnetic field in a ZFC protocol.  It is known that magnetic-field change can induce chaos \cite{kondor:89,billoire:03}, so that it may have interfered with their extraction of the chaos exponent.

The problem, therefore, is how to extract $1/(\Ds/2-y)$ from experiment without changing the magnetic field. A  solution was developed by \citet{zhai:22}, who utilized the (small but finite) time dependence of the field-cooled magnetization, $M_{\text {FC}}(t)$.  In this protocol, the magnetic field remains constant, so that only a change in temperature can contribute to chaos.

The protocol was first to establish a reference curve.  A CuMn 6 at.\% single crystal ($\Tg=31.6$ K) was cooled from 40 K to the measurement temperature $T_1=18$~K in a constant field $H=40$ Oe at 10 K/min.  After the temperature was stabilized, the reference magnetization $M_{\text {FC}}^{\text {ref}}(t,T_1)$ was recorded.  In the temperature-cycling protocol, the sample was cooled from 40 K to $T_1=18$~K at 10 K/min, where the magnetization was recorded for a duration of $10^4$~s.  The temperature was then lowered to $T_2=T_1-\Delta T$.  After aging for $10^3$~s, the temperature was raised back to $T_1=18$ K, and the magnetization $M_{\text {FC}}^{\text {cyclic}}(t)$ recorded.

If there is no chaos, the two time-dependent magnetizations, $M_{\text {FC}}^{\text {ref}}(t,T_1)$ and $M_{\text {FC}}^{\text {cyclic}}(t)$, can be overlapped directly by shifting $M_{\text {FC}}^{\text {cyclic}}(t)$ by \changes{a time interval} $\delta t$ [see Fig.~\ref{fig:MFC_chaos_evidence}, curves (a) and (b)], and aging can be taken as cumulative and reversible.  However, if chaos is present, $M_{\text {FC}}^{\text {cyclic}}(t)$ cannot be made to overlap with $M_{\text {FC}}^{\text {ref}}(t,T_1)$ by shifting $M_{\text {FC}}^{\text {cyclic}}(t)$ \changes{by any} $\delta t$ [see Fig.~\ref{fig:MFC_chaos_evidence}, curves (c) and (d)].

Figure \ref{fig:MFC_chaos_evidence} displays the results of these experiments.  Comparing Fig. \ref{fig:MFC_chaos_evidence} (b) with Fig. \ref{fig:MFC_chaos_evidence} (c) shows that temperature chaos begins between 17.55K and 17.54 K, or between $\Delta T = 0.45$ K and 0.46 K, for an unprecedented accuracy of 10 mK.  The amplitude of the lack of overlap increases with increasing $\Delta T$, as can be seen in Fig.~\ref{fig:MFC_chaos_evidence} (d) where $\Delta T=12$ K.  \citet{zhai:22} display additional values for $\Delta T$ for $T_1=18$ K, and for different $T_1=17-14$ K.  Having these different $T_1$, and the concomitantly different $\Delta T$ for the onset of temperature chaos, enables the extraction of $1/(\Ds/2-y) \approx 1.1$, well within the range of accepted values for $(\Ds/2-y)$  from theoretical calculations and simulations for the 3D Ising model \cite[see][for a list of these values]{zhai:22}.

\subsubsection{The numerical point of view}\label{subsubsect:off-equilibrium-in-field-several-T-1-numerical}
The crucial elements to successfully observe temperature chaos in nonequilibrium simulations are three: the capacity to reach long times (enabled by specialized
hardware, see Sec.~\ref{subsect:intro-dedicated_computers}), the idea that temperature chaos is a rare event \cite[][see Sec.~\ref{sect:Equilibrium}]{fernandez:13}, and the statics-dynamics equivalence~\cite[][see Secs.~\ref{subsect:off-equilibrium-no-field-2} and~\ref{subsect:off-equilibrium-no-field-4}]{barrat:01,janus:08b,janus:10b,janus:17}.

\citet{janus:21} performed numerical simulations of the EA model in three dimensions with a lattice size of $L=160$. The numerical protocol consisted of a direct quench from a random configuration, effectively at infinite temperature, to a working temperature $\Tm<\Tg$, where the system is left to relax for a time \tw.

\begin{figure}[t]
    \centering
    \includegraphics[width=8.5cm]{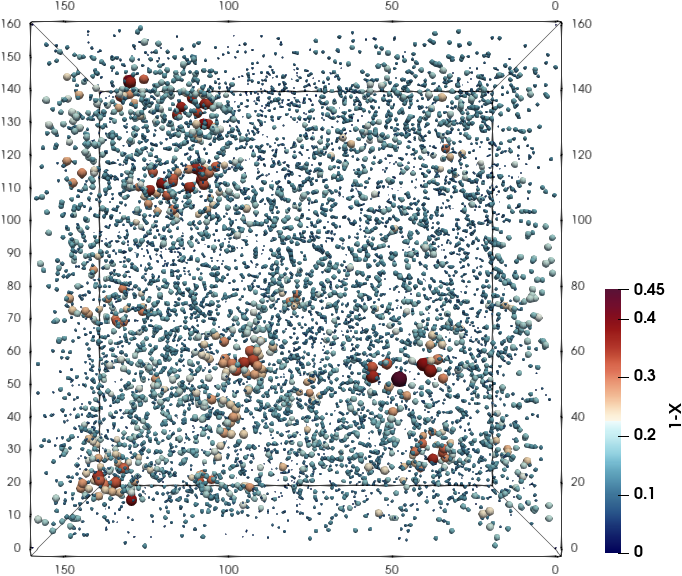}
\caption{Randomly chosen spheres of a sample of size $L=160$ are depicted with a color code depending on $1-X$ ($X$ is the chaoticity parameter computed for spheres of radius $r=12$ and $\xim=12$ and temperature $T_1=0.7$ and $T_2=1.0$). For visualization purposes, spheres are represented with a radius $12(1-X)$, so that only fully chaotic spheres (\emph{i.e.}, $X = 0$) have their real size. Figure from~\cite{janus:21}.} 
\label{fig:janus_2021_chaotic_spheres}
\end{figure}

At this point, it is necessary to, somehow, compare configurations with different temperatures $T$ of the thermal bath. The natural choice is to use the chaoticity parameter $X^J_{T_1,T_2}$, see Eq.~\eqref{eq:chaotic_parameter}. While computing the chaoticity parameter at equilibrium is straightforward, in off-equilibrium dynamics there is an extra parameter to play with: it is necessary to select the time $t$ of the compared configurations. The statics-dynamics equivalence suggests that, if in equilibrium the relevant length scale is the linear size $L$, the nonequilibrium system should be governed by the coherence length $\xim(\tw)$, whose growth strongly depends on $T$ (see Sec.~\ref{subsect:off-equilibrium-no-field-3}). Thus, the authors used the coherence length of the system as a proxy of its aging state and compared configurations at different times by keeping $\xi_{T_1}(\twone) \approx \xi_{T_2}(\twtwo)$. This strong condition will be discussed again in Sec.~\ref{subsubsec:memory_and_rejuvenation_num}.

When studying the chaoticity parameter as a global observable, the authors observed almost no signal. Again, the statics-dynamics equivalence suggests that the nonequilibrium system with $L \gg \xim(\tw)$ can be considered as a collection of equilibrated systems with $L\sim \xim(\tw)$. Therefore, computing global observables implies averaging over equilibrated regions. As explained in Sec.~\ref{subsect:Equilibrium-1}, this averaging washes the chaotic signal away.

Hence, the study of temperature chaos has to be local. The authors defined spherical regions of size $\xim(\tw)$ and computed within them a chaoticity parameter
\begin{equation}\label{eq:def-X-out-eq-local}
    X^{s,r}_{T_1,T_2} (\xi) = \frac{\bigl\langle \bigl[q^{s,r}_{T_1,T_2}\bigr]^2 \bigr\rangle_J} {\bigl(\bigl \langle \bigl[q^{s,r}_{T_1,T_1}\bigr]^2 \bigr \rangle_J \bigl \langle \bigl[q^{s,r}_{T_2,T_2}\bigr]^2 \bigr \rangle_J\bigr)^{1/2}} \, ,
\end{equation}
where the superindex $\lbrace s,r \rbrace$ labels the specific sphere $s$ and the radius of the considered sphere $r$ for which we are computing the chaoticity parameter and the thermal averages $\langle \cdots \rangle_J$ are as in Eq.~\eqref{eq:chaotic_parameter}. This local observable emphasizes the heterogeneity of the phenomenon, as shown in Fig.~\ref{fig:janus_2021_chaotic_spheres}, in which there is a vast majority of nonchaotic spheres (blue-like colors) and specific regions with very chaotic spheres (red-like colors).

\begin{figure}[t]
    \centering
    \includegraphics[width=8.5cm]{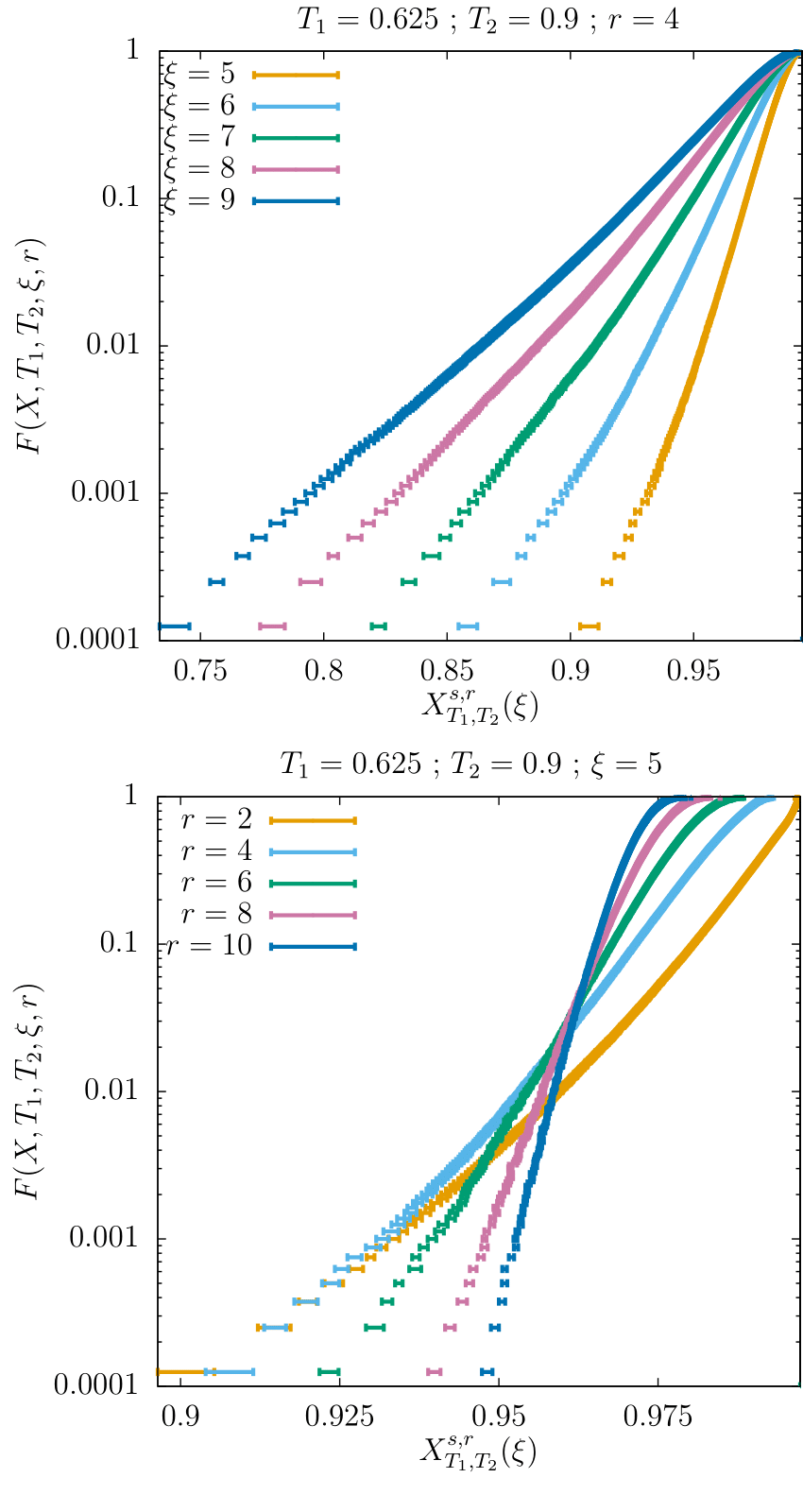}
\caption{Distribution function $F(X,T_1,T_2,\xim,r)$ for $T_1=0.625$ and $T_2=0.9$, for spheres of fixed radius $r=4$ (top) and fixed coherence length $\xim=9$ (bottom). Data taken from \citet{janus:21}.} 
\label{fig:janus_2021_distribution_function}
\end{figure}

A more quantitative study was conducted in~\citet{janus:21} by computing the probability distribution function of the chaoticity parameter of the spheres,
\begin{equation}\label{eq:prob_chaotic}
F(X,T_1,T_2,\xi,r) = \mathrm{Prob} \bigl[ X^{s,r}_{T_1,T_2} (\xi) < X \bigr]. 
\end{equation}
Three main results can be extracted from this computation. First, the vast majority of the spheres exhibit no chaos. See, for example, Fig.~\ref{fig:janus_2021_distribution_function} (top), $\xim=9$ curve, approximately $90 \%$ of the spheres have $X^{s,r}_{T_1,T_2}>0.9$. Next, the chaotic signal monotonically increases with the coherence length $\xim$. Last, for each coherence length, there exists a value for the radius of the  sphere that maximizes the chaotic signal [see Fig.~\ref{fig:janus_2021_distribution_function} (bottom)].

This analysis, though very rich and interesting, is also afflicted by an excess of information: there are a lot of variables that can be explored. To clarify the situation and focus on the strong chaotic signal, the authors fixed the probability level $F$ and the coherence length $\xi$, studying the change of the chaotic signal with the size of the sphere. Full details of the procedure can be found in~\citet{janus:21} and \citet{moreno-gordo:21}.

As sketched above, it was found that the chaotic signal peaks for a specific value of the radius of the studied sphere. The analysis of the peaks found two main results. First, the optimum radius to study temperature chaos is proportional to the coherence length $\xim$. Second, there exists a characteristic coherence length $\xim^*$ that points to a crossover between a weakly chaotic regime and a strongly chaotic regime. Moreover, since $\xim^*$ indicates the crossover between weak and strong chaos, it must be the nonequilibrium analog of the equilibrium chaotic length $\ell_\text{c}$ (see Sec.~\ref{subsect:intro-6}). The authors found that the crossover length $\xim^*$ admits a scaling with the temperature difference $T_2-T_1$ of the form
\begin{equation}\label{eq:chaos_scaling}
    \xim^*(T_1,T_2,F) \propto (T_2-T_1)^{-1/(\Ds/2-y)_{\mathrm{NE}}} \, 
\end{equation}
very similar to the scaling of the chaotic length $\ell_\text{c}$. \citet{fernandez:13} found a value of the exponent $(\Ds/2-y)=1.07(5)$ at equilibrium while \citet{janus:21} found $(\Ds/2-y)_{\mathrm{NE}}=1.19(2)$ out of equilibrium. This agreement supports the idea of a crossover length.

The crossover length $\xi^*$ turned out to be fundamental to understanding temperature chaos  out of equilibrium  and allows the estimation of the necessary conditions to observe effects that deeply depend on the presence of a strong chaotic regime, such as memory and rejuvenation.

In summary, temperature chaos is real, at least from an experimental point of view.  The onset is clear from Fig.~\ref{fig:MFC_chaos_evidence}, and the extraction of $(\Ds/2-y)$ gives a value consistent with theory and simulations.  What remains is the nature of temperature chaos: is it a rapid rise from onset, or does it smoothly increase across a broad range of $\Delta T$?  Another unanswered question is its magnitude.  Does the chaotic state fully replace  the reversible or cumulative aging state, or is it small compared to it? The next section, on memory, will provide
at least a partial answer.

\subsection{Rejuvenation and memory}\label{subsec:memory_and_rejuvenation}
\subsubsection{An experimental approach}\label{subsubsec:memory_and_rejuvenation_exp}
\begin{figure}[t]
\centering
    \includegraphics[width=.85\linewidth]{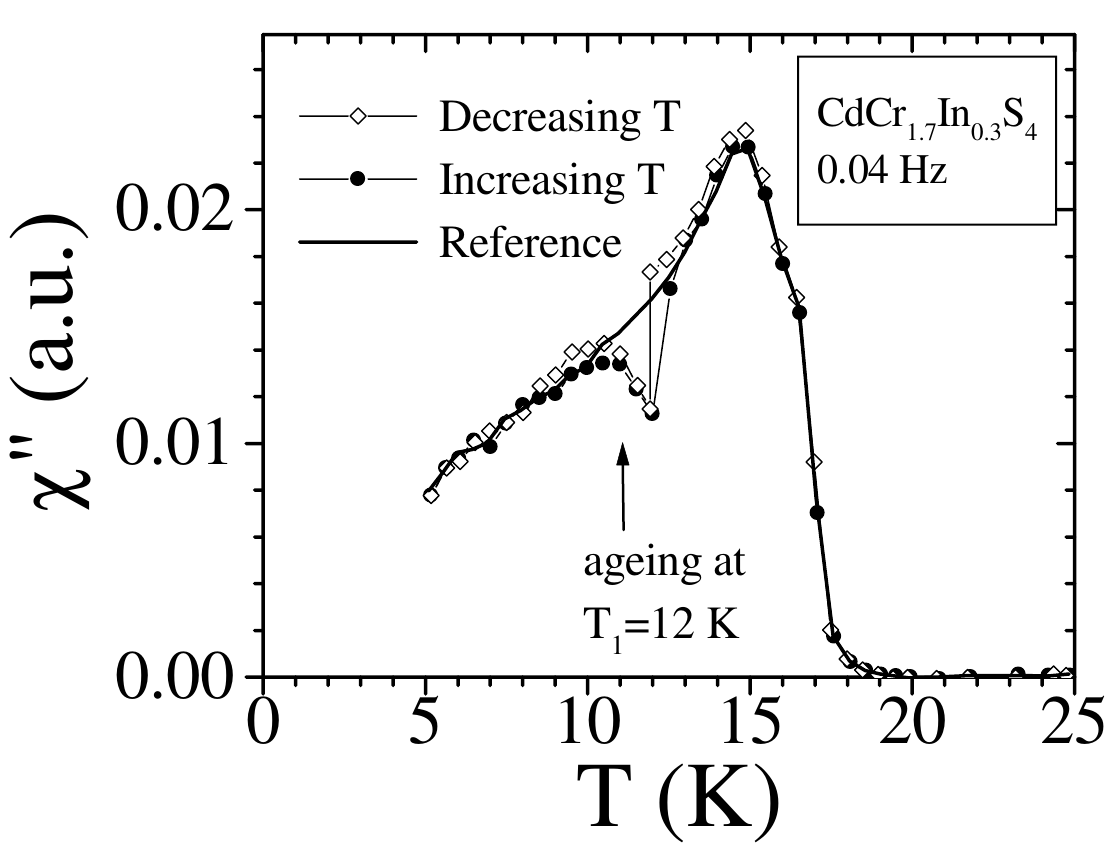}
\caption{The out-of-phase susceptibility $\chi^{\prime\prime}$ of the CdCr$_{1.7}$In$_{0.3}$S$_4$ spin glass.  The solid line is measured upon heating the sample at a constant rate of 0.1 K/min (reference curve).  Open diamonds: the measurement is done during cooling at this same rate, except that the cooling procedure has been stopped at 12 K during 7~h to allow for aging.  Cooling then resumes down to 5 K: $\chi^{\prime\prime}$ is not influenced and goes back to the reference curve (chaos).  Solid circles: after this cooling procedure, the data is taken while reheating at the previous constant rate, exhibiting memory of the aging stage at 12 K.  Reproduced from Fig. 1 of \citet{jonason:00}.}  
    \label{fig:jonason_rejuvenation_memory}
\end{figure}

One of the remarkable features of spin-glass dynamics is their behavior under aging and temperature cycling, namely rejuvenation and memory.  These properties were first exhibited by \citet{jonason:98} and are reproduced in Fig.~\ref{fig:jonason_rejuvenation_memory}.  The imaginary part of the ac magnetic susceptibility, $\chi^{\prime\prime}(\omega)$, of the insulating spin glass CdCu$_{1.7}$In$_{0.3}$S$_4$ ($\Tg=16.7$ K) as a function of temperature $T$ was first measured without aging, providing the reference curve (solid line in Fig. \ref{fig:jonason_rejuvenation_memory}).  Then the experiment was repeated, but this time as the temperature was lowered, it was stopped at $T=12$ K, where the system aged for 7~h.  Then the temperature was lowered to the final $T=5$ K.  Remarkably, $\chi^{\prime\prime}(\omega)$ {\it returned} to the reference curve after a small change in $T$.  This behavior was termed ``rejuvenation'': The measured $\chi^{\prime\prime}(\omega)$ behaved as if it had never been aged at $T=12$ K! This effect was attributed to temperature chaos, but there remained need for a definitive proof of this conjecture.

What was seen next was even more unexpected.  As the system was heated from its lowest temperature, $\chi^{\prime\prime}(\omega)$ retraced its behavior, reproducing the drop seen during aging when the temperature was decreasing.  This was termed ``memory''.  That it occurred after the system had rejuvenated makes it remarkable.  For if rejuvenation is (and was) ascribed to a transition to a chaotic state, how could it ``remember'' what had occurred at a previous higher temperature after aging?

\begin{figure}
    \centering
    \includegraphics[width=8.5cm]{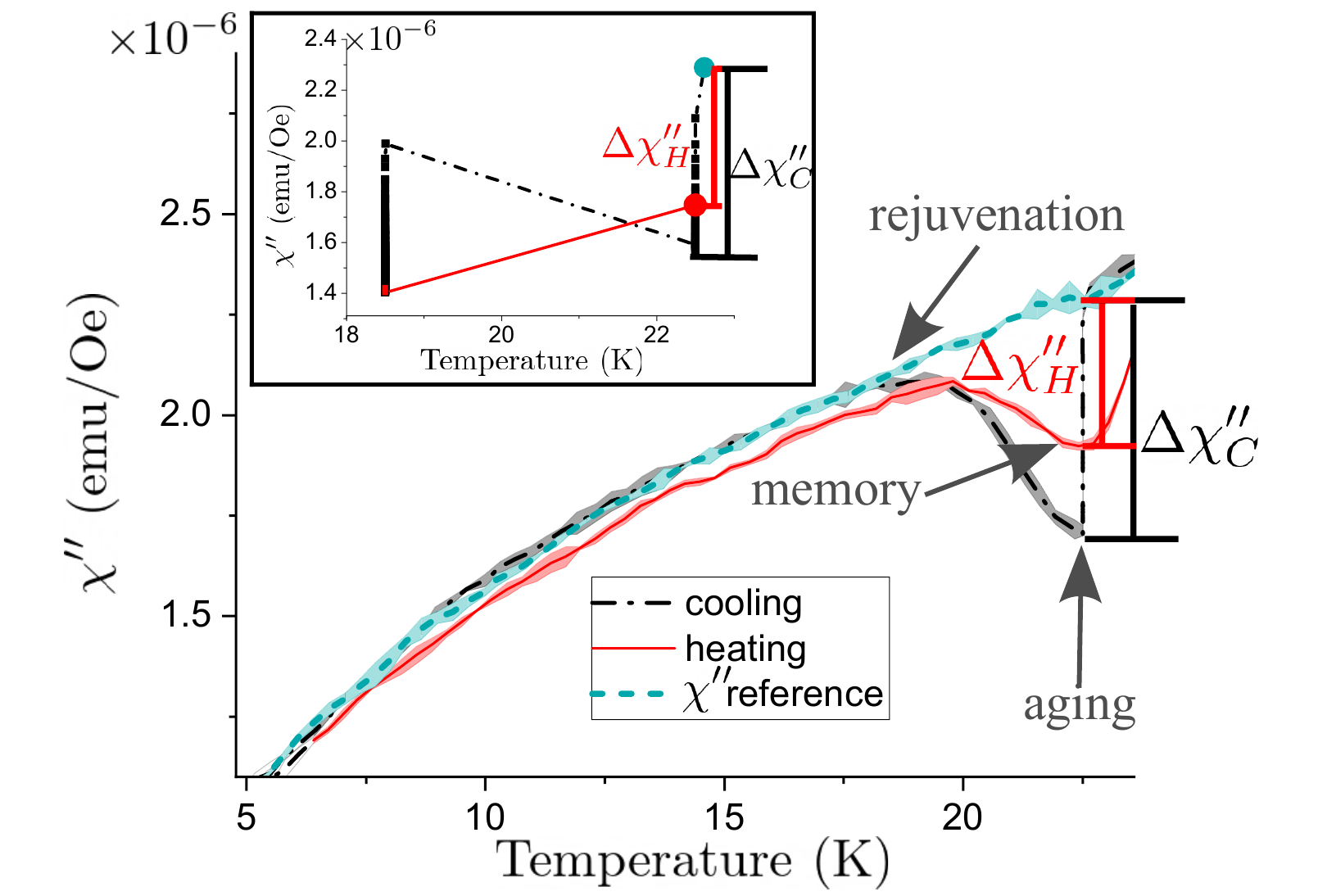}
\caption{Measurements of the ac susceptibility displaying memory as a function of temperature.  The imaginary part of the 1 Hz magnetic susceptibility, $\chi^{\prime\prime}(\omega)$.  The dashed data is the reference curve measured while continuously cooling the sample at 1 K/min.  The dot-dashed data includes waiting at $T_2=22.5$ K for a time $\twone= 1$ h and rejuvenation upon lowering the temperature.  The temperature-cooling rate is also 1 K/min.  The solid curve is the heating data taken at a rate of 1 K/min that exhibits the memory effect at $T_1$. The inset shows an example of a double-waiting-time experiment.  Reproduced from Fig. 1 of \citet{freedberg:24}.}  
    \label{fig:non-linear_chi_measure}
\end{figure}

The breakthrough in our understanding of the coexistence of rejuvenation and memory originated with the experiments of \citet{freedberg:24} and the simulations of the Janus Collaboration \cite{janus:23,paga:23b}.  The former quantified memory through
\begin{equation}
{\mathcal {M}}={\Delta \chi_H^{\prime\prime}}/{\Delta \chi_C^{\prime\prime}}.
\end{equation}
Using this definition for the amount of memory, see Fig.~\ref{fig:non-linear_chi_measure}, \citet{freedberg:24}
presented a quantitative analysis of the effects of two wait times $\twone$ and $\twtwo$ at various temperatures $T_1$ and lower temperatures $T_2$ based on a real-space model for the growth of the correlations at the two temperatures.

The outstanding element of this work was the development of a phenomenological model of memory with three adjustable parameters: a scaling constant, a length scale, and an exponent.  The scaling constant was of order unity and the exponent, though not fully understood, was reasonable.  After developing the physics of the model below, more will be said regarding the length scale.

\begin{figure}[t]
    \centering
    \includegraphics[width=8.5cm]{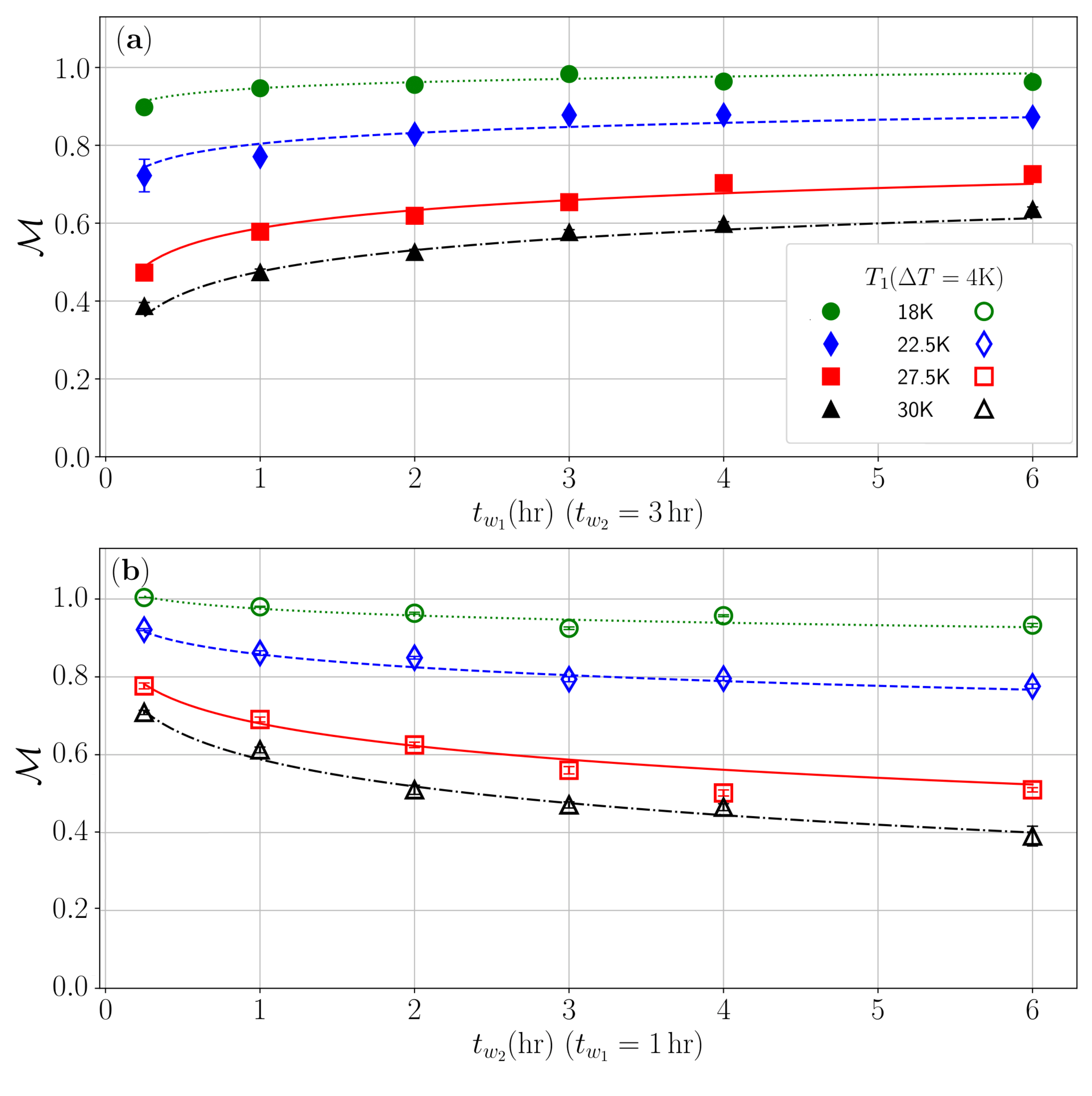}
\caption{${\mathcal {M}}$ vs waiting time with four different first waiting temperatures.  For both (a) and (b), the first and second waiting temperatures are the same: $T_1=18,~22.5,~27.5$, and 30 K and $T_1-T_2=$ 4 K.  In both cases either the first (a) or second (b) waiting times were varied from 0 - 6 h (the horizontal axis), while the other waiting time was fixed at 3 and 1 h, respectively.  Closed (open) markers indicate a variation of the first (second) waiting time.  The lines are fits to their phenomenological model with the same set of three adjustable parameters. 
 Reproduced from Fig. 2 of \citet{freedberg:24}.}  
    \label{fig:memory_chi2_vs_time}
\end{figure}
\begin{figure}[t]
    \centering
        \includegraphics[width=\linewidth]{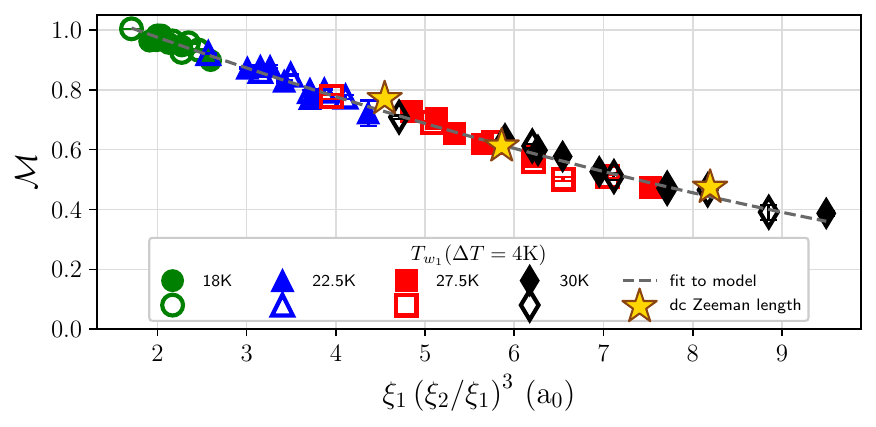}
    \caption{Memory vs waiting time data plotted against a model for memory loss predicted by \cite{freedberg:24}.  Additionally, the three starred data points use coherence lengths extracted from dc experiments with the ac values of memory.  The data collapse supports the model.  $T_1$ (18, 22.5, 27.5 and 30 K) and $T_1 - T_2 = 4$ K. In both cases either the first (closed markers) or second (open markers) waiting times are varied from 0 to 6 h, while the other waiting time is fixed at 3 and 1 h, respectively.  Statistical error bars are present on each data point, though they are typically smaller than the marker.  Reproduced from Fig. 3 of \citet{freedberg:24}.} 
\label{fig:memory_vs_tw}
\end{figure}

The model quantitatively describes how the correlations at the first waiting temperature grow, resulting in a decreasing susceptibility.  Quenching to the second wait temperature, those correlations remain but new correlations at the second temperature begin to grow and ``eat'' into the higher-temperature correlations.  Upon returning to the first waiting temperature, the original correlations are present but now reduced in magnitude.  The remarkable agreement of this model with the data is illustrated in Fig.~\ref{fig:memory_chi2_vs_time}, where  it is important to note that the same adjustable parameters are used for all the exhibited data.

As to the length scale, their data show there must be another length scale at play other than the coherence length that is on the order of 100 nm.  The origin of this length scale is not understood but may be related to the equilibrium correlation length.

All the data is considered at once in Fig.~\ref{fig:memory_vs_tw}, which shows how the memory scales with $\xi_1(\xi_2/\xi_1)^3$, where 1 and 2 refer to the higher and lower temperatures, respectively.  This remarkable data collapse shows the memory depends explicitly on the coherence length at both temperatures.  It also shows the importance of another length scale in the problem.  The data in this figure are from 52 double-waiting-time experiments. 

This work shows a physically reasonable model that explains memory and the interference between two different waiting-time temperatures. Complementary data were reported by \citet{paga:23b} and by \citet{he:24} ``through the `prism' of the coherence length.'' 

\begin{figure}[t]
    \centering
    \includegraphics[width=8.5cm]{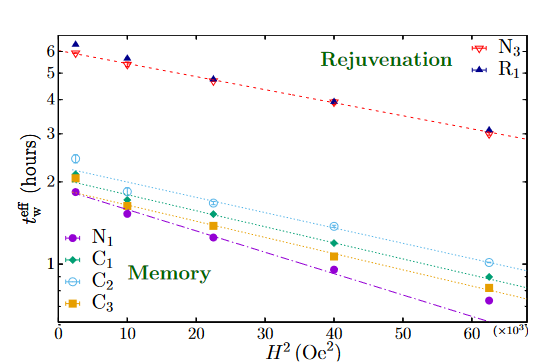}
\caption{The abbreviations N, R, and C stand for native, rejuvenation, and cycle, respectively.  In N$_3$, the temperature is lowered from above $\Tg=41.6$ K, for a CuMn single crystal 8 at.\%, to the lower temperature $T_2$ (here, $T_2=26$ K, $\twtwo=3$ h), and the effective waiting time, given by the peak in $S(t)$ defined by Eq. (5), is measured for different magnetic fields.  Next, the system is cooled from above $\Tg$ to $T_1=30$ K and aged for 1 h.  The temperature is then dropped to $T_2=26$ K, and aged for 3 h.  The points are labeled $R_1$.  As can be seen, the two procedures yield nearly exactly the same \teff for all values of $H^2$, {\it independent of the aging at $T_1$}.  This is a clear demonstration of rejuvenation.  Memory is measured through a full temperature cycle, from $T_1,\twone\rightarrow T_2,\twtwo\rightarrow T_1$ where \teff is measured.  The text discusses the physical meaning for the three protocols $C_1,~C_2$ and $C_3$. See Tab.~\ref{tab:exp_hue_memory} for  parameter details. Reproduced from Fig. 2 of \citet{paga:23b}.}  
    \label{fig:he_experiments_for_rejuvantion}
\end{figure}

\begin{table}[t]
\begin{ruledtabular}
\begin{tabular}{c c c c c c c}
Protocol&$T_1$\ (K)&$\twone$\ (h)&$T_2$\ (K)&$\twtwo$\ (h)&\Tmeas(K)&$\xi/a$\\\hline
N$_1$&30&1  &---&---&30&13.075\\
N$_2$&26&1/6&---&---&26&8.1011\\
N$_3$&26&3  &---&---&26&11.961\\
N$_4$&16&3  &---&---&16&6.3271\\
R$_1$&30&1  &26 &3  &26&11.787\\
  C$_1$&30&1  &26 &1/6&30 &12.621\\
C$_2$&30&1  &26 &3  &30 &12.235\\
C$_3$&30&1  &16 &3  &30 &12.846\\
\end{tabular}
\end{ruledtabular}
\caption{Parameters of the  experiments shown in Fig. \ref{fig:he_experiments_for_rejuvantion}. Table taken from \citet{paga:23b}.}\label{tab:exp_hue_memory}
\end{table}

Upon cooling the spin glass from above $\Tg$ to the first measuring temperature $T_1$, the system ages for a time $\twone$.  As a consequence, the spin-glass coherence length grows from nucleation to $\xi(\twone,T_1)$.  When the temperature is then lowered to $T_2$, the correlations created at $T_1$ are frozen.  This concept was first introduced by \citet{bouchaud:01}.  In accord with \citet{freedberg:24}, when the system is aged at $T_2$ for a time $\twtwo$, the system is now in a chaotic state that has nothing to do with the state created at $T_1,\twone$.  Its coherence length encompasses a coherent volume that interferes with the coherent volume created at $T_1,\twone$.  Hence, when heating back to $T_1$, the net correlated volume is the {\it difference} between the two correlated volumes.

This is exhibited in the lower part of Fig.~\ref{fig:he_experiments_for_rejuvantion}.  The $C_n$ are  separate temperature and waiting-time cycles, and illustrate unequivocally the relationship between memory and competing correlation volumes.  Each of the $C_n$ has three steps: 1) the system is ``prepared'' at $T_1=30$ K by waiting for the same time $\twone= 1$~h.  2) The temperature is dropped to $T_2$ and the system is aged for $\twtwo$. 3) The system is heated back to $T_1=30$ K and \teff measured.  $C_1$ sets $T_2=26$ K and $\twtwo=1/6$~h.  $C_2$ sets $T_2=26$ K and $\twtwo=3$~h.  $C_3$ sets $T_2=16$ K and $\twtwo=3$~h.

Memory is quantified by comparing the magnitude of the correlation volume measured at step 3) with that of the native protocol \citep{paga:23b}:
\begin{equation}\label{eq:C-Zeeman}
{\cal C}_{\text{Zeeman}}={\Ns^{\text{cycle}}}/{\Ns^{\text{native}}}\,,
\end{equation}
where $\Ns$ is the number of correlated spins, as in Eq.~\eqref{eq:Nc_correlated_spin}.
If the two volumes are the same, memory is perfect.  If, after step 3), the measured correlation volume is smaller than the initially prepared volume at step 1), memory is lessened, a direct result of the interference between the two states.  The slopes in Fig. \ref{fig:he_experiments_for_rejuvantion} are steeper the larger the coherent volume [Eqs. \eqref{eq:Zeeman_energy_def} and \eqref{eq:Nc_correlated_spin}].

Considering $C_1$, the system has transitioned from the prepared state at $T_1$ into a chaotic state at $T_2$, but only for a short time (1/6 h).  The coherent volume in the chaotic state has grown during this time, so that its interference with the initially prepared coherent volume is significant.  Hence, the memory is diminished, as exhibited by the shallower slope as compared to the native slope exhibited in Fig. \ref{fig:he_experiments_for_rejuvantion}.

Now, $C_2$ {\it increases} $\twtwo$ to 3~h, so that the coherent volume in the chaotic state can grow beyond its value in $C_1$.  This should lead to a larger interference, hence smaller memory, leading to a more shallow slope than found for $C_1$.  Its behavior is seen in Fig. \ref{fig:he_experiments_for_rejuvantion}.

Finally, the $C_3$ protocol has the same $\twtwo$ as $C_2$, but the temperature drop to $T_2$ is much greater.  At such a low temperature, the growth of the coherent volume of the chaotic state is very slow (almost none in this experiment), so there should be almost no interference with the coherent volume created at $T_1$.  This is exhibited in Fig.~\ref{fig:he_experiments_for_rejuvantion}, where the slope of $C_3$ is close to the native slope.

\begin{figure*}[t]
 \centering
 \includegraphics[width=0.86\textwidth]{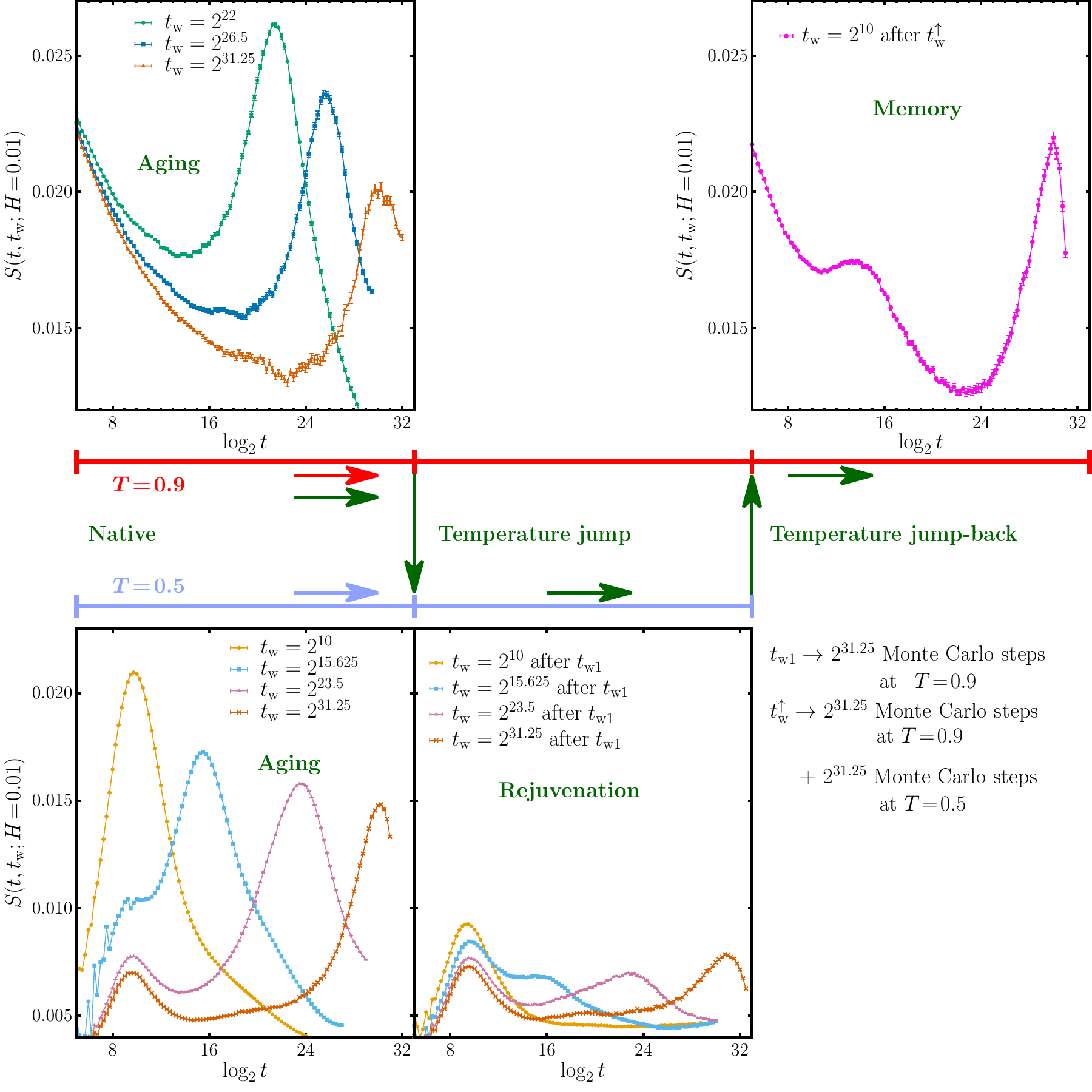}
   \caption{The ZFC numerical experiment
       measuring rejuvenation and memory.  At time $\tw$, a
     magnetic field $H=0.01$ is applied and the magnetic density,
     $M_\mathrm{ZFC}(t,\tw;H)$, is recorded.  Left panels show aging.
     The relaxation function $S_\mathrm{ZFC}(t,\tw;H)$ is displayed for the \emph{native} runs at the warmer, $T_1=0.9$, and colder, $T_2=0.5$, temperatures  . In the temperature-varying protocol (schematized by the green arrows),  after a waiting time \changes{$\twone= 2^{31.25}$ } at the hot temperature $T_1=0.9$, the system  is quenched to the cold one $T_2=0.5$ Then, the system relaxes at $T_2$   for an additional time, after which the magnetic field is switched   on and the function $S_\mathrm{ZFC}(t,\tw;H)$ is measured (bottom-center). Finally, after the  waiting time $\tw^\uparrow= \twone+\twtwo=2^{32.25}$
            ({\it i.e.}, the system has spent half its life at the
              initial temperature $T_1$ and half at the colder
              temperature $T_2$ without a field), the spin glass is
            suddenly heated back to $T_1$. 
            The system is allowed to relax for
            a short time, $\tw=2^{10}\ll\twone$, after which
              the magnetic field is switched on. The
            $S_\mathrm{ZFC}(t,\tw;H)$ measured after the \emph{jump
              back} (top-right) has a
            peak very similar to the one before the first jump (top-left), evincing the \emph{memory} of the
            aging at the initial temperature $T_1$, notwithstanding the
            rejuvenation observed when staying at the lower temperature
            $T_2$. Figure taken from \citet{janus:23}.}
  \label{fig:thermal_protocol_num}
\end{figure*}
\begin{figure*}[t]
\centering
   \includegraphics[width=0.86\textwidth]{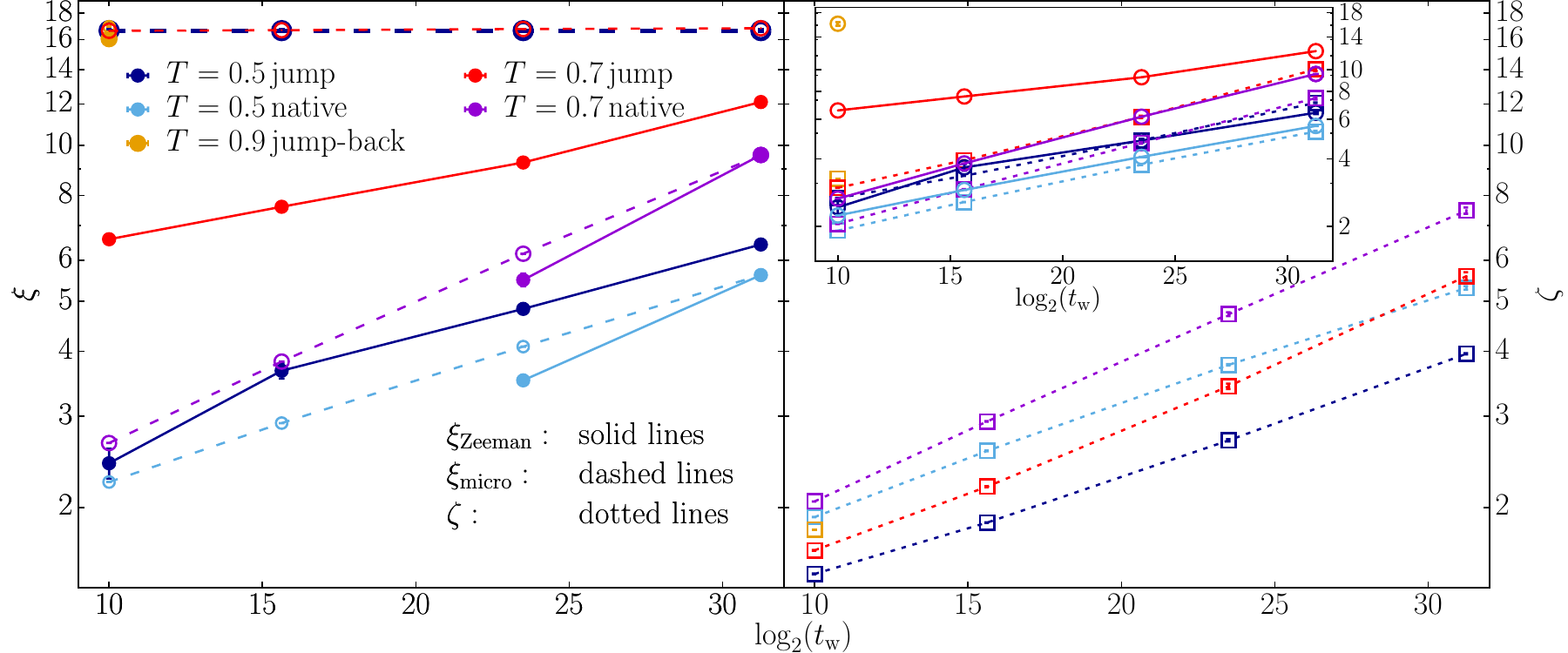}
  \caption{Aging dynamics is controlled by three length scales. The solid lines and   filled circles are for $\xi_{\mathrm{Zeeman}}(\tw,T)$, the dashed lines and empty circles are for $\xi_{\mathrm{micro}}(\tw,T)$, and the dotted lines and empty squares are for $\zeta(t_1,t_2)$. Left: for the \emph{native} protocols and \emph{jump} runs at $T_2=0.5$, $\xi_\mathrm{Zeeman}(\tw)$ follows quite closely the behavior of $\xi_\mathrm{micro}(\tw)$.  The jump to $T_2=0.5$ satisfies the chaos requirement in Eq.~\eqref{eq:chaos_requirement_exp_to_sim} and hence exhibits \emph{rejuvenation}.  
  When the system jumps back to $T_1=0.9$ (\textit{i.e.}, $T_1=0.9 \to T_2=0.5 \to T_1=0.9$), $\xi_\mathrm{Zeeman}^\mathrm{jump-back}(\tw)$ goes back to its original value $\xi_\mathrm{micro}^\mathrm{native}(\tw)$ after an extremely short time (\textit{memory}). Instead, $\xi_\mathrm{Zeeman}(\tw)$ never becomes small for the jump protocols with $T_2=0.7$. The $\xi_\mathrm{micro}(\tw)$ for the \emph{jump} runs are superimposed (the spins are frozen for this length scale). Right: the size of the regions undergoing coherent rearrangements when evolving from the initial to the final time, $\zeta(t_1,t_2)$, is much smaller than $\xi_\mathrm{micro}(\tw)$ for all jump protocols. The two times at which  the correlation length, $\zeta(t_1,t_2)$, is evaluated are $t_1=\tw$ and $t_2=2\tw$. Inset: comparison of $\zeta(t_1=\tw,t_2=2\tw)$ and $\xi_\mathrm{Zeeman}(\tw)$. For \emph{native} runs, the two lengths are approximately equal. Instead, for the jump protocols $1.8 \, \zeta(t_1=\tw,t_2=2\tw)$ is shown. Indeed, for $T_2=0.5$ only, and using an appropriate scaling factor of approximately 2, it is clear that $\zetaJ(t_1=\tw,t_2=2\tw)$ can be made to coincide with $\xiZ^\text{jump}$. Figure taken from \citet{janus:23}.} \label{fig:three_scales_memory}
\end{figure*}
These three temperature cycles are strong evidence for \changes{our} interpretation of memory.  This \changes{is more quantitative than \citet{bouchaud:01}, who arrive at the same conclusion but only state that ``memory is not perfect if \twtwo is large enough.''} Adjusting $\twtwo$, one can change the value of the coherent volume when returning to $T_1$ at will, from memory loss to complete loss of memory.  This behavior is born out in Fig.~\ref{fig:he_experiments_for_rejuvantion}, and also in the independent experiments and model of \citet{freedberg:24}.

\subsubsection{Rejuvenation and memory in computer experiments}
\label{subsubsec:memory_and_rejuvenation_num}
This section presents the full numerical evidence for the phenomena of rejuvenation and memory, which builds
on all the physical insight derived from the results in the previous sections of this review, most notably 
the scaling law in Sec.~\ref{subsect:off-equilibrium-in-field-fixed-T-5} and the formalization
of temperature chaos out of equilibrium (see Sec.~\ref{subsubsect:off-equilibrium-in-field-several-T-1-numerical}).

Inspired by experiments on single-crystal CuMn, the Janus Collaboration \cite{janus:23} conducted extensive numerical simulations to mimic ZFC protocols with temperature cycles, as in Fig.~\ref{fig:jonason_rejuvenation_memory}.
They implemented a fixed-temperature or \emph{native} protocol 
as well the \emph{jump} or \emph{jump-back} temperature-varying protocols. The native runs were already described in Sec.~\ref{sect:off-equilibrium-in-field-fixed-T}, the other two are as follows:
\begin{itemize} 
\item  \emph{Jump} run: The starting random spin configuration is instantaneously placed at the hot temperature $T_1<\Tg$ and allowed to relax for time $\twone$ without a field, achieving a large $\xim(\twone) \sim  17 \, \mathrm{a}_0$. Then, the temperature abruptly drops to the cold temperature $T_2$. After an additional time $\twtwo$, the magnetic field is switched on and the relaxation function $S_\mathrm{ZFC}$, Eq.~\eqref{eq:St_def}, is measured.
\item \emph{Jump-back} run: After the system relaxes at the cold temperature $T_2$, it returns to the hot temperature $T_1$ before turning on the magnetic field.
\end{itemize}
Recall that in \emph{native} runs the peak of $ S_\mathrm{ZFC}(t,\tw;H) $ defines $\teff \simeq \tw$ [compare Fig.~\ref{fig:PRL_scaling_tricks}
with Fig.~\ref{fig:thermal_protocol_num} (left)]. 
\changes{The numerical relaxation functions also exhibit a spurious peak at $t \sim 2^{10}$, which lacks any physical significance. We believe this peak arises from a spin-coupling effect that occurs as soon as the external magnetic field is applied. This hypothesis is supported by the observation that the short-time peak does not shift with $\tw$. Moreover, for small $\tw \sim 2^{10}$, this spurious peak merges with the physical one, making the extraction of $\xiZ$ even more challenging ---unless one employs the new strategy introduced in Sec.~\ref{subsect:off-equilibrium-in-field-fixed-T-5}}.

The experimental results of \citet{li:24} suggest that the temperature chaos in a CuMn sample is weak whenever $T_1-T_2$ is small. It follows that in a simulation, chaotic spheres will be too rare to affect the overall sample relaxation substantially unless
\begin{equation}
    \label{eq:chaos_requirement_exp_to_sim}
    \left. \frac{T_1}{T_2} \right|_\mathrm{sim} \approx \left. \frac{\Delta T_\mathrm{sim}}{\Tg} \right|_\mathrm{CuMn} \, \left[ \frac{\xi_\mathrm{CuMn}(\tw) }{\xim(\tw)} \right]^{1/(\Ds/2-y)}\, ,
\end{equation}
where the scaling law is based on Eq.~\eqref{eq:chaos_scaling}. Consider two cases with $T_1=0.9$ that differ in the cold temperature $T_2$: $T_2=0.5$ meets the chaos requirement of Eq.~\eqref{eq:chaos_requirement_exp_to_sim};  $T_2=0.7$ does not [recall that the glass temperature is $\Tg=1.102(3)$].

\begin{figure}[t]
\centering
  \includegraphics[width=8.5cm]{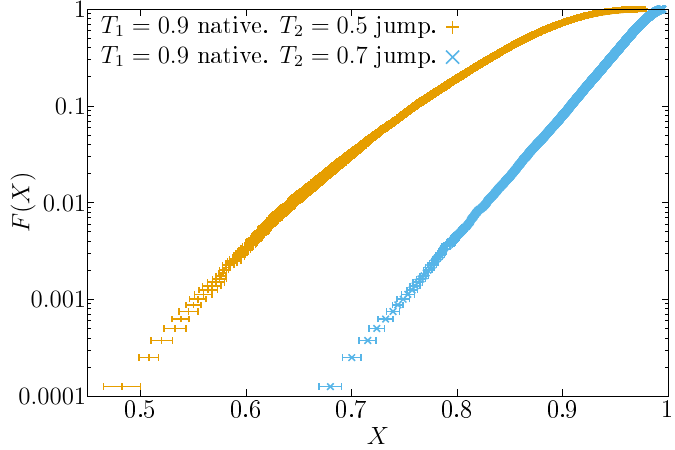}
  \caption{Comparison of probability distributions for the chaotic parameter defined in Eq.~\eqref{eq:prob_chaotic}, recall Sec.~\ref{subsubsect:off-equilibrium-in-field-several-T-1-numerical}, computed with the $T_1=0.9$ native protocol and the $T_2=0.5$ jump protocol (yellow curve) or the $T_2=0.7$ jump protocol (blue curve). Strong chaos correlates with the rejuvenation phenomenon. Figure taken from \citet{janus:23}.} 
  \label{fig:janus_2023_chaos}
\end{figure}

The main result of~\citet{janus:23} is reproduced in Fig.~\ref{fig:thermal_protocol_num}.
Rejuvenation is well visible in the central-bottom panel, as the peak of $S_\mathrm{ZFC}$  occurs at time $\teff \ll (\twone+\twtwo)$. This analysis is phenomenological and closely related to the experimental one, but the advantage of ``numerical experiments'' is that they give access to the full configurations and, hence, to microscopic correlation functions.
To explain this phenomenon, we need three length scales: $\xim$, $\zeta$ and $\xiZ$ (see Sec.~\ref{subsect:glossary}).
Let us focus on Fig.~\ref{fig:three_scales_memory}. In a fixed-temperature protocol (native runs) $\xiZ$   follows the behavior of $\xim$, recall Fig.~\ref{fig:ximicmac}. In contrast, in a jump run, the situation is more intricate. The size of the glassy domains, \emph{i.e.}, $\xim$, is frozen, but the system responds to the external field with a shorter $\xiZ$ (rejuvenation).
In particular, if the cold temperature $T_2$ satisfies Eq.~\eqref{eq:chaos_requirement_exp_to_sim}, we have full chaos and  $\xiZ^\mathrm{jump}(T_2,\twtwo) \sim \xiZ^{\mathrm{native}}(T_2,\twtwo-\twone)$; otherwise, we have an intermediate situation, as the $T_2=0.7$ case shows.
The last ingredient is that $\xiZ^\mathrm{jump}$ values can be matched by a simple factor to the values of $\zeta^\mathrm{jump}$. This implies that, in the presence of strong chaos, the \changes{rearrangements} are the ones responding to the magnetic field. This observation will be vital for the quantification of memory, as it allows us to perform numerical experiments without an external field and simply calculate $\zeta$ to quantify $\xiZ$ in temperature-varying protocols.

Physical intuition suggests that temperature chaos is key for understanding the origin of memory and rejuvenation. In Eq.~\eqref{eq:chaos_requirement_exp_to_sim}, \citet{janus:23} implicitly used the main results of \citet{zhai:22,janus:21} to translate what a strong chaotic regime is from experiment  to  simulation. $T_2=0.5$ meets the requirement for strong chaos, while $T_2=0.7$ does not. 

To quantitatively check this statement, the authors computed the chaotic parameter locally in a similar manner as explained in Sec.~\ref{subsubsect:off-equilibrium-in-field-several-T-1-numerical}. However, the authors did not impose the strong criterion of $\xi_{\text{micro},T_1}(\twone) = \xi_{\text{micro},T_2}(\twtwo)$, but just took the configurations with the largest coherence length $\xim$ for each temperature, resulting in the largest chaotic signal ever reported in a numerical simulation. The probability distributions of Eq.~\eqref{eq:prob_chaotic} were computed with native configurations at $T_1=0.9$ and jump configurations  at $T_2=0.5,0.7$. Fig.~\ref{fig:janus_2023_chaos} shows that the $T_2=0.5$ curve exhibits strong chaos while the $T_2=0.7$ curve does not.

These results point to temperature chaos as a necessary condition for rejuvenation.

\subsubsection{Simulation and experiment: the memory coefficients}
\label{subsec:memory_effects}

This section introduces a numerical observable for quantifying memory in temperature-varying protocols. 
It will be shown that, if the relevant length scales are known, a simple scaling law  can describe all the different memory coefficients (numerical and experimental).

\begin{figure}[t]
 \includegraphics[width=\columnwidth]{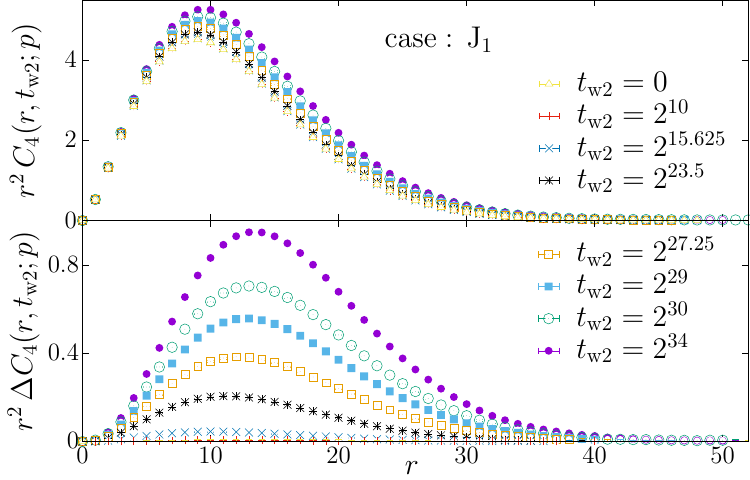}
  \caption{{Top}: the curves $r^2 C_4(r,\twtwo;p)$ vs $r$. {Bottom}: as in Top, but subtracting the native $C_4$ from the correlation function of the temperature-cycled system. As the time spent at the colder temperature $T_2$ increases, the system loses memory. The memory loss is evinced by the increasing signal in $r^2\Delta C_4$. Figure taken from \citet{paga:23b}.}
  \label{fig:memory_definition_num}
\end{figure}

The equilibrium nonlinear susceptibility is proportional to the integral of $r^2C_4$ for $0<r<\infty$~\cite{binder:86}. Thus, following \citet{janus:17,janus:17b,zhai-janus:20a,zhai-janus:21}, one can generalize this equilibrium relation by computing these integrals using the nonequilibrium correlation function  $C_4$ of Eq.~\eqref{eq:C4_def}.
As shown in Fig.~\ref{fig:memory_definition_num}, there is a small but detectable difference in the behavior of $r^2 C_4$ as $\twtwo$ varies in temperature cycles. \citet{paga:23b} considered the curve with $\twtwo=0$ as the reference curve and proposed as numerical memory coefficient
\begin{equation}
\label{eq:Memory_coefficient}
\mathcal{C}_\mathrm{num} = 1-  \frac{ \int_{0}^{\infty} \mathrm{d}r\, r^2 \,\Delta C_4(r,\twtwo;p)}{\int_{0}^{\infty} \mathrm{d}r\, r^2 \, C_4(r,0;p)}\,, 
\end{equation}
where $\Delta C_4(r, t_\mathrm{w2}; p) = C_4(r,t_\mathrm{w2};p) - C_4(r, 0; p)$ and $p$ denotes the different 
protocols.

Given the accuracy achieved for $\Delta C_4(r,\twtwo, p)$, $\mathcal{C}_\mathrm{num}$ is sensitive to even tiny differences in the state of the system just before and just after the temperature cycle.

The natural choices for the scaling variables are
\begin{equation}\label{eq:def-x-and-y}
x\!=\! \Big[\frac{\zeta(T_2,\twtwo)}{\ximN(T_1,\twone)}\Big]^{D-\theta/2}\!\!,\
y=\frac{T_1}{\Tg} \, \zeta(T_2,\twtwo)\,.
\end{equation}

\begin{figure}[t]
	\centering
	\includegraphics[width = .95\columnwidth]{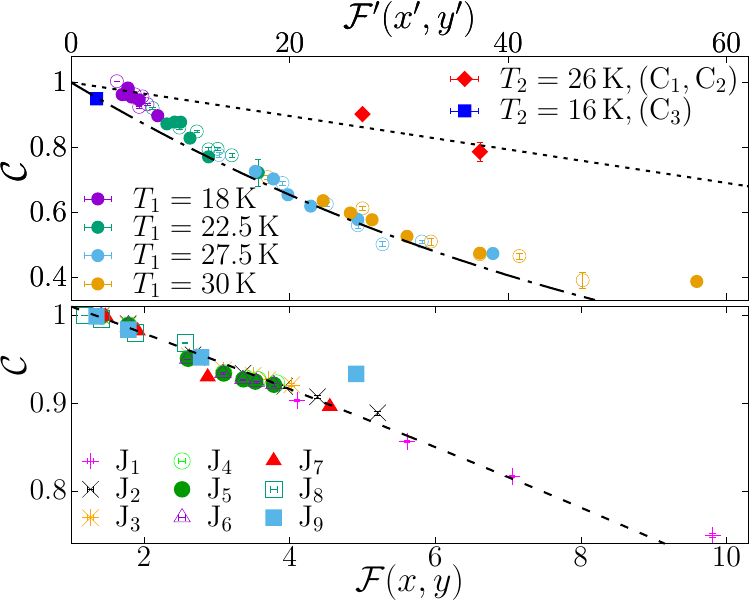}
	\caption{{Top:} Experimental memory coefficients $\mathcal{C}_{\chi''}$ (circles) and $\mathcal{C}_\mathrm{Zeeman}$ (rhombus and squares) vs the scaling function in Eq.~\eqref{eq:scaling_law_gen}. Here, $x'$ and $y'$ are experimental proxies 
    for the $x$ and $y$ scaling variables of Eq.~\eqref{eq:def-x-and-y}.   
Empty circles refer to $\twone=1 \, \mathrm{h}$ (full circles for
$\twtwo= 3 \, \mathrm{h})$. The dotted line is a fit to a straight
$\mathcal{C}_\mathrm{Zeeman}=1-\alpha \mathcal{F}'$.  The dashed line is $(1-\alpha \mathcal{F}')^{3.9}$ and represents the experimental findings for $\mathcal{C}_{\chi''}$ surprisingly well.  {Bottom:} As in Top, for coefficient $\mathcal{C}_\mathrm{num}$. See Tab. \ref{tab:memory_quantification_NUM} for parameter details. Figure taken from \citet{paga:23b}; data for the $\mathcal C_{\chi''}$ coefficient from \citet{freedberg:24}.}
	\label{fig:scaling_law_memory}
 \end{figure}

Both $x$ and $y$ are approximately accessible to experiment through $\Ns^{\text{jump}}$ and $\Ns^{\text{native}}$.
\citet{paga:23b} proposed a simple scaling law,
\begin{equation}\label{eq:scaling_law_gen}
    \mathcal{F}(x,y)= y \,  [1+a_1 x+a_2 x^2]\,,
\end{equation}
which, with fitted $a_1$ and $a_2$ coefficients,  describes the behavior of  both experimental and numerical memory coefficients.
The scaling representation, dashed line in Fig.~\ref{fig:scaling_law_memory} (top), shows that, away from  the $\mathcal{C}\approx 1$ region, the relation $\mathcal{C}_{\chi''}\approx [\mathcal{C}_{\text{Zeeman}}]^{K}$
holds with $K\approx 3.9$. This  shows that the
different memory coefficients carry the same physical information. Although the scaling in Eq.~\eqref{eq:scaling_law_gen} could be
feared to be accurate only near ${\cal C}\approx 1$, the ansatz turns out to cover all reachable values of ${\cal C}$.

In these sections, a quantitative formulation has been developed for memory in rejuvenated glasses utilizing measured and calculated lengths (both correlation lengths and coherence lengths) and confirming that rejuvenation and temperature chaos are strongly related.

\subsection{A different kind of memory: the Mpemba effect}
\label{subsec:mpemba}

\begin{table}[t] 
\begin{ruledtabular}
\begin{tabular}{c c c c c}
             $T$-{drop} & $T_1$ & $T_2$ & $t_\mathrm{w1}$ & $\xi_\mathrm{micro}(T_1,t_\mathrm{w1})$\\ [1pt]
                        \hline  \\ [-1.9ex]
                        J$_1$ & 1.0 & 0.625   & $2^{19.5}$  & 8.038(1)\\
                        J$_2$ & 1.0 & 0.625   & $2^{22}$  & 10.085(15)\\
                        J$_3$ & 1.0 & 0.625   & $2^{24.625}$  & 12.75(3)\\
                        J$_4$ & 1.0 & 0.625   & $2^{27.25}$  & 16.04(3)\\
                        J$_5$ & 1.0 & 0.625   & $2^{28.625}$  & 18.08(5)\\
                        J$_6$ & 1.0 & 0.625   & $2^{29.875}$  & 20.20(8)\\
                        J$_7$ & 1.0 & 0.700   & $2^{29.875}$  & 20.20(8)\\
                        J$_8$ & 0.9 & 0.500   & $2^{31.25}$  & 16.63(5)\\ 
                        J$_9$ & 0.9 & 0.700   & $2^{31.25}$  & 16.63(5)\\ [0pt] 
\end{tabular}
\end{ruledtabular}
\caption{Parameters of the numerical simulations shown in Fig. \ref{fig:scaling_law_memory}. Table taken from \citet{paga:23b}.}
   \label{tab:memory_quantification_NUM}
\end{table}

Experimental and numerical spin glasses exhibit strong memory effects that are deeply connected with the aging state of the system (or equivalently, the characteristic length scales of the system) as has been explained above. This section follows \citet{janus:19} to present a different kind of memory: the Mpemba effect.

Consider two beakers of water that are almost identical except for the fact that one is hotter than the other. If we use a thermal reservoir (for example, a freezer) at a temperature below freezing point, it can be observed under some circumstances that the initially hotter beaker freezes faster than the colder one. This is known as the Mpemba effect~\cite{mpemba:69}.

The interesting story of the discovery of this effect by Mpemba can be found in \citet{mpemba:69}, although versions of this phenomenon have been reported since ancient history~\cite{aristotle-ross:81} and have been widely studied in systems like nanotube resonators~\cite{greaney:11}, clathrate hydrates~\cite{ahn:16}, granular fluids~\cite{lasanta:17}, and colloidal systems~\cite{kumar:20}. 

\begin{figure}[t]
    \centering
    \includegraphics[width=8.5cm]{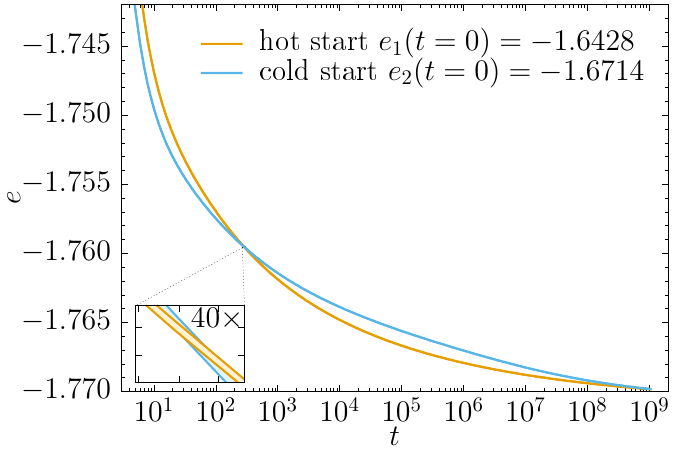}
\caption{Time evolution of the energy of two spin-glass systems, one initially at a higher temperature $T=1.3$ and the other at a lower temperature $T=1.2$, both quenched to $T=0.7 \approx 0.64 \Tc$. The initially hotter system (\emph{i.e.}, having a higher energy $e$) colds faster than the initially colder (lower-energy) one in agreement with the original Mpemba experiment. Inset: Closeup of the crossing between energy-time curves. Error bars are equal to the thickness of the lines. Figure from~\citet{janus:19}.} 
\label{fig:janus_2019_mpemba}
\end{figure}

\citet{janus:19} observed and explained this effect in a well-controlled system by taking advantage of numerical simulations in spin glasses.  The authors simulated a 3D EA model in a lattice of system size $L=160$. 
As long as simulations are performed out of equilibrium, it is necessary to identify what faster cooling means, since temperature is only well defined in equilibrium. The natural candidate to take the place of the temperature seems to be the energy density $e(t)$ because it is the observable conjugate with (the inverse of the) temperature. 

Two different off-equilibrium systems are simulated. In the first one, the temperature of the thermal reservoir is quenched from infinite temperature to $T_1=1.3$, which is above the critical temperature [$\Tc=1.102(3)$, see \citet{janus:13}]. The system is left to evolve until it reaches the energy $e_1 \approx -1.6428$ and this is the starting point of the first system in the numerical experiment. The second simulated system is prepared in a very similar way but the temperature of the thermal reservoir is now $T_2=1.2$ and the initial energy is lower, $e_2=-1.67174$.

At this point, the temperature of the thermal reservoir of both systems is quenched to $T=0.7 \approx 0.64 \Tc$, and the energies are recorded for several times. The results of this numerical experiment are shown in Fig.~\ref{fig:janus_2019_mpemba}.

The key to understanding what is happening is the control parameter of the phenomenon: the coherence length. Although the \textit{hot-start} system has a higher energy density in the initial setup, it is also \textit{older} in the aging sense of the coherence length $\xi_{\text{micro},1}(\tw=0)=12$. The \textit{cold-start} system has a lower initial energy density but it is \textit{younger}, with a coherence length $\xi_{\text{micro},2}(\tw=0)=5$.

Two ingredients are responsible for the Mpemba effect in spin glasses. The first is the temperature dependence of the growth of the coherence length; for the system initially at $T=1.3$, $\xim$ grows faster than in the system at initial temperature $T=1.2$. The second one is that, while the coherence length $\xim$ is a slow variable that remains essentially constant when changing the temperature, the energy density is a fast variable and, as a first approximation, when a quick temperature change takes place, it changes instantaneously to the value of the energy density corresponding to its new thermal reservoir.

\section{Issues for future investigations}\label{sect:conclussions}

We collect here a few unanswered questions whose answers, in our opinion, could lead to significant  progress. We shall distinguish the more theoretical aspects (\ref{subsect:conclussions-1}) from questions  of an experimental nature (\ref{subsect:conclussions-2}).

\subsection{What is yet to be nailed down theoretically}\label{subsect:conclussions-1}

\changes{
\subsubsection{On the low-temperature phase.}}


The renormalization group (RG), which has served us so well to clarify the nature of phases and phase transitions in clean systems \cite{wilson:75,cardy:96,amit:05}, needs to be extended to spin glasses. This will not be easy. Numerical simulations clearly hint that universality ---perhaps the most important consequence of the RG framework--- extends to disordered systems \cite{ballesteros:98b,fytas:13,fytas:16,fytas:19}, including classical  spin glasses \cite{hasenbusch:08b,fernandez:16b}. For quantum spin glasses we even have direct experimental confirmation \cite{king:24} of critical exponents obtained with numerical simulations \cite{bernaschi:24b}.

On the other hand, our favorite tool, the perturbative RG, does not lead to unambiguous predictions. Even seemingly simple questions still lack an answer. Think, for instance, of the stability of the spin-glass phase under an externally applied magnetic field (see Sec.~\ref{subsect:Equilibrium-2}).

Some new ideas to carry out nonperturbative computations would clearly be welcome. These might include further development of real-space tools such as the Monte Carlo RG~\cite{parisi:01b} or the $M$-expansion around the Bethe lattice \cite{altieri:17,angelini:22}. More speculatively, we could think of the metastate ---the mathematically rigorous way of building the low-temperature phase \cite{aizenman:90,newman:92,newman:96c,newman:03,read:14}, see \citet{newman:22} for a recent review. It is conceivable that the De Dominicis RG program at zero field~ \cite{dedominicis:98,dedominicis:06} could  blend with the metastate approach to build a rigorous theory of the spin-glass phase~\cite{read:14}.

\changes{\subsubsection{On numerical simulations}}
As discussed in Sec.~\ref{subsect:intro-dedicated_computers}, NP-completeness is making spin glasses increasingly popular in an engineering context.  This has driven the development of  specific hardware to solve spin-glass problems. One may easily speculate that the main difficulties hampering spin-glass numerical simulations are common to both the physics and the engineering contexts. It follows that eventual advances in simulation algorithms (or hardware) in physics may have far-reaching consequences.

Progress in this problem has been absent for the last two decades. Our best algorithm for sampling equilibrium configurations, parallel tempering, was introduced in the 1990s~\cite{geyer:95,hukushima:96,tesi:96}. The largest system equilibrated deep in the spin-glass phase in $D=3$ has been an $L=32$ cubic lattice for 15 years~\cite{janus:10}. We have some evidence hinting to a physical origin for these difficulties, namely temperature chaos~\cite{fernandez:13,billoire:18}. 

Besides new hardware developments (see Sec.~\ref{subsect:intro-dedicated_computers}), new simulation algorithms are needed. Temperature chaos suggests that these should work at a \emph{single} temperature. Alternatively, ingenious ways of circumventing chaotic  phenomena should be devised. Quantum annealing could be such an approach, but it may be hampered by chaotic phenomena related to the transverse field~\cite{knysh:16}.\\


\changes{\subsubsection{The chiral-glass scenario and the problem of the critical exponents}\label{sec:chiral}

The alert reader will have noticed an important paradox. On the one hand, as discussed in detail in this review, the Ising Edwards-Anderson model beautifully describes the nonequilibrium dynamics of CuMn samples below $\Tg$, even if CuMn is a Heisenberg material (more on this below). On the other hand, the critical exponents of the three-dimensional  Ising Edwards-Anderson model \cite[which are quite well known by now, see][]{janus:13}, do not agree with their experimental counterpart. Take, for instance, the nonlinear magnetization exponent $\gamma$. Its value for the Ising Edwards-Anderson model is $\gamma=6.33\pm 0.11$.  Probably the most accurate experimental determination of the same critical exponent was obtained from zero-field measurements on the Heisenberg spin glass 0.5 at.\% AgMn~\cite{levy:88} and yielded a very different $\gamma = 2.3\pm 0.2$. The authors are unaware of other experimental reports for values of $\gamma$ in metallic spin glasses that come close to the value obtained from simulations.  Similar differences between simulation and experiment are present for other critical exponents (\emph{e.g.}, $\beta,~\delta,~{\text {and}}~\nu$).

The failure in reproducing the critical exponents to describe CuMn or AgMn  at temperatures near to, but above, $\Tg$ forces one to consider what could be the simplest, yet adequate, model to describe this temperature region. In fact, microscopic models of spin glasses include several interactions,
such as the RKKY interaction, which is invariant over rotations
\cite{ruderman:54,kasuya:56,yosida:57} and the Dzyaloshinsky-Moriya (DM)
interaction that breaks the rotational symmetry
\cite{dzyaloshinsky:58,moriya:60}. The DM interaction introduces a certain
degree of random anisotropy, which results in real spin glasses that are never fully
isotropic (this theoretical limit is named the Heisenberg spin glass).
It is possible to
classify materials according to the degree of anisotropy of their interactions
\cite{petit:02}. Perhaps the best example of an Ising spin glass is the
extremely anisotropic Fe$_{0.5}$Mn$_{0.5}$TiO$_3$. At the other end of the
spectrum, we have very isotropic alloys such as AgMn or CuMn.\footnote{It has been emphasized that
modeling AgMn or CuMn is  notoriously difficult \cite{peil:09}, because of the presence of
short-range spin-density wave ordering \cite{cable:82,cable:84,lamelas:95}.} The experimental finding $\gamma=4.0$ in the strongly uniaxial Fe$_{0.5}$Mn$_{0.5}$TiO$_3$~\cite{gunnarsson:91}, midway between  the Ising Edwards-Anderson model and the
AgMn values, namely  $\gamma_\text{IEA}=6.33\pm 0.11$ and $\gamma_\text{AgMn}=2.3\pm 0.2$, gives weight to the idea that symmetries may play a fundamental  role in clarifying this problem.

Indeed, in the context of universality and the renormalization group, simplified symmetry-motivated Hamiltonians are crucial to compute critical exponents. The numerical simulation of microscopically motivated interactions such as RKKY is simply too costly. Hence, the Edwards-Anderson model, recall Eq.~\eqref{eq:1C:EAHam}, was introduced with $n$-component spins $\vec\sigma_i=(S_{1,i},\ldots,S_{n,i})$ subjected to the constraint $\vec\sigma_i\cdot\vec\sigma_i=1$
(with $n=3$ we have the Heisenberg case, $n=2$ is known as XY, while with $n=1$ one recovers the Ising spins).
In this way, the distinctive features of the RKKY interaction (quenched disorder, frustration, a three-dimensional lattice structure and $n$-dimensional rotational and reflection symmetries) are preserved.
This Heisenberg Edwards-Anderson (HEA) model was hoped to be a streamlined representation  for highly isotropic materials, such as CuMn  or AgMn, yet sufficient for computing critical exponents. Unfortunately, the story turns out not to be that simple.

All attempts to identify a phase transition at a finite 
temperature in the HEA model carried out during the
1980s and 1990s failed 
\cite{mcmillan:85,olive:86,morris:86,matsubara:91}. 
However, \citet{matsubara:91} found a phase transition, provided that a small, random and anisotropic 
term is  added to the HEA Hamiltonian.
The crucial importance of the anisotropic interactions 
was also
emphasized by~\citet{gingras:93}. The accepted picture 
at the time was that the lower critical dimension (\emph{i.e.}, 
the spatial dimension below which there is no phase 
transition) for the HEA model lay
somewhere between 3D and 4D \cite{coluzzi:95}. This 
theoretical picture, when confronted with the experimental 
finding of a finite-temperature transition in materials 
with only small anisotropies~\cite{levy:88},
suggests that anisotropic interactions (no matter how 
small) play a fundamental role. 

A different, and provocative viewpoint was suggested 
by \citet{mauger:90}: although there was no spin-glass transition in the fully isotropic HEA model, a different order parameter called 
chirality might be present.  Chirality 
is a scalar observable that describes vorticity and 
alignment between neighboring spins. This idea was 
elaborated upon by \citet{kawamura:92} in his 
\emph{spin-chirality decoupling} scenario. In the ideal case of 
a purely isotropic system, only chiralities would order at a nonvanishing temperature $T_\text{CG}>0$,
but any small anisotropy would couple spins and chiralities (resulting in spins
ordering at $T_\text{SG}=T_\text{CG}$). Yet, at the turn of the century, the scenario changed quite abruptly. \citet{lee:03} 
employed more efficient simulation algorithms to equilibrate larger systems at lower temperatures than was 
possible in the 1980s. By combining these new results with modern finite-size
scaling techniques, they showed that, contrary to all previous expectations,  
$T_\text{SG}>0$ for the purely isotropic HEA model. The main open question today 
regards a comparison of $T_\text{SG}$ and $T_\text{CG}$ for the isotropic model. While some groups find that the 
most economic interpretation of their simulations for the HEA model is the \emph{absence}
of spin-chirality decoupling, \emph{i.e.}, $T_\text{CG}=T_\text{SG}$
\cite{campos:06,fernandez:09b,nakamura:19}, Kawamura and coworkers found that
$T_\text{CG}$ is larger than ---yet very close to---  $T_\text{SG}$
\cite{viet:09,ogawa:20}.

Real materials, however, always have some degree of anisotropy, 
hence one needs to 
understand what happens when some random (albeit small) anisotropic interactions are added  
to the HEA model. An analogy with ferromagnetic systems suggests that the anisotropic term will 
be a relevant perturbation in the renormalization group sense. This scenario has some consequences. One starts by modeling a real material through an HEA model, plus random anisotropies. We quantify these anisotropies through a coupling constant $g_\text{micro}$ when the correlation length is $\xi\sim 1$ lattice spacing ($g_\text{micro}$ will admittedly be very small for CuMn or AgMn). Now, even a 
tiny $g_\text{micro}$ gets amplified to $g_\text{effective}
=g_\text{micro}\xi^\phi$ in an effective Hamiltonian for a correlation length $\xi$ \cite[$
\phi$ is a positive crossover exponent, see, \emph{e.g.}, ][]{amit:05}. Hence, a new renormalization group fixed-point takes over and dictates the asymptotic critical exponents as $\xi\to\infty$. Presumably, the relevant fixed point at large-enough $\xi$ will be the one of the Ising Edwards-Anderson model. Two known facts strongly hint at the validity of this scenario: (i) $T_\text{SG}$ grows inordinately for even 
a tiny $g_\text{micro}$~(\citealp{martin-mayor:11b}; this could be guessed from \citealp{matsubara:91}), and (ii) for large-enough $g_\text{micro}$ the critical exponents turn out to be those 
of the Ising Edwards-Anderson model~\cite{parisen:06,liers:07,baityjesi:14}. It is an unfortunate fact, however, that we know essentially nothing 
about the details of this crossover from Heisenberg to Ising behavior. Investigating the crossover experimentally will also 
be a serious challenge, because the dynamic critical exponent appropriate for $T>\Tg$ 
is large ($z\approx  6$, recall Fig.~\ref{fig:xi_many_T}), which means that doubling $\xi$ ---by 
lowering  $T>\Tg$--- while keeping the sample in thermal equilibrium requires extending 
the time duration of experiments by a factor of 64.

Kawamura and his collaborators have advanced a phenomenological 
interpretation of the existing experiments that somehow circumvents this frustrating 
lack  of knowledge  about the crossover that arises when $g_\text{micro}>0$. They start with simulations of the HEA model with 
$g_\text{micro}=0$ \cite{kawamura:15,ogawa:20}. According to their view, 
$T_\text{CG}>T_\text{SG}$ for the purely isotropic HEA model but they conjecture that,
as soon as
$g_\text{micro}>0$, there is a single phase transition (\emph{i.e.}, $T_\text{CG}=T_\text{SG})$.
Although in typical experiments the magnetization (a spin-related quantity) is measured, the relevant critical exponents would be those of the chiralities (as computed for  $g_\text{micro}=0$). In addition, \changesbis{\citet{kawamura:03} proposed ways to access the chiralities that have been demonstrated experimentally~\cite{taniguchi:04,pureur:04}}. This  approach was followed in \citet{taniguchi:07} by  measuring the exponent $\delta$ from the chiralities, finding results in agreement with determinations of $\delta$ from the magnetization. As for exponent $\gamma$, one has $\gamma_{SG}^{g_\text{micro}=0}\approx 3$ and
$\gamma_{CG}^{g_\text{micro}=0}\approx 2$~\cite{fernandez:09b,ogawa:20}. Of the two $\gamma$ exponents, both of them computed for the purely isotropic HEA model, the 
chiral one  lies closer to the experimental determination for 0.5 at.\% AgMn,
$\gamma = 2.3\pm 0.2$~\cite{levy:88}. More comparisons of critical exponents can be found  in \citet{kawamura:10,kawamura:15,ogawa:20}.

\changesbis{
There have been as well claims of different nonequilibrium behavior for Ising and Heisenberg materials in (at least) two different aspects:
\begin{enumerate}
    \item \citet{dupuis:01} and \citet{bert:04} described the aging dynamics as an activated process with free-energy barriers $\sim \xiZ^\varPsi$, with an exponent $\varPsi$ that would be different for Ising and Heisenberg materials. However, \citet{bert:04} extracted $\xiZ$ in different ways for Heisenberg and Ising materials, which has been recently shown to be unnecessary~\cite{zhai-janus:21}. Whether or not the overall dynamical differences in Heisenberg or Ising materials were simply caused by the different ways of extracting $\xiZ$ is an open question. 
    \item \citet{herisson:04} noted in their extended account of  their experiment \cite{herisson:02} that the  measured fluctuation-dissipation ratio (FDR)  ---recall Sec.~\ref{subsect:off-equilibrium-no-field-2}--- was surprisingly curvature-less. This feature is reminiscent of the FDR predicted for systems with one-step RSB \cite[the lack of curvature had been independently related to chiralities by][]{kawamura:03}. We should note, however, that the  Ising Edwards-Anderson model displays changes even in the \emph{sign} of the curvature for its FDR obtained at finite times ---see Fig.~\ref{fig:3D:ko} and Sec.~\ref{subsect:off-equilibrium-no-field-4}. It is, hence, clear that one can find measuring-time combinations in which the FDR is also curvature-less for an Ising system, a fact that calls for caution. 
\end{enumerate}
}

In our opinion, the important theoretical challenge for the future will be developing an 
appropriate field-theoretical framework that can explain  
the crossover when the random-anisotropy $g_\text{micro}$ becomes different from zero. One 
would like to understand as well how the chiralities mix with the spin operators. 
Another important open problem would be to understand why  the Ising Edwards-Anderson model reproduces quantitatively the 
experimental aging dynamics at $T<\Tg$, while the same model does not make a good job 
of explaining the critical exponents. Perhaps the fact that critical exponents are equilibrium quantities (which enforces carrying out experiments at $T>\Tg$) will play a role in solving this conundrum. Clearly enough, further numerical investigations of the HEA model with random anisotropies  can be instrumental in understanding the crossover.  From the experimental side, new 
determination of the critical exponents on single crystals 
that allow reaching large values of $\xi$  could shed light on the crossover, 
particularly by focusing on changes of the effective values for the critical exponents 
as $\xi$ grows.
}
\changes{\subsubsection{Experiments with potential for strong impact on theory}}

As explained in Sec.~\ref{subsect:off-equilibrium-no-field-2}, important experiments in the 1990s and the 2000s convinced the community about the far-reaching nature of the generalized fluctuation-dissipation relations. In the last ten years, single-crystal samples supporting  large spin-glass coherence lengths $\xim$ have been obtained. A more detailed experimental investigation of the fluctuation-dissipation relations at larger $\xim$ would be clearly welcome.

Experiments on an ensemble of small spin-glass grains, where $\xim$ would be of the order of the grains' size, open up another avenue. Under these conditions, large sample-to-sample fluctuations may unveil exotic phenomena such as multifractality \cite{janus:24}. Reliably measuring single-grain magnetizations would be a (not minor) technical challenge but even the collective magnetization of a (small) set of grains may be useful to investigate finite-size effects experimentally. This will bring us closer to the broader field of structural glasses, where nonlinear responses unveil length scales akin to $\xim$ \cite[\changes{see also Sec.~\ref{sec:non-linear-susc}}]{adam:65,albert:16}.

\subsection{Experimental opportunities and challenges}
\label{subsect:conclussions-2}
 The remaining issues that experiments can address arise from our current understanding of spin-glass dynamics.  As is so often the case with spin glasses, there will be surprises that will dictate other opportunities for investigation.  And, of course, the interaction with simulations and theoretical advances will suggest even other lines of inquiry.  For now, we briefly describe a few of the more obvious directions for experimental research.

\paragraph*{The onset of temperature chaos.} 
Temperature chaos is thought to be the origin of rejuvenation (see Sec.~\ref{subsec:memory_and_rejuvenation}), but  the question of its onset remains.  Is it a sharp or broad transition as a function of temperature change? Is there a mixture of reversible and chaotic states in the transition region? The jury is still out on these fundamental issues.  The experiments of \citet{guchhait:15b} exhibit a finite temperature window of about 100 mK between complete reversibility and complete chaos, while those of \citet{zhai:22} find a very narrow temperature range of 10 mK for the onset of temperature chaos.  Further, the latter find evidence for enhanced chaos as the temperature change increases.  The two experiments are on different systems, but the question remains: what happens in the transition region?

\paragraph*{Probing thin-film dynamics for times greater than $t_{\text {co}}$.}  As pointed out in Sec.~\ref{subsect:off-equilibrium-in-field-fixed-T-3}, experiments on thin films were interpreted in terms of two time regimes.  The first was the growth of the correlation length from nucleation to the film thickness $\mathcal {L}$ at a time defined as $t_{\text {co}}$, with a rate of growth appropriate to 3D.  The second was an apparent instantaneous value reached in the parallel direction, appropriate to 2D. The simulations of Sec.~\ref{subsect:off-equilibrium-in-field-fixed-T-3} however, found four time regimes for the growth of the correlation length.  In particular, after the correlation length reached the vicinity of $\mathcal {L}$, the correlation length {\it continued to grow} more rapidly than in 3D, finally reaching an equilibrium value appropriate to 2D. It would be of interest to measure the magnitude of the correlation length experimentally well after $t_{\text {co}}$.  This will require much longer time measurements than have been taken heretofore.

\paragraph*{The nature of rejuvenation.} Rejuvenation is nearly always attributed to temperature chaos.  Is this true?  An alternate view has been proposed by \citet{berthier:02,berthier:03}, who ``demonstrated in a numerical simulation of the 4D Ising spin glass that a temperature change $\Delta T/\Tg$ induces {\it strong} rejuvenation effects, whereas the direct observation of the overlap between configurations reveals {\it no sign of chaos} on the dynamically relevant length scales.'' Making reference to the claim of \citet{jonsson:02} that temperature chaos is the origin of rejuvenation, they state that this analysis ``cannot be viewed as definitive evidence for temperature chaos.'' Note that in Sec.~\ref{subsect:off-equilibrium-in-field-several-T-1} we point out that \citet{jonsson:02} obtained a value $1/(\Ds/2-y)\approx 2.6$ whereas \citet{berthier:02} use $\/(\Ds/2-y) \approx 1$, the value found in most simulations and by \citet{zhai:22}. This issue may be illuminated by the experiments suggested above.

\paragraph*{Spin glasses in a field: crossover or condensation?}
The vexing question of whether there is a spin-glass transition in the presence of a magnetic field, and the allied question of whether the de Almeida-Thouless line \cite{dealmeida:78}, predicting a magnetic-field-dependent transition temperature $\Tg(H)$, exists has  bedeviled theorists and experimenters alike. 
This issue was discussed in Sec.~\ref{subsect:Equilibrium-2}, but we briefly restate it.

The mean-field theory predicts a phase transition to an ordered spin-glass phase for $T < \Tg(H)$.
The droplet picture, however, finds that ``a magnetic field destroys the spin-glass phase so that there is no de Almeida-Thouless transition'' \cite{fisher:86b}.  The long-distance dependence of the correlation function between spins located at position $i$ and $j$ is $C_{ij}\sim q_{\text {EA}}^2T/(Y|i-j|^{\theta})$, with $\theta \simeq 0.2$ for $D = 3$, and $Y$ a generalization of the twist modulus or inter-facial tension.  This very slow fall off may explain why the spin-glass system appears critical for $T<\Tg$.

Thus, there are two very different pictures for the behavior of a spin glass in the presence of a magnetic field. The numerical picture for $D=3$ 
is muddled (Sec.~\ref{subsec:dynamics_in_a_field}), while a transition has been found in higher dimensions (Sec.~\ref{subsect:Equilibrium-2}). Experimentally, there is evidence for a de Almeida Thouless line \cite{kenning:91} consistent with the mean-field picture.  The fundamental question is whether this is indicative of a phase transition to a spin-glass state in the presence of a magnetic field, or simply a crossover to the droplet-style phase.  A future experimental investigation of ac susceptibility  compared to dc magnetization measurements may shed light on this fundamental issue.

\paragraph*{Spin-glass magnetization in high magnetic fields.}
\changes{\begin{figure}[tb]
    \centering
    \includegraphics[height=0.68\linewidth, angle=-0]{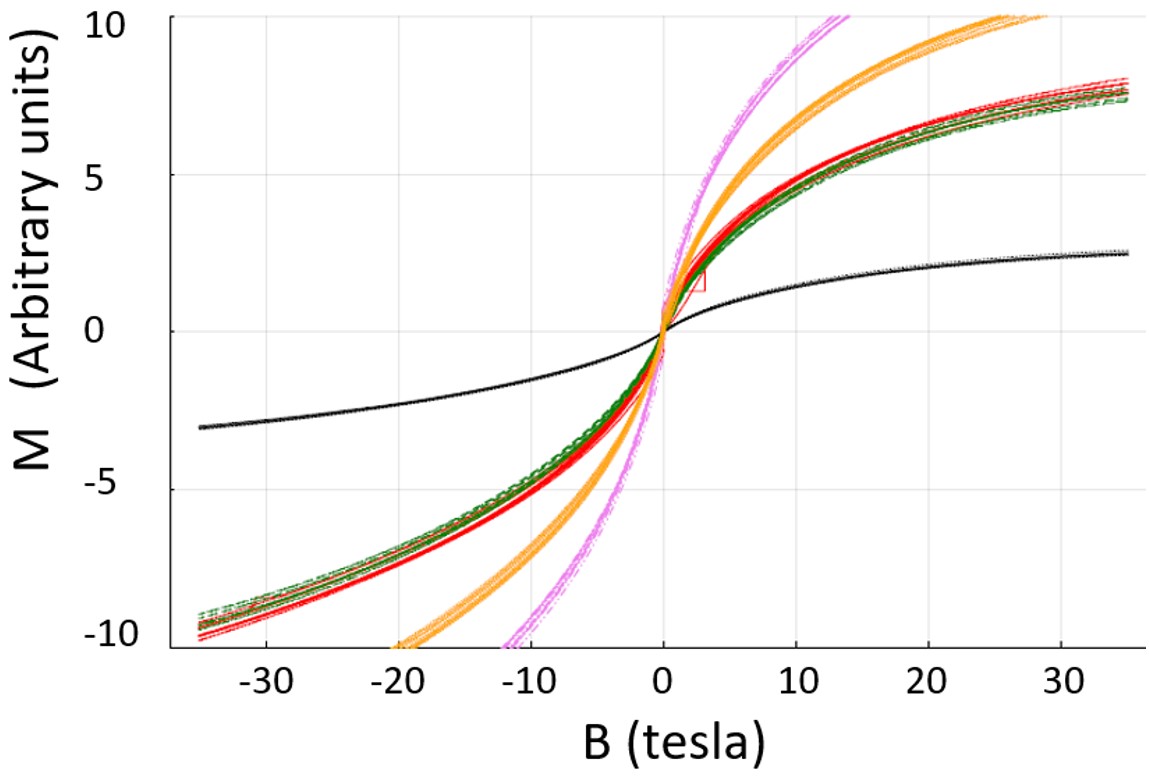}
    \caption{\changes{Magnetization against the applied field for five CuMn alloys.  The Mn concentrations are 2.6\% (violet), 4\% (gold), 7\% (red), 7.92\% (green) and 13.5\% (black).  Each sample has several temperatures plotted with temperatures ranging from at least 5 K above $\Tg$ to about 10 K.  For example, the 13.5\% sample ($\Tg=58$ K) temperature range was from 70 K to 4.3 K and the 2.6\% sample ($\Tg = 19$ K) temperature range was from 25 K to 4.3 K. Data from \citet{dahlberg:25}.}}  
    \label{fig:dahlberg-55}
    \end{figure}}
\changes{With the development of high magnetic field facilities such as the High Magnetic Field Laboratory at Florida State University, it is possible to fully saturate a dilute spin glass that could give insight into the interactions between spins.  There has been a preliminary unpublished effort on CuMn to explore such.  The surprise is there is minimal temperature dependence in the $M$ vs $H$ data for the five CuMn samples investigated with concentrations varying from 2.6\% Mn to 13.5\% Mn (see Fig.~\ref{fig:dahlberg-55} and previous related work in Fig.~\ref{fig:tholence_tournier}).  This includes temperatures from at least 5 K above $\Tg$ to the order of 10 K.  This work is in its infancy with an extensions of measurements on much lower Mn concentrations in progress.}

\paragraph*{Dynamics of spin glasses using aging.}
The ac susceptibility has been seen to be a powerful tool to investigate aging, rejuvenation and memory.  It is not often discussed that these intrinsic spin-glass features disappear with increasing the frequency of ac susceptibility measurements.  At least in the case of the recent work on CuMn it was found that aging and therefore rejuvenation and memory were unmeasurable at frequencies on the order of 10~Hz.  Considering this in terms of an Arrhenius law, it implies there is a minimum energy barrier relevant to the spin-glass state.  What has not been investigated is the frequency dependence of aging as a function of temperature.  This would require measurements of both the real and imaginary parts of the susceptibility.  For example, if one interprets the disappearance of aging as a measure of a minimum energy barrier then at the disappearance of aging the imaginary part would go to zero since, without transitioning over energy barriers, there would be no dissipation. 

An opposite perspective would take note of the absence of rejuvenation and memory for measurement frequencies greater than 10 Hz; that is, for response times faster than $10^{-1}$~s.  Because of ultrametric symmetry, the response time of the vast majority of states is set by the highest barrier created in a waiting time $\tw$.  ``Asking'' the spin glass to respond more rapidly than that time (\emph{i.e.}, shorter times) would involve states with lower barrier heights, but their number is exponentially smaller than those at the largest barrier height.  Concomitantly, the response at a higher frequency would be exponentially smaller.

To test this perspective, one needs make measurements at different $\tw$.  For longer $\tw$, the response time would be longer, so that one would have to go to lower frequencies to observe rejuvenation and memory.  


\section{Summary and Conclusions}\label{sect:summary}
The authors have attempted to inform the reader of this review of the early history of spin glasses, the subsequent quantitative developments, both theoretical and experimental, coalescing into our current understanding of their many facets.  We feel it is important to illustrate the dynamics of this ``useless'' material because of its remarkable general importance across a wide spectrum of scientific research. Indeed, the 2024 physics Nobel prize went in part to J.J. Hopfield for his application of the structure of spin-glass dynamics to neural networks.

We have led the reader through the very early experimental manifestations of spin glasses, well before their general nature was understood.  The crucial experiments were presented that defined the nature of the spin glass.  This was followed by a description of the various models developed to elucidate the nature of the spin-glass state.  We did not include many important ancillary developments associated with spin-glass dynamics because of space limitations, but rather attempted to focus on a perhaps narrow perspective, but one that we believe provides a rather complete picture within its boundaries.

We hope that the reader will be encouraged by our presentation to join with us in this pursuit of Anderson's cornucopia of opportunities.

\section*{Acknowledgments}
\changes{The authors thank Jean-Philippe Bouchaud and the anonymous referees for their helpful suggestions.}
This work was partially supported by Ministerio de Ciencia, Innovación y Universidades (MICIU, Spain), Agencia Estatal de Investigación \changes{(AEI, Spain, MCIN/AEI/10.13039/501100011033)}, and European Regional Development Fund (ERDF, A way of making Europe) through Grants no. PID2020-112936GB-I00 and PID2022-136374NB-C21.
EM acknowledges the hospitality of Mark Bowick at UCSB (USA), where part of this work was done, and the support by grant NSF PHY-2309135 to the Kavli Institute for Theoretical Physics (KITP). His research was supported by the 2021 FIS (Fondo Italiano per la Scienza) funding scheme (FIS783 - SMaC - Statistical Mechanics and Complexity) and the 2022 PRIN funding scheme (2022LMHTET - Complexity, disorder and fluctuations: spin glass physics and beyond) from Italian MUR (Ministry of University and Research). EDD and RLO were supported by the U.S. Department of Energy, Office of Science, Basic Energy Sciences, Materials Science and Engineering Division, under Award No. DE-SC0013599.  The single CuMn crystals were grown by Dr. D.L. Schlagel at Ames Laboratory, which is operated by the U.S. DOE by Iowa State University under Contract No. DE-AC02-07CH11358. IGAP and FRT acknowledge the support of ICSC - Italian Research Center on High-Performance Computing, Big Data, and Quantum Computing, funded by the European Union - NextGenerationEU.

\bibliographystyle{apsrmp4-2}
%

\end{document}